\documentclass[12pt]{article} 

\usepackage[reqno]{amsmath}
\usepackage{amssymb,amsthm,graphicx,url,verbatim,longtable,accents,nicefrac}
\allowdisplaybreaks
\usepackage{times,subcaption,dcolumn,booktabs,setspace,soul,latexsym,wasysym,titling}
\usepackage[export]{adjustbox}
\usepackage[encapsulated]{CJK}
\usepackage[utf8]{inputenc}
\usepackage{enumitem}

\usepackage{caption}
\usepackage[natbib=true,backend=biber,style=authoryear,citestyle=authoryear,ibidtracker=false,maxcitenames=2,uniquename=false,uniquelist=false]{biblatex}
\DeclareDelimFormat[cbx@textcite]{nameyeardelim}{\addspace}
\AtEveryBibitem{\clearlist{language}}

\makeatletter
\def\@biblabel#1{}
\makeatother

\usepackage[all]{xy}
\usepackage[letterpaper,margin=1in]{geometry}
\newcolumntype{.}{D{.}{.}{-1}}\newcolumntype{d}[1]{D{.}{.}{#1}}
\usepackage[dvipsnames,usenames]{color}\definecolor{spot}{rgb}{0.6,0,0}

\graphicspath{{figs/}}

\newcommand{\E}{\mathbb{E}}
\newcommand{\Var}{\operatorname{Var}}

\newcommand{\Cov}{\operatorname{Cov}}
\newcommand{\Cor}{\operatorname{Cor}}
\newcommand{\IV}{\operatorname{IV}}

\usepackage[titletoc,title]{appendix}
\newtheorem{proposition}{Proposition}

\newcommand{\plim}{\operatornamewithlimits{plim}}

\newcommand{\tildX}{\tilde{X}}

\usepackage{algorithmicx}
\usepackage{algorithm} 

\algrenewcommand\alglinenumber[1]{
    {\sf\footnotesize\addfontfeatures{Colour=888888,Numbers=Monospaced}#1}}
\usepackage[noend]{algpseudocode}

\newcommand{\titl}{
  Correcting Bias When Using Latent Regressors}

\usepackage[pdftex, bookmarksopen=true, bookmarksnumbered=true,
pdfstartview=FitH, breaklinks=true, urlbordercolor={0 1 0},
citebordercolor={0 0 1}, colorlinks=true, citecolor=spot,
linkcolor=spot, urlcolor=spot, 
pdfauthor={Connor T.\ Jerzak, Stephen A.\ Jessee},
pdftitle={\titl}]{hyperref}

\newcommand{\blind}{0} 

\ifnum\blind=0
\title{\titl\thanks{
Forthcoming in \textit{Political Analysis}. Authors are listed in alphabetical order.
We thank Devin Caughey, Guilherme Duarte, Robert Kubinec, Jeff Lewis, Umberto Mignozzetti, Jacob Montgomery, Aaron Pancost, Erik Snowberg, Zeynep Somer-Topcu, Chris Warshaw, and participants at MPSA, PolMeth, and the University of Texas Political Methodology Workshop for helpful comments. We thank Major Valls for excellent research assistance. Package link: 
\href{https://github.com/cjerzak/lpmec-software}{\url{GitHub.com/cjerzak/lpmec-software}}.
}}
\author{
Connor T.\ Jerzak\thanks{
Assistant Professor, Government Department, the University of Texas at Austin, email:
\href{mailto:connor.jerzak@austin.utexas.edu}{\texttt{connor.jerzak@austin.utexas.edu}} URL:
\href{https://www.connorjerzak.com}{\texttt{ConnorJerzak.com}} ORCID: 0000-0003-1914-8905.
}
\and
Stephen A.\ Jessee\thanks{Professor, Government Department, the University of Texas at Austin, email:
\href{mailto:sjessee@utexas.edu}{\texttt{sjessee@utexas.edu}} URL: \url{https://laits.utexas.edu/\~sjessee/} ORCID: 0000-0002-3937-6123.
}%
}
\fi

\ifnum\blind=1
\title{\titl}
\fi

\newcommand{\SoftwareURL}{
\ifnum\blind=0
\href{https://github.com/cjerzak/lpmec-software}{\url{GitHub.com/cjerzak/lpmec-software}}
\fi
\ifnum\blind=1
\url{GitHub.com/blinded}
\fi
}

\newcommand{\TutorialURL}{
\ifnum\blind=0
\texttt{GitHub.com/cjerzak/lpmec-software/blob/main/tutorial/IntroTutorial.R?raw=true}
\fi
\ifnum\blind=1
\url{GitHub.com/blinded}
\fi
}

\begin{document}
\newpage 
\maketitle
\begin{abstract}
 \noindent
Many core concepts in political science are latent and therefore can only be measured with error. Measurement error in a predictor attenuates slope coefficient estimates in regression, often biasing them toward zero. We show that widely used strategies for correcting attenuation bias---including instrumental variables and the method of composition---are themselves biased when applied to latent regressors, often performing worse than simple regression ignoring the measurement error altogether. We derive a correlation-based correction using split‑sample measurement strategies. Rather than assuming a particular estimation strategy for the latent trait, our approach is modular and can be easily deployed with a wide variety of latent variable measurement strategies, including additive score, factor, or machine learning models; it requires no joint estimation while yielding consistent slopes under standard assumptions. Simulations and applications show stronger relationships after our correction, sometimes by as much as 50\%. Open‑source software implements the procedure. Results underscore that latent predictors demand tailored error correction; otherwise, conventional practice can exacerbate bias.
  \vspace{0.1cm}
\\ \textbf{\noindent Word count: } 5,991
\vspace{0.1cm}
\end{abstract}

\singlespacing
\newpage 

\section*{Introduction}

Many key concepts in political science, such as political knowledge, ideology, and democracy, are fundamentally unobservable and can only be estimated based on sets of observed indicators. A large body of research relies on measures of these latent variables as predictors in regression models to test theories and make empirical claims. For example, studies have employed proxies for political knowledge to explain patterns of political participation \citep{carpini1996americans}, used estimates of individuals' ideology to predict vote choice and other political behaviors \citep{tausanovitch2013measuring,jessee2009spatial}, and used democracy indices to assess the relationship between democratic institutions and economic growth  \citep{acemoglu2019democracy} or the risk of conflict \citep{democracy2001}.

Because latent variables are unobserved and must be inferred from indicators, any empirical estimate of them (e.g., an index score or latent‑trait estimate) inevitably contains measurement error. While the problems of measurement error and attenuation bias are widely recognized \citep{fuller2009measurement,carroll2006measurement,benoit2009treating, pancost2022measuring, fong2021machine}, the additional challenges posed by latent predictors are often underemphasized in applied work. Indeed, a survey of top political science journals finds that while a growing number of recent articles mention latent variables or measurement error, a small and relatively constant number discuss the two in combination (see Appendix V). 

In this paper, we examine the bias introduced by using estimated latent variables as predictors and present methods to mitigate it under specific assumptions. We build on previous work, including \citet{Achen_1975} and \citet{ansolabehere2008strength}, while deriving new results and introducing tools that account for key features of latent variables and their estimation. We show that two commonly used techniques for adjusting regression estimates with latent predictors can and often do exacerbate bias, sometimes markedly. Most notably, the method of composition \citep{treier2008democracy}, which is the most prominent method used in this area, typically makes the bias due to measurement error \emph{worse} than it would be if the measurement error had simply been ignored altogether.

Building on split-sample methods \citep{angrist1995split}, we use multiple estimates of latent traits based on disjoint sets of indicators to correct standard OLS and 2SLS estimators, which are each biased in opposing directions in the latent predictor case. The method is modular—compatible with a wide range of latent trait estimation approaches and allows for measurement approaches that do not specify a full statistical model for the indicators as a function of the latent trait (e.g., additive/averaging scores, factor models, and machine‑learning approaches).

In applications, we show that accounting for latent-predictor measurement error can shift estimates by as much as 50\%. But in other cases, measurement error may be so mild that corrections are negligible. We also present theory and simulation results. These show that with latent predictors, the standard OLS estimator (ignoring measurement error in the latent predictor) is biased toward zero, and that previously used methods for addressing measurement error can result in even more severely biased estimates. By contrast, our corrected estimator greatly reduces bias and produces estimates similar to those obtained using the more involved full joint estimation approach. Finally, we provide open-source software---an R package, \texttt{lpmec}---for implementing these methods, helping applied researchers account for measurement error when using latent predictors in practice.

\section{Measurement Error and Bias in Linear Regression}

\subsection{Standard Classical Measurement Error}

We define the \emph{standard (non-latent) measurement-error} case as the scenario where the regressor $X$ has an intrinsic scale but is observed with additive noise. Here, and throughout, our parameter of interest is the slope $\beta_X$ in a linear regression of an outcome $Y$ on a predictor $X$:
\begin{equation} 
Y_i = \beta_0 + X_i \beta_X + \varepsilon_i.
\label{eq:regression}
\end{equation} 
We use $\beta_X$ to denote the true regression coefficient that we seek to estimate, and $X$ is the true predictor variable, measured without error; $i$ indices are suppressed where unambiguous. 

Under the classical measurement error setup, the true predictor $X$ is not directly observed, but we instead observe  
\begin{equation}
\tilde{X}=X+U,
\label{eq:measurement-error}
\end{equation}
where $\E[U]=0$, $U$ is independent of both $X$ and $\varepsilon$, and $\Var(U) = \sigma_{U}^2$ so that $\Var(\tildX)=\Var(X)+\Var(U)$. When estimating a linear regression using noisy $\tildX$, rather than $X$ itself, as our predictor, the least squares estimate of $\beta_{X}$ is biased toward zero according to the standard measurement error attenuation factor \citep{fuller2009measurement,bound2001measurement}: 
\begin{equation}
\label{eq:StandardCaseAttenuation}
\plim\hat{\beta}_{\tildX}=\frac{\sigma^2_{X}}{\sigma^2_{X} + \sigma^2_{U}} \; \beta_X = \lambda \; \beta_X,
\end{equation} 
where $\lambda \equiv \nicefrac{\sigma^2_X}{(\sigma^2_X+\sigma^2_U)}$ is called the attenuation factor. 

In the case of multivariate regression that includes covariates $C$ in addition to $X$, which is measured with error, all slope coefficients will typically be biased unless $X$ is uncorrelated with the other covariates (see \cite{carroll2006measurement}, Chapter 3). The attenuation factor for the slope coefficient on the variable measured with error becomes 
\begin{equation}
\label{eq:attenuation-x-multivar}
\lambda_1 = \frac{\sigma^2_{X|C}}{\sigma^2_{X|C} + \sigma^2_U}
\end{equation}
where $\sigma^2_{X|C}$ is the residual variance from regressing $X$ on $C$. Note that $\lambda_1 \leq \lambda$ with the equality holding only when $X$ is uncorrelated with all other covariates. Measurement error in $X$ will also tend to bias coefficient estimates for $C$, with this bias becoming more severe as the variance of the measurement error in $X$ increases. In the case of non-classical measurement error, these results are important but less straightforward to derive,  and often require simulation-based adjustment (see \cite{carroll2006measurement} p. 49).

\subsection{Measurement Error in Latent Predictors}

When a predictor $X$ is latent and must be estimated, measures of $X$ necessarily include some uncertainty. Additionally, its scale must be defined (\textit{identified}), and as a consequence, regression coefficients are attenuated by a different factor than in the standard measurement error case above (see Table \ref{tab:classical-vs-latent}). To understand this, note that one reason classical measurement error in $X$ biases coefficient estimates is that it changes the scale of the regressor, with the variance of the observed $\tildX$ being larger than that of the true \(X\). When estimating latent traits, however---unless the scale is identified such that the variance of the estimates equals the variance of the true $X$ plus the variance of the typically unknown measurement error---then the resulting attenuation bias will differ from the result in Equation \ref{eq:StandardCaseAttenuation}. 

To show how identification rescaling changes measurement-error bias, we start with the noisy measure $\tildX$ in Equation \ref{eq:measurement-error} and then impose the usual latent-trait normalization: mean zero, unit variance, and a fixed direction. Together, this identifies the scale in the common one-dimensional setting.\footnote{See \citet{rivers2003identification} for a formal discussion.} As is typically done in latent traits estimation, we assume that the estimates of the latent trait have the same identifying restriction, i.e., $\hat{X}$ has been estimated or transformed so that it also has mean 0 and standard deviation 1 in the sample, with the same directional restriction. The fact that this identification restriction applies both to the true latent trait and to whatever point estimates of the latent trait are used in subsequent analyses is typically overlooked in applied work, but it is central to understanding how measurement error operates in the case of latent predictors. 

Formally, we assume that our point estimates of a latent variable predictor are
\begin{equation}
\label{eq:xhat-identified}
\hat{X}=\frac{\tilde{X} - \text{mean}(\tilde{X})}{\text{sd}(\tilde{X})},
\end{equation}
which are the transformed (i.e., identified) set of estimates of the latent trait. Importantly, researchers do not observe the intermediate $\tildX$ in this case, since this would require knowing the measurement error variance. These observed, identified estimates of $X$, which we call $\hat{X}$, can be represented as an affine transformation of $\tildX$, which implies that the estimated slope coefficient will be changed by a compensating linear transformation. 

\begin{table}[t]
\centering
\caption{Classical measurement error versus latent-predictor measurement error.}
\label{tab:classical-vs-latent}
\begin{tabular}{p{0.26\linewidth}p{0.34\linewidth}p{0.34\linewidth}}
\toprule
 & \textbf{Observable predictor with classical error} & \textbf{Estimated latent predictor} \\
\midrule
\textsc{Target Regressor} & $X$ has an intrinsic scale & $X$ has an identified scale, e.g. $\E[X]=0$, $\Var(X)=1$ \\
\hfill \\ 
\textsc{Observed Regressor} & $\tilde X = X+U$ & $\hat X=[\tilde X-\text{mean}(\tilde X)]/\text{sd}(\tilde X)$ \\
\hfill \\ 
\textsc{Variance} & $\Var(\tilde X)=\Var(X)+\Var(U)$ & $\Var(\hat X)=1$ by construction \\
\hfill \\ 
\textsc{OLS Attenuation} & $\beta_X$ is multiplied by $1/(1+\sigma_U^2)$ when $\Var(X)=1$ & $\beta_X$ is multiplied by $1/\sqrt{1+\sigma_U^2}$ \\
\hfill \\ 
\textsc{Consequence for IV} &  IV can correct classical attenuation &  IV over-corrects unless the identifying rescaling is addressed \\
\bottomrule
\end{tabular}
\end{table}

In the latent-predictor case, identification sets $\sigma^2_X=1$. Because the regression uses $\hat{X}=\tilde{X}/\sqrt{1+\sigma^2_U}$ up to centering, which affects only the intercept, the slope on $\hat{X}$ equals the slope on $\tilde{X}$ multiplied by $\sqrt{1+\sigma^2_U}$:
\begin{align}\label{eq:latent-attenuation-factor}
\plim \hat{\beta}_{\hat{X}} = \frac{1}{\sqrt{1 + \sigma_U^2}} \beta_X.
\end{align}
In other words, the slope coefficient estimate in the latent predictor case is still attenuated, but not as severely as in the standard case since \(0 < \frac{1}{1 + \sigma_U^2} < \frac{1}{\sqrt{1 + \sigma_U^2}} < 1\) whenever \(\sigma_U^2 > 0\). 

As shown in Appendix Section~\ref{s:alternative-identification-restrictions}, the logic of our results below applies to other identifying restrictions (e.g., setting two observations' latent trait values at -1 and 1, respectively, or even assuming different identifying restrictions for the true latent variable and for its estimates) in a straightforward way. Since mean 0, standard deviation 1 restrictions are common in the literature and also produce easily interpretable measures, we adopt them in all of our discussions below.

\section{Existing Approaches and Their Limitations}

\subsection{No Adjustment}

The most common approach to dealing with measurement error in latent predictors has been to simply ignore the problem and use point estimates of the latent trait as the predictor without any further adjustment. In cases where measurement error is small relative to the variance of the predictor, using this standard approach may be reasonable. But with significant measurement error, slope estimates will be biased toward zero, sometimes dramatically so.

\subsection{Full Joint Estimation}

When attenuation from treating estimated latent scores as error-free is a concern, a natural solution is to estimate the measurement and outcome models jointly \citep{richardsongilks1993}. Examples include Bayesian factor--outcome models \citep{kubinec_2019,battaglia2024inference} and SEM/LISREL-style approaches. Joint estimation is attractive because the outcome and indicators inform inference about the latent trait within a single model. 

This joint approach requires a full statistical model for the indicators as a function of the latent trait and also for the outcome as a function of the latent trait. While modern computational tools, including \texttt{brms} \citep{burkner2017brms}, have made some forms of joint modeling easier, estimation can still be somewhat computationally demanding and, more importantly, potentially difficult for ``black‑box'' scores (e.g., PCA, additive indices, or ML/text embeddings) that lack an explicit likelihood or one that can be tractably deployed via MCMC. Because joint estimation is the gold-standard benchmark whenever it is feasible, we treat it as the reference point for evaluating other methods.

\subsection{Instrumental Variables}

When full joint estimation is infeasible, a standard errors-in-variables workaround is instrumental variables (IV), which aims to recover $\beta_X$ by isolating variation due to the signal rather than measurement noise (cf. \citet{gillard2006historical}). In the classical setting with $\tilde X=X+U$, IV/2SLS corrects attenuation by using an instrument $Z$ that satisfies: (i) $\Cov(X,Z)\neq 0$ (relevance); (ii) $\Cov(Z,U)=0$; and (iii) $\Cov(Z,\varepsilon)=0$ (exclusion). Under these conditions, 2SLS effectively leverages the part of the observed regressor that co-moves with $Z$---the component attributable to $X$, not $U$---to identify the target slope.\footnote{Although we focus on measurement error here, IV can also address simultaneity and omitted-variable bias (see \cite{pancost2022measuring}).}

This logic is appealing in latent-variable applications because one can often construct internal instruments based on different sets of indicators for the same latent variable. In the classical errors-in-variables case, this ``two-measures'' strategy can plausibly satisfy the IV conditions: the alternate measure is correlated with $X$ and, if built from independent information, can be approximately uncorrelated with the measurement error in the primary measure (and unrelated to $\varepsilon$).

However, applying this standard IV logic directly to identified latent scores $\hat X$ is generally inappropriate. Unlike $\tilde X$, $\hat X$ is a standardized (identified) transformation of $\tilde X$: identification forces $\Var(\hat X)$ to match $\Var(X)$ by construction. As a result, unadjusted IV tends to ``correct'' as if operating on $\tilde X$ even though the analyst is regressing on $\hat X$. Thus, unadjusted IV is biased away from zero: for any nonzero \(\beta_X\),
\[
\plim \hat{\beta}_{\IV,\hat X^{(1)} \mid \hat X^{(2)}}
=
\beta_X\sqrt{1+\sigma_U^2},
\]
so IV inflates the coefficient magnitude whenever \(\sigma_U^2>0\) (Appendix I).

\subsection{Method of Composition (MOC)}

A different---and in applied latent-trait work, common---approach is the method of composition (MOC), which aims to propagate uncertainty from the measurement stage, where $X$ is estimated based on a set of observed indicators $W$, into the second-stage regression. In the two-stage MOC implementation commonly applied with latent predictors, one first draws a latent-score vector from $p(X\mid W)$, then re-anchors it to a common identified scale and regresses $Y$ on the resulting standardized draw. We denote the pre-standardized draw by $\tilde X_{\mathrm{MOC}}^{(t)}$ and the standardized regressor by $\hat X_{\mathrm{MOC}}^{(t)}$. Crucially, these draws condition on $W$ but not on $Y$, even though $Y$ contains information about $X$ whenever $\beta_X\neq0$.\footnote{\citet[p.~216]{treier2008democracy} state this as the assumption, $p(x\mid Z)=p(x\mid w,y,Z)$; here their $Z$ corresponds to our indicator matrix $W$ and, suppressing additional covariates, this is the restriction $p(X\mid W,Y)=p(X\mid W)$.}

Many high-impact works have used MOC with the goal of appropriately accounting for the uncertainty in latent predictors \citep[e.g.][]{bernhard2020parties, caughey2018policy, caughey2019public, jessee2010partisan, kastellec2015polarizing, lax2019party, TAI_HU_SOLT_2024}. The Varieties of Democracy (V-Dem) project even provides, in addition to point estimates of their democracy measures, a series of posterior samples so that researchers can employ MOC in subsequent analyses \citep{pemstein2018varieties}.\footnote{Our critique concerns using MOC to fit second-stage regression models with latent predictors. V-Dem's use of MOC to combine individual measures into more general indices and account for the resulting uncertainty \citep{mcmann2016strategies} does not necessarily suffer from the same issues.} 

More concretely, MOC proceeds according to the following steps for posterior draws $t=1,\ldots,T$: 
\begin{enumerate}[leftmargin=*]
    \item Fit the measurement model using $W$ alone and retain pre-standardized posterior latent-trait draws from $p(X\mid W)$, denoted $\tilde X_{\mathrm{MOC}}^{(t)}$ for $t=1,\ldots,T$.
    \item Re-anchor each posterior draw to a common identified scale before outcome analysis, since the latent trait is identified only up to location, sign, and scale; denote the standardized draw by $\hat X_{\mathrm{MOC}}^{(t)}$. 
    \item For each draw, estimate $Y_i=\beta_0^{(t)}+\beta_{\mathrm{MOC}}^{(t)}\hat X_{\mathrm{MOC},i}^{(t)}+\varepsilon_i$, recording the slope estimate $\hat{\beta}_{\mathrm{MOC}}^{(t)}$ and its estimated variance-covariance matrix $\hat V^{(t)}$.
    \item Summarize either the draw-specific slopes $\{\hat{\beta}_{\mathrm{MOC}}^{(t)}\}_{t=1}^T$ or coefficient draws from a normal approximation centered at $\hat{\beta}_{\mathrm{MOC}}^{(t)}$ with covariance $\hat V^{(t)}$.
\end{enumerate}

This two-stage sampling-and-refitting procedure is typically justified as a way to ``propagate" measurement-stage uncertainty forward into the regression and avoid overstating precision. However, when the regressor is measured with error, the dominant first‑order problem is attenuation bias: measurement error pulls slope estimates toward zero.

Indeed, this use of MOC averages over a set of latent-score draws, each of which will typically contain more random variation than the posterior-mean latent scores. In the stylized additive model of Appendix I, the point estimates are equal to the true latent traits plus some measurement error, and the individual draws from the posterior distribution for $X$ given $W$ are approximated as pre-standardized draws around the point estimates:
\[
\tilde X_{\mathrm{MOC}}^{(t)}=\tilde X+\xi^{(t)}=X+U_{\mathrm{MOC}}+\xi^{(t)},
\]
where $\tilde X=X+U_{\mathrm{MOC}}$ is the pre-standardized point estimate and $\xi^{(t)}$ is an additional source of mean-zero draw variation. 

Appendix I formalizes this intuition in a stylized additive-error representation in which a standardized MOC draw can be written as
\[
\hat X_{\mathrm{MOC}}^{(t)}
=
\frac{X+U_{\mathrm{MOC}}+\xi^{(t)}}{\sqrt{1+\sigma_{U,\mathrm{MOC}}^2+\sigma_\xi^2}},
\]
with $\xi^{(t)}$ mean-zero and orthogonal to $X$, $U_{\mathrm{MOC}}$, and $\varepsilon$. In that setting, MOC draw-level regressions are attenuated by $1/\sqrt{1+\sigma_{U,\mathrm{MOC}}^2+\sigma_\xi^2}$ rather than $1/\sqrt{1+\sigma_{U,\mathrm{MOC}}^2}$. Thus, when posterior draws are used in this two-stage way, draw-level variation can worsen attenuation relative to using posterior means. In the multivariate case, this additional draw-level noise can also exacerbate biases on the slope estimates for other predictors that are correlated with $X$.

\subsection{Multiple Over-Imputation}

A compromise between MOC and full joint estimation is multiple over-imputation \citep{rubin1988overview,blackwell2017unified}. Multiple over-imputation proceeds in this setting as follows:
\begin{enumerate}[leftmargin=*]
    \item Fit the measurement model to $W$ alone and obtain an approximation to $p(X_i\mid W)$, e.g., via $\mu_i=\E(X_i\mid W)$ and $\sigma_i=\operatorname{SD}(X_i\mid W)$.
    \item Form an imputation data set containing the observed outcome $Y$ and a latent-score column. The latent-score entries are over-imputed rather than fixed at $\mu_i$, with $(\mu_i,\sigma_i)$ supplied as unit-specific priors ($Y$ included in the imputation model).
    \item For each completed data set $r=1,\ldots,M_{\mathrm{imp}}$, re-anchor the completed latent-score vector $X_{\mathrm{imp}}^{(r)}$ to the identified latent scale, fit $Y_i=\beta_0^{(r)}+\beta_X^{(r)}X_{\mathrm{imp},i}^{(r)}+\varepsilon_i$, and pool the resulting slope estimates across completed data sets.
\end{enumerate}
The second and third steps make multiple over-imputation more similar to full joint estimation MOC because the outcome can enter the imputation step. It remains distinct from full joint estimation, however, because the original measurement model is not re-estimated jointly with the outcome equation; instead, the first-stage measurement distribution is approximated through unit-specific summaries supplied to the imputation model. Consequently, the completed-data regressions should generally be interpreted as estimates from a staged approximation to $p(X\mid W,Y)$, not as draws from the full joint posterior $p(\beta_0,\beta_X,X\mid W,Y)$, except under strong specification and compatibility conditions. We therefore include multiple over-imputation as a comparison method in the simulations below, with details in Appendix~\ref{s:Bayesian}.

\section{A Correlation-Corrected Estimator Using Split Indicators}

Taken together, these limitations---na\"ive IV's tendency to over-correct once identification rescaling is imposed, and MOC's tendency to add noise and worsen attenuation---motivate a different tack: estimate the latent-scale measurement error and then undo the identification-aware attenuation implied by Equation~\ref{eq:latent-attenuation-factor}. To do this, we need a way to estimate $\sigma^2_U$, since the attenuation factor is determined by this measurement-error variance. Importantly, this is measurement error variance on the scale of the true latent trait $X$, not the transformed (standardized) scale for the estimates of the latent trait that we can observe. 

Indeed, one way to learn about $\sigma_U^2$---and therefore about the latent-predictor attenuation factor in Equation \ref{eq:latent-attenuation-factor}---is to obtain (at least) two independent measurements of the same latent trait $X$ on the \emph{latent scale} (i.e., before the identification/standardization step that produces $\hat{X}$). Intuitively, if two measurements of $X$ give similar values for each observation, then the measurement error variance $\sigma_U^2$ must be small; if the two measurements often disagree substantially, then $\sigma_U^2$ must be larger.

In indicator-based latent-trait settings, we can usually construct such repeated measures by partitioning the indicators into two disjoint sets and estimating the trait separately within each set. For example, with 10 indicators for a latent trait $X$, we could use 5 indicators to obtain estimates $\hat X^{(1)}$ and the other 5 to obtain estimates $\hat X^{(2)}$. If we split the indicators evenly, it may be reasonable to assume that the two resulting estimates have the same measurement-error variance.\footnote{For an odd number of indicators, consider all partitions where the two sets differ by one item.} Following the notation above, let $\tilde X^{(1)}$ and $\tilde X^{(2)}$ denote the two \emph{unobserved} (unidentified) measurements on the same scale as the true latent trait:
\[
\tilde X^{(1)} = X + U^{(1)} \qquad\text{and}\qquad \tilde X^{(2)} = X + U^{(2)},
\]
where we assume $U^{(1)}$ and $U^{(2)}$ each have mean zero and the same variance $\sigma_U^2$, and are independent of each other and of $X$.\footnote{This independence assumption is similar to the conditional independence assumption commonly used in latent traits measurement strategies, including many item response models.} Finally, let $\hat X^{(1)}$ and $\hat X^{(2)}$ denote the corresponding \emph{identified} versions of these split-indicator estimates---i.e., the quantities we actually observe after imposing the usual mean-zero, unit-variance normalization (consistent with Equation \ref{eq:xhat-identified}).
In this section, $\sigma_U^2$ denotes the common error variance of the split-specific unstandardized measures $\tilde X^{(1)}$ and $\tilde X^{(2)}$; when the full-indicator score is discussed separately, we write $\sigma_{U,\mathrm{full}}^2$.

Although we do not observe $\tilde X^{(1)}$ or $\tilde X^{(2)}$ directly, their correlation is identical to the observed split-score correlation, $\operatorname{Cor}(\hat X^{(1)},\hat X^{(2)})$, because each observed split score is an affine (mean/variance) transformation of its latent-scale counterpart. The idea of using correlation between multiple sets of estimates to correct for measurement error has been around since at least \citet{spearman1904proof}. In that context, we can readily show that
\begin{equation} 
\begin{aligned}
    \rho\equiv\Cor(\hat X^{(1)}, \hat X^{(2)}) &= \Cor(\tilde X^{(1)}, \tilde X^{(2)}) 
    = \frac{\Cov(\tilde X^{(1)},\tilde X^{(2)})}{\sqrt{\sigma^2_{\tilde X^{(1)}}\sigma^2_{\tilde X^{(2)}}}} \\
    &= \frac{\Cov(X+U^{(1)}, X+U^{(2)})}{\sqrt{(\sigma^2_X+\sigma^2_U)^2}} \\
    &= \frac{\sigma^2_X}{\sigma^2_X+\sigma^2_U}.
\end{aligned}
\label{eq:correlation-correction}
\end{equation}
In our setup, $\sigma^2_X=1$ by identification so that $\rho=\nicefrac{1}{(1+\sigma^2_U)}$. Thus $\rho$ equals the classical attenuation factor $\lambda$ only in this standardized equal split-error special case.

In our latent variable predictor setup, the attenuation factor for the slope coefficient is $\frac{1}{\sqrt{1+\sigma^2_{U}}}$ (see Equation \ref{eq:latent-attenuation-factor}). In the standard observable variable case, it is $\frac{1}{1+\sigma^2_{U}}$ (see Equation \ref{eq:StandardCaseAttenuation}). This suggests using the square root of the sample correlation between the two separate estimates of $X$ to correct attenuation in the estimated slope coefficient, $\beta_X$.\footnote{In some situations, this split-half may not be appropriate. For example, if indicators are designed to tap different aspects of a latent variable rather than with internal consistency in mind (e.g., \cite{gosling2003very}), it may be preferable to consider test-retest reliability.} To do this, we first estimate the standard linear regression coefficient using $\hat X^{(1)}$ as our predictor to obtain $\hat{\beta}_{\hat X^{(1)}}$ and then correct this estimate:
\begin{equation}
    \hat{\beta}^{*}_{\hat X^{(1)}} = \frac{\hat{\beta}_{\hat X^{(1)}}}{\sqrt{\widehat{\Cor}(\hat X^{(1)}, \hat X^{(2)})}}. 
\end{equation}
We define this square-root correction for partitions with a positive estimated split-half correlation; nonpositive split-half correlations are flagged or discarded as weak-measurement partitions.

This correction will rely on several maintained assumptions. First, the target parameter is the coefficient on the identified latent trait in the linear regression of $Y$ on $X$, with $\E[X]=0$ and $\Var(X)=1$. Second, each split-indicator estimate can be represented, before standardization, as $\tilde X^{(j)}=X+U^{(j)}$ for $j\in\{1,2\}$, where the split-specific errors are mean-zero and independent of $X$ and of the regression disturbance. Third, the two split-specific errors are independent of one another, which is the split-indicator analog of conditional independence assumptions common in latent-trait models. Fourth, for the simplest derivation, the split-specific error variances are equal; Appendix II evaluates the sensitivity of the estimator to violations of this equality. Finally, the split estimates must be sufficiently correlated for the correction to be empirically informative since low split-half correlations are a special case of the ``weak instruments problem'' and should be interpreted as a diagnostic indicating that the indicators contain limited information about the latent trait. Under these assumptions, we find Proposition \ref{prop:bivariate-consistency}. 
\begin{proposition}[Consistency of the split-indicator correction]
\label{prop:bivariate-consistency}
Suppose $\{(Y_i,X_i,U_i^{(1)},U_i^{(2)})\}_{i=1}^n$ are i.i.d. with finite second
moments and
\[
Y_i=\beta_0+X_i\beta_X+\varepsilon_i,\qquad
\E[X_i]=0,\qquad \Var(X_i)=1,\qquad \Cov(X_i,\varepsilon_i)=0.
\]
For a fixed, sign-aligned split of the indicators, suppose
\[
\tilde X_i^{(j)}=X_i+U_i^{(j)},\qquad j\in\{1,2\},
\]
where $\E[U_i^{(j)}]=0$, $\Var(U_i^{(1)})=\Var(U_i^{(2)})=\sigma_U^2$,
$U_i^{(1)}$ and $U_i^{(2)}$ are independent of one another, and
$(U_i^{(1)},U_i^{(2)})$ are independent of $(X_i,\varepsilon_i)$. Let $\hat X_i^{(j)}$
denote the sample-standardized version of $\tilde X_i^{(j)}$, with directions aligned
to the target latent direction and across splits. Let $\hat X^{(s)}=(\hat X_{1}^{(s)},\ldots,\hat X_{n}^{(s)})^\top$ for $s\in\{1,2\}$. If the population split-half
reliability $\rho=\Cor(\tilde X_i^{(1)},\tilde X_i^{(2)})$, equivalently the probability
limit of $\widehat{\Cor}(\hat X^{(1)},\hat X^{(2)})$, is positive, then
\[
\hat{\beta}^{*}_{\hat X^{(1)}}
=
\frac{\hat{\beta}_{\hat X^{(1)}}}{\sqrt{\widehat{\Cor}(\hat X^{(1)},\hat X^{(2)})}}
\quad\text{satisfies}\quad
\plim \hat{\beta}^{*}_{\hat X^{(1)}}=\beta_X.
\]
The correction is applied on the event
$\widehat{\Cor}(\hat X^{(1)},\hat X^{(2)})>0$, whose probability tends to one under
$\rho>0$.
The same result holds using $\hat X^{(2)}$ as the predictor. See Appendix I for proof. 
\end{proposition} 

\noindent Here and below (suppressing again dependence on $i$ in notation), we use the asterisk superscript to denote our correlation-corrected estimators. Because we have no reason to privilege either split-half estimate, we compute both $\hat{\beta}^{*}_{\hat X^{(1)}}$ and $\hat{\beta}^{*}_{\hat X^{(2)}}$ and average them within a partition.\footnote{We do not use $\hat X$---the point estimates of the latent trait using all of the indicators---because our correction factor is estimated from the split-half reliability of $\hat X^{(1)}$ and $\hat X^{(2)}$, not the full-indicator score; applying the same correction to $\hat X$ would require additional assumptions or estimation.} This averaging reduces variability from the choice of partition.\footnote{Our averaging approach, and in particular the equivalent averaged instrumental variables (IV) approach we describe in the next section, shares commonalities with \cite{gillenexperimenting}.} Averaging across the two split-half estimates also makes the method less sensitive to unequal error variances. If one split yields a correction that is too small, the other tends to yield a correction that is too large. As we show in Appendix II, this averaging mitigates but does not eliminate imbalance bias when $\sigma_{U,1}^2 \neq \sigma_{U,2}^2$; over plausible ranges of $\sigma_{U,1}^2$ and $\sigma_{U,2}^2$, the remaining multiplicative factor appears close to 1.

Finally, because there are typically multiple ways to partition the indicators into two equally sized sets, we define $\hat{\beta}^{*}$ as the median over partitions after first averaging $\hat{\beta}^{*}_{\hat X^{(1)}}$ and $\hat{\beta}^{*}_{\hat X^{(2)}}$ within each partition. The median is less sensitive to unstable partitions with low split-half correlations. For a fixed finite set of valid partitions, the median-aggregated estimator remains root-$n$ consistent when all partition-specific estimators share the same probability limit, but its limiting distribution is generally the median of a joint Gaussian vector rather than Gaussian. We therefore use a row bootstrap to capture uncertainty in both the coefficient and the correction factor. Because the finite-partition median is nonsmooth, additional complexities are involved (see Appendix I), but Winsorized means are another reasonable, mathematically smoother aggregation approach, yielding similar performance (Appendix VIII). 

The same split-indicator approach above can be applied to the IV framework. In our case, with two different measures of the latent trait, one split estimate can be used as the regressor and the other as the instrument. As shown above, standard IV adjustments for measurement error will \emph{over}-correct for attenuation, resulting in estimates that are inflated, i.e., biased away from zero by a factor of $\sqrt{1+\sigma^2_U}$. This suggests using the correlation between the two sets of latent trait estimates to adjust the standard IV estimate. In contrast to the bivariate OLS case considered above, in the IV case we multiply, rather than divide, our slope estimate by $\sqrt{\widehat{\Cor}(\hat X^{(1)}, \hat X^{(2)})}$ for partitions with positive estimated split-half correlations. 

In fact, as shown in Appendix I, the correlation-corrected IV and correlation-corrected OLS estimators are equivalent in the bivariate case. Despite this equivalence, one advantage of the IV approach is its ease of extension to multivariate regression (see Appendix I). In multivariate regressions, the error-prone variable may be the main predictor or merely a ``control''; in either case, failing to account for its measurement error can bias the estimated coefficients on all predictors in the regression.

Appendix I formalizes this extension. In brief, consider a regression of $Y$ on a vector of latent predictors $X$ and observed covariates $C$. Standard IV correction will produce consistent slope estimates for $C$, but estimates for $\boldsymbol{\beta}_X$ will be biased away from zero. Appendix Section~\ref{s:multivariate-corrected-iv} derives this result formally, showing that the population IV coefficient on the standardized latent predictor converges to $D\boldsymbol{\beta}_X$ rather than $\boldsymbol{\beta}_X$, where
\[
D=\operatorname{diag}\!\left(d_1,\ldots,d_q\right),
\qquad d_k=\sqrt{1+\sigma_{U,k}^2}.
\]
If each latent predictor has two split-indicator estimates, the first split can be used as the endogenous regressor and the second as its instrument, with $C$ included as exogenous controls in both stages. Under the same split-error independence assumptions, outcome-disturbance exogeneity, and standard rank conditions, the uncorrected IV coefficient on each standardized latent predictor is inflated by the corresponding scale factor $d_k$. Multiplying the IV coefficient for latent component $k$ by $\sqrt{\widehat{\Cor}(\hat X_{\cdot k}^{(1)},\hat X_{\cdot k}^{(2)})}$ therefore recovers the coefficient on the identified latent scale when the componentwise split-half correlations are positive. 

As with other IV applications, estimates can become unstable when the split-half correlations are close to zero or nonpositive. Therefore, we recommend conducting standard diagnostics. In the bivariate case, this includes the first-stage $F$-statistic from regressing $\hat X^{(1)}$ on $\hat X^{(2)}$ and any observed covariates. With multiple latent components, researchers should inspect componentwise first-stage diagnostics, e.g., for each latent component $k$, the partial first-stage $F$-statistic from regressing $\hat X_{\cdot k}^{(1)}$ on $\hat X_{\cdot k}^{(2)}$ and $C$, or an appropriate multivariate weak-IV diagnostic when $q>1$. In most latent variable applications, the indicators would be expected to have strong correlations since they are assumed to be driven by the same latent variable, so these problems seem unlikely.

\section{Illustration: Political Knowledge Predicting Duty to Vote}\label{s:Application}

To make the identification and measurement-error issues concrete, we use the 2024 ANES to compare several approaches in a familiar latent-predictor setting. The latent predictor is political knowledge, measured with a four-item battery covering federal spending, Senate term length, and party control of the House and Senate; the outcome is a seven-point item measuring whether respondents view voting as a duty or a choice. We code knowledge as the number of correct answers and standardize it to mean zero and variance one, matching the identifying restriction used above. Table~\ref{tab:knowledge-voteduty-simpleplot-reduced} shows estimates of $\beta_X$ using several different approaches. The all-item unadjusted OLS estimate is 0.419. As shown in Appendix Figure~\ref{fig:knowledge-voteduty-simple}, the uncorrected OLS slope estimates increase as more indicators are used---a pattern consistent with attenuation. 

\begin{table}[!htbp]
\centering
\caption{Political knowledge and duty-to-vote estimator comparisons.}
\label{tab:knowledge-voteduty-simpleplot-reduced}
\begin{tabular}{llr}
\toprule
\textbf{Measure} & \textbf{Estimator} & \textbf{Slope} \\
\midrule
Standardized \% Correct & OLS & 0.419 \\
  & Uncorrected IV & 1.143 \\
  & Correlation-Corrected & 0.620 \\
\addlinespace
IRT & OLS & 0.420 \\
  & Uncorrected IV & 1.139 \\
  & MOC & 0.200 \\
  & Correlation-Corrected & 0.624 \\
  & Full Joint & 0.668 \\
\bottomrule
\end{tabular}
\par\vspace{0.5em}
\parbox{0.92\linewidth}{\footnotesize \textit{Note:} $N = 3,059$. Slopes are from regressions of the seven-point duty-to-vote measure on standardized political knowledge estimates; higher outcome values indicate a stronger sense that voting is a duty.}
\end{table}

Applying the split-indicator correction to the four knowledge items yields $\hat{\beta}^{*}=$0.62, roughly 50\% larger than the unadjusted estimate. The result is similar when knowledge is estimated with an item-response model rather than an additive score: the uncorrected OLS slope estimate is 0.42 and the correlation-corrected estimate is 0.624. By contrast, the MOC yields a slope of 0.20, while uncorrected split-indicator IV produces estimates above 1 for both approaches of estimating knowledge. 

\section{Comparing the Performance of Estimators with Simulation}\label{sec:Simulation}

To compare the performance of our correlation-corrected estimator with other approaches, in this section, we conduct Monte Carlo simulations, generating simulated data ($W$ and $Y$) with parameter values set to point estimates from real data. Specifically, we rely on the 2016 Cooperative Congressional Election Study (CCES) ($N${=}38{,}851), a survey that includes 30 agree/disagree policy items that we use as indicators to estimate respondent ideology ($X$). We assume a commonly used item response model 
\begin{equation}\label{eq:idealpoint}
 \Pr(W_{ij}=1)=\Phi(X_i \eta_j+\alpha_j),
 \end{equation}
and our dependent variable ($Y$) in the regression model is reported presidential vote (0=Democratic, 1=Republican; other responses dropped).

Although the type of data and the approach to estimating the latent trait here are quite common in the literature, both the number of observations ($N$) and the number of indicators ($M$) in the CCES are much larger than in most latent trait applications in political science and related literatures. This allows us to assess the performance of estimators across a range of $N$ and $M$ values, from small to large. Specifically, we set parameter values equal to estimates from the full CCES data. We then simulate data for $N\in\{500,1000,5000\}$ and $M\in\{6,12,30\}$, sampling observations and indicators randomly for each value of $N$ and $M$. We then estimate $\beta_X$ with these simulated datasets using each of several approaches discussed above (see Appendix III for more details).

Figure~\ref{fig:CombinedVaryND} summarizes the simulation
results. Uncorrected OLS is attenuated toward zero, and the bias is
largest when the latent predictor is estimated from few
indicators. The uncorrected IV estimator (not shown; see Appendix Figure~\ref{fig:BayesianComparison}) moves in the opposite direction, over-correcting, often dramatically, because it ignores the identifying rescaling of the latent scores. MOC also performs poorly in this design because it fits outcome regressions to noisier posterior draws of the latent predictor. MO performs well with a large number of indicators, but struggles somewhat with smaller $M$. By contrast, the split-indicator correction has low bias across the simulated values of $N$ and $M$ and performs similarly to the full joint benchmark. The gains are most substantively important in the common setting where latent traits are measured with modest numbers of indicators. Details are in Appendix~\ref{s:SimDetails}--\ref{s:Bayesian}, which also reports simulation results with multiple latent predictors and covariates, where the same patterns hold (with the correction reducing bias and RMSE relative to uncorrected OLS).

\begin{figure}[H]
\centering
\includegraphics[width=0.88\linewidth]{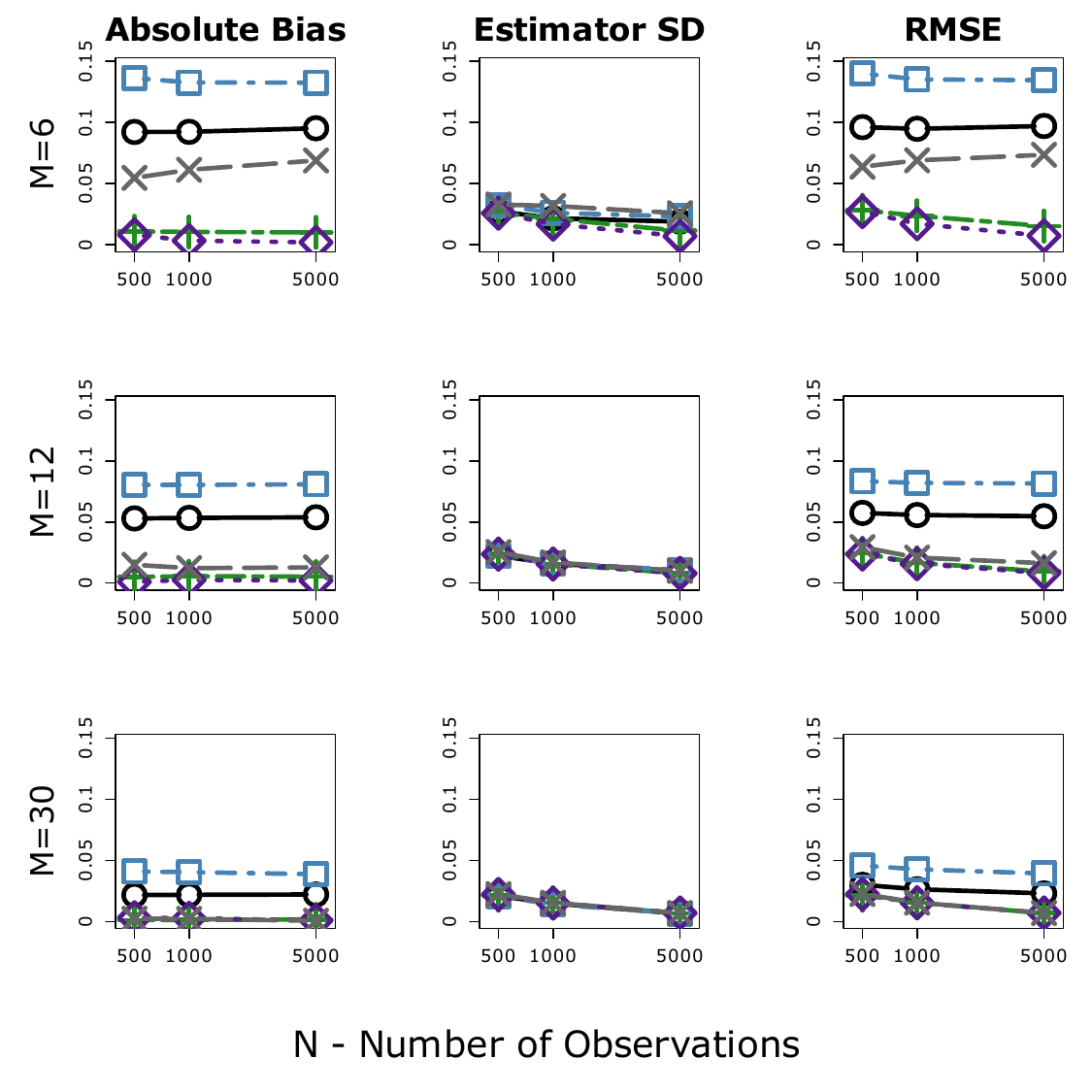}
\par\vspace{0.25em}
\includegraphics[width=0.95\linewidth]{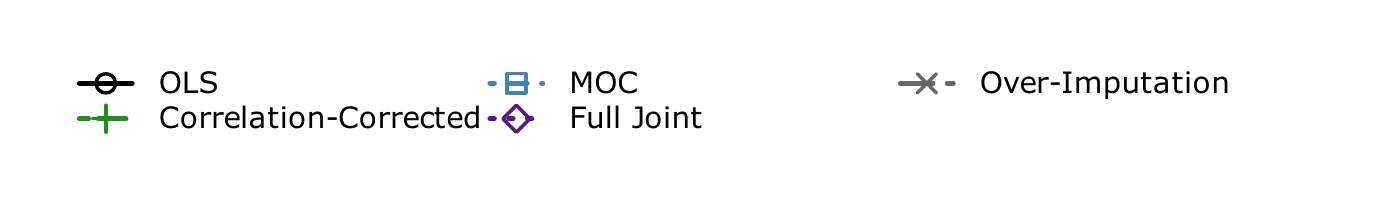}
\caption{Simulation results varying $N$ and $M$. Results compare OLS, method of composition (MOC), over-imputation, full joint estimation, and the split-indicator correlation-corrected estimator across the simulation designs described in Section~\ref{sec:Simulation}.}
\label{fig:CombinedVaryND}
\end{figure}

Appendix Table \ref{tab:cOLSvOLS_ANES-Partisanship_Part1} shows the percentage change in coefficient estimates when moving from standard OLS to our corrected estimator, for a finer appendix grid of $N$ and $M$ values. While these differences are relatively small when using 30 indicators, we see changes around 20\% when $M=12$, 25\% when $M=10$, and 35\% when $M=6$. Because estimating latent traits from smaller indicator sets is far more common in practice than using 30 or more indicators, these results for $M$ between 6 and 12 are often the empirically relevant regime. 

Appendix III, Section~\ref{s:Coverage}, reports a coverage analysis: the corrected estimator improves coverage over unadjusted approaches, though it undercovers for small $M$; for large $N$ and $M$, it reaches nominal coverage, unlike the unadjusted alternatives.

\section{Guidance, Limitations, Extensions}

These results show that latent predictors create distinctive measurement-error problems: standard regressions attenuate slopes, MOC here worsens bias, and standard IV tends to overcorrect. Our correlation-corrected estimator reduces bias, performs comparably to full joint estimation, and works with any latent-trait estimator, including methods without explicit models or likelihoods.

Applied researchers using latent predictors should check standard estimates against our corrected estimates. In some cases, we may expect attenuation bias to be negligible. For example, estimates of legislator ideology based on hundreds of roll call votes tend to be estimated extremely precisely, at least conditional on the assumed measurement models \citep{lewis2004measuring}. NOMINATE legislator ideology scores estimated separately with the first and second sessions of the 117th U.S. Senate are correlated at 0.99, implying that standard OLS regressions using NOMINATE scores would be attenuated by less than one percent. Appendix \ref{s:DemGrowth} presents an application of democracy predicting economic growth that shows less than a 1\% change in the estimated coefficient after using our corrected estimator.

The estimated coefficients on other latent predictors may show non-trivial, albeit not huge, changes when applying our corrected methods. For example, \citet{ansolabehere2008strength} finds that correlations between various issue scales when estimated from the two halves of ANES items range from 0.61 to 0.84. This implies that standard OLS slope estimates in these situations may need to be adjusted upwards in magnitude by between 9 and 28 percent. 

But in other applications for latent predictors, attenuation bias may be much more severe. It is quite common for latent traits to be estimated based on a small number of indicators. The political knowledge example presented above demonstrates that correcting standard regression estimates using our proposed method can yield a roughly 50\% increase in the estimated slope coefficient. With 10–20 policy questions for ideology, the corrected estimates are often 10–25\% larger than standard estimates. When using 6 indicators, correcting estimates can increase their magnitude by over one-third.

Future work should optimize indicator splits to reduce estimator variance, and extend the correction beyond linear models. \hfill $\blacksquare$

\newpage

\printbibliography

\addtocontents{toc}{\protect\setcounter{tocdepth}{1}}
\tableofcontents

\singlespacing

\renewcommand{\thefigure}{A.I.\arabic{figure}}
\setcounter{figure}{0}  
\renewcommand{\thetable}{A.I.\arabic{table}}
\setcounter{table}{0}  
\renewcommand{\theequation}{A.I.\arabic{equation}}
\setcounter{equation}{0}
\renewcommand{\thesection}{A.I.\arabic{section}}
\setcounter{section}{1}

\renewcommand{\thesubsection}{A.I.\arabic{subsection}}
\setcounter{subsection}{0}

\renewcommand{\thesubsubsection}{A.I.\arabic{subsection}.\arabic{subsubsection}}
\setcounter{subsubsection}{0}

\section*{Appendix I: Technical Derivations and Estimator Properties}\addcontentsline{toc}{section}{Appendix I: Technical Derivations and Estimator Properties}

This section illustrates how the standard IV correction for measurement error yields estimates biased upward in magnitude in the case of latent predictors. It also demonstrates how, in the split-indicator setup, the correlation between $\hat X^{(1)}$ and $\hat X^{(2)}$ can be used to adjust the IV estimator to provide consistent estimates. Finally, it is shown that the correlation-corrected IV estimate is equivalent to the correlation-corrected OLS estimate in the bivariate case before moving on to an analysis of multivariate corrected IV, a stylized analysis of the Method of Composition, and a discussion of alternative identification restrictions.

\subsection{Consistency of the Bivariate Corrected Estimator}

Assume the true latent trait $X$ has zero mean and unit variance. Additionally, assume the split-specific measurement errors $U^{(1)}$ and $U^{(2)}$ are mean-zero, mutually independent, possess a common variance $\sigma_U^2$, and are independent of both $X$ and the regression error $\varepsilon$.

For a fixed split, write $\tilde X^{(j)}=X+U^{(j)}$ for $j\in\{1,2\}$. After aligning each split to the
target latent direction, consider the population-standardized representation
\[
\hat X^{(j)}=\frac{\tilde X^{(j)}}{\sqrt{1+\sigma_U^2}}.
\]
The implemented scores in a given sample of units $i\in\{1,...,N\}$ use sample centering and scaling,
\[
\hat X_i^{(j)}=\frac{\tilde X_i^{(j)}-\bar{\tilde X}^{(j)}}{s_{\tilde X^{(j)}}}.
\]
Under the maintained i.i.d. finite-second-moment assumptions, $\bar{\tilde X}^{(j)}\overset{p}{\to}0$ and $s_{\tilde X^{(j)}}\overset{p}{\to}\sqrt{1+\sigma_U^2}$, so replacing the sample-standardized score by the population-standardized expression changes the relevant OLS slopes and split-half correlations only by $o_p(1)$, meaning their difference converges in probability to zero as the sample size approaches infinity. Then (assuming the bivariate OLS regression model $Y = \beta_0 + X\beta_X + \varepsilon$):
\begin{align*} 
\plim \hat\beta_{\hat X^{(1)}}
&=
\frac{\Cov(Y,\hat X^{(1)})}{\Var(\hat X^{(1)})}
=
\Cov\!\left(\beta_0+X\beta_X+\varepsilon,\;\;
\frac{X+U^{(1)}}{\sqrt{1+\sigma_U^2}}\right)
\\ &=
\frac{\beta_X\Var(X)+\Cov(\varepsilon,X)+\Cov(\varepsilon,U^{(1)})}
{\sqrt{1+\sigma_U^2}}
\\ &=
\frac{\beta_X}{\sqrt{1+\sigma_U^2}},
\end{align*} 
where the second equality uses $\Var(\hat{X}^{(1)})=1$ and final equality uses $\Var(X)=1$, $\Cov(\varepsilon, X)=0$, and independence of $U^{(1)}$ from $(X,\varepsilon)$.
Similarly,
\[
\rho\equiv\Cor(\tilde X^{(1)},\tilde X^{(2)})
=
\frac{
\overbrace{\Cov(X+U^{(1)},X+U^{(2)})}^{\substack{=\Var(X)=1,\\\text{cross terms again vanish}}}
}
{\sqrt{\Var(X+U^{(1)})\Var(X+U^{(2)})}}
=
\frac{1}{1+\sigma_U^2}.
\]
Equivalently, $\rho$ is the probability limit of the sample correlation
$\widehat{\Cor}(\hat X^{(1)},\hat X^{(2)})$ of the standardized split scores.
Therefore,
\[
\plim
\frac{\hat\beta_{\hat X^{(1)}}}
{\sqrt{\widehat{\Cor}(\hat X^{(1)},\hat X^{(2)})}}
=
\frac{\beta_X/\sqrt{1+\sigma_U^2}}
{\sqrt{1/(1+\sigma_U^2)}}
=
\beta_X,
\]
By demonstrating that the probability of the adjusted slope estimate equals the true parameter $\beta_X$, the equation establishes that this correlation-corrected approach yields a statistically consistent estimator.

Now, now, another common method for correcting for measurement error in a predictor is the use of instrumental variables (IV). Although its use today is often motivated by causal identification, IV can also, under certain assumptions, produce consistent estimates of slope coefficients when a predictor is measured with error.\footnote{Although we focus on measurement error here, note that in a causal framework, IV can be used to address measurement error as well as simultaneity and omitted variable bias under suitable assumptions (see \cite{pancost2022measuring}).} 

In the standard observable predictor setting, if $X$ is measured with error following the setup above and we seek to estimate the slope coefficient on $X$ in a linear regression predicting some other variable $Y$, we can correct for the resulting attenuation bias by leveraging an instrument $Z$ with the following properties: (i) $\Cov(X,Z)\neq 0$; (ii) $\Cov(Z,U)=0$; (iii) $\Cov(Z,\varepsilon)=0$. In other words, a valid instrument in this setting must be correlated with the true predictor $X$ and be uncorrelated with both the measurement error in $X$ and the error term in the regression predicting $Y$ with $X$. 

Following \citet{angrist2009mostly}, the two-stage least squares (2SLS) IV estimator of the slope if we were able to observe $\tildX$ could be written as (in this observable [non-latent] predictor set up): 
\begin{equation} \label{eq:IV-tilde-x-def}
\begin{aligned}
\hat{\beta}_{\IV,\tildX}
&=
\frac{\hat{\beta}_{Y\sim Z}}{\hat{\beta}_{\tildX\sim Z}}
=
\frac{\nicefrac{\Cov(Y,Z)}{\Var(Z)}}{\nicefrac{\Cov(\tildX,Z)}{\Var(Z)}}
=
\frac{\Cov(X\beta_X+\varepsilon,Z)}{\Cov(X+U,Z)} \\
&=
\underbrace{
\frac{\beta_X\Cov(X,Z)+\Cov(\varepsilon,Z)}
     {\Cov(X,Z)+\Cov(U,Z)}
}_{\substack{\text{under valid-IV orthogonality}\\
\Cov(\varepsilon,Z)=\Cov(U,Z)=0}}
=
\frac{\beta_X\Cov(X,Z)}{\Cov(X,Z)},
\end{aligned}
\end{equation}
since by assumption $\Cov(Z, \varepsilon)=0$ and $\Cov(Z,U)=0$. Hence it follows that in this observable predictor setup 
\begin{equation}  \label{eq:IV-tilde-x-plim}
    \plim \hat{\beta}_{\IV,\tildX} = \beta_X.
\end{equation}

Conversely, in the latent predictor case, we can obtain a suitable instrumental variable by splitting the indicators used to estimate the latent variable into two sets and estimating the latent trait separately using each of these two sets of indicators, calling these estimates $\hat X^{(1)}$ and $\hat X^{(2)}$ as above. We can then use $\hat X^{(1)}$ as the regressor and $\hat X^{(2)}$ as the instrument, or vice-versa. 

Because these are both measures of the same latent trait, they should be highly correlated with each other (which we can also check empirically) and also highly correlated with the true latent trait $X$ (which we cannot check empirically).\footnote{In the latent predictor case, different indicators of the same latent trait, and especially scales constructed based on multiple indicators, are typically highly correlated with each other \citep[see e.g.][]{ansolabehere2008strength}, suggesting that the weak instruments problem should rarely be severe in practice.} 

We assume that the split-specific measurement errors are orthogonal across splits on the unstandardized latent scale.\footnote{This is analogous to the conditional independence assumption used in many latent traits models, i.e., that conditional on $X$, the indicators are independent of each other.} For the direction with $\hat X^{(1)}$ as the predictor and $\hat X^{(2)}$ as the instrument, the IV conditions require
\[
\Cov(\hat X^{(2)},U^{(1)})=0,\qquad
\Cov(\hat X^{(2)},\varepsilon)=0,\qquad
\Cov(\hat X^{(2)},\hat X^{(1)})\neq 0.
\]
The reverse direction requires the analogous conditions
$\Cov(\hat X^{(1)},U^{(2)})=0$, $\Cov(\hat X^{(1)},\varepsilon)=0$, and relevance of
$\hat X^{(1)}$ for $\hat X^{(2)}$. With $\tilde X^{(1)}=X+U^{(1)}$ and
$\tilde X^{(2)}=X+U^{(2)}$, these conditions follow from the stronger maintained model
in which $\Cov(X,\varepsilon)=0$ and $(U^{(1)},U^{(2)})$ are mean-zero, mutually
independent, and independent of $(X,\varepsilon)$. Here, $U^{(1)}$ is the error in
the unstandardized true-scale measure $\tilde X^{(1)}$, not $X-\hat X^{(1)}$, because
$\hat X^{(1)}$ has been rescaled by the identifying normalization.

In contrast to the IV setup with an observable predictor, our estimates $\hat X^{(1)}$ and $\hat X^{(2)}$ are not on the same scale as $X$ itself. If we had $\tildX^{(1)}=X+U^{(1)}$ to use as our predictor, then the IV results in Equations \ref{eq:IV-tilde-x-def} and \ref{eq:IV-tilde-x-plim} would hold. In our case, however, we use as our predictor $\hat X^{(1)}=\frac{\tildX^{(1)}}{\sqrt{1+\sigma^2_{U}}}$, i.e. a rescaled version of the unstandardized split measure since it is standardized by dividing by $\sigma_{\tildX^{(1)}}=\sqrt{1+\sigma_{U}^2}$. This implies that our IV estimate using $\hat X^{(1)}$ will have the following relationship with the IV estimate using $\tildX^{(1)}$:
\begin{equation}\label{eq:CorrectedIV}
\begin{aligned}
\hat{\beta}_{\IV,\hat X^{(1)} \mid \hat X^{(2)}}
&=
\frac{\Cov(Y,\hat X^{(2)})}
     {\Cov(\hat X^{(1)},\hat X^{(2)})} \\
&=
\hat{\beta}_{\IV, \tildX^{(1)} \mid \tildX^{(2)}}
\underbrace{
\frac{1/\sqrt{1+\sigma_U^2}}{1/(1+\sigma_U^2)}
}_{\substack{=\sqrt{1+\sigma_U^2}\\\text{} }} \\
&=
\hat{\beta}_{\IV, \tildX^{(1)} \mid \tildX^{(2)}}\sqrt{1+\sigma_U^2}.
\end{aligned}
\end{equation}
and hence
\begin{equation}
    \plim \hat{\beta}_{\IV,\hat X^{(1)} \mid \hat X^{(2)}} = \beta_X\sqrt{1+\sigma_{U}^2}.
\end{equation}
In this way, using IV to correct for measurement error in this latent predictor setting actually \emph{over}-corrects for the attenuation bias in standard linear regression, resulting in coefficient estimates that are inflated in magnitude on average since $\sqrt{1+\sigma_{U}^2} > 1$ whenever $\sigma_{U}^2>0$, i.e., whenever there is any measurement error in the predictor. Note that here, we are using the notation $A \mid B$ to indicate that $A$ is our predictor and $B$ is our instrumental variable.

This implies a further correction that can be accomplished by multiplying the IV estimate by the reciprocal of this factor, so that 
\begin{equation}
    \hat{\beta}^*_{\IV, \hat X^{(1)} \mid \hat X^{(2)}}=\frac{\hat{\beta}_{\IV,\hat X^{(1)} \mid \hat X^{(2)}}}{\sqrt{1+\sigma_{U}^2}}=\hat{\beta}_{\IV,\hat X^{(1)} \mid \hat X^{(2)}}\sqrt{\rho}.
\end{equation}
With this correction in place, the same asymptotic logic as earlier implies  
\begin{equation}
    \plim  \hat{\beta}^*_{\IV, \hat X^{(1)} \mid \hat X^{(2)}}=\beta_X.
\end{equation}
Hence, by adjusting the split-half IV estimate to account for the identifying restriction used for the estimated latent traits (which itself is based on the measurement error variance $\sigma^2_U$), we can obtain a consistent estimate of the linear regression slope coefficient on the true latent $X$ predicting $Y$.

Of course, the correction used in $ \hat{\beta}^*_{\IV, \hat X^{(1)} \mid \hat X^{(2)}}$ is based on the variance of the measurement error (on the true latent variable's scale) $\sigma_{U}^2$ which we do not typically know. As discussed in the previous section, however, we use these replicate measures, $\hat X^{(1)}$ and $\hat X^{(2)}$, to learn about $\sigma^2_U$ and, consequently, about the split-half reliability $\rho$. Since the sample correlation between these replicate measurements consistently estimates $\rho$, we can calculate our corrected estimator as the product of the IV slope estimate and the square root of the correlation:
\begin{equation}\label{IV-cor-corrected}
    \hat{\beta}^*_{\IV, \hat X^{(1)} \mid \hat X^{(2)}}=\hat{\beta}_{\IV,\hat X^{(1)} \mid \hat X^{(2)}} \; \sqrt{\widehat{\Cor}(\hat X^{(1)}, \hat X^{(2)})}.
\end{equation}

Another way of understanding the corrected IV estimator is to note that if we transformed our measures $\hat X^{(1)}$ and $\hat X^{(2)}$ from having variance 1 to having variance $1+\sigma^2_U$, which would effectively mean we are in the standard measurement error setup, then the standard IV estimator would provide consistent estimates.\footnote{We actually would not need to apply this transformation to whichever one is used as the instrument, since any affine transformation of the instrument will cancel out in the expression for the IV estimator.} A convenient way to do this is to estimate $\sigma^2_U$ with $[\nicefrac{1}{\widehat{\Cor}(\hat X^{(1)}, \hat X^{(2)})}-1]$, because because $\widehat{\Cor}(\hat X^{(1)}, \hat X^{(2)})$ consistently estimates $\frac{1}{1+\sigma_U^2}$. Then, note that since we want the variance of our predictor to be $1+\sigma^2_U$ we can multiply it by $\nicefrac{1}{\sqrt{\widehat{\Cor}(\hat X^{(1)}, \hat X^{(2)})}}$ and use this transformed predictor in a standard IV analysis.\footnote{Since the ratio of the uncorrected OLS and uncorrected IV estimators provides information about the split-half reliability $\rho$ in this standardized split model (see \cite{pancost2022measuring}), yet another equivalent estimator can be constructed based on the square root of the product of the baseline OLS and uncorrected IV estimators using the partitioned $\hat X^{(1)}$ and $\hat X^{(2)}$. Since this approach is less intuitive, we do not discuss it here.}
Obviously, multiplying the predictor by $\nicefrac{1}{\sqrt{\widehat{\Cor}(\hat X^{(1)}, \hat X^{(2)})}}$ has the result of dividing the estimated slope coefficient by the same factor, which is equivalent to multiplying it by $\sqrt{\widehat{\Cor}(\hat X^{(1)}, \hat X^{(2)})}$ as in Equation \ref{IV-cor-corrected} above.

Also as above, we may want to calculate our corrected IV estimator over all equally sized partitions of the indicators for estimating $\hat X^{(1)}$ and $\hat X^{(2)}$. For each such partition, take the average of the corrected estimate with $\hat X^{(1)}$ as the regressor and $\hat X^{(2)}$ as the instrument and the corrected estimate with $\hat X^{(2)}$ as the regressor and $\hat X^{(1)}$ as the instrument, then choose as our estimator the median of all such partition averages. In fact, this estimator, which we call $\hat{\beta}^{*}_\text{IV}$, is equivalent to $\hat{\beta}_\text{OLS}^{*}$ (see below). In other words, the correlation-adjusted OLS estimator and the correlation-adjusted IV estimator are equivalent and below we refer to them together as the corrected estimator.

Despite the equivalence shown between $\hat{\beta}_\text{OLS}^{*}$ and $\hat{\beta}_\text{IV}^{*}$ in the bivariate case, one advantage of the IV approach is the ease with which it can be extended to multivariate regression (discussed more later). If one predictor is measured with error, we may either have other ``control'' variables or the predictor measured with error may itself be a control variable in which case adjusting for this may be important to prevent bias in the other coefficient(s) of interest.\footnote{Note in the latter case we may not care about appropriately adjusting the IV coefficient on the predictor that is measured with error.} We may even have multiple predictors measured with error, and believe that we have instruments that are suitable to address this. In both the OLS and IV cases, we employ the bootstrap to assess uncertainty, as this row-level resampling allows us to incorporate uncertainty not only in the coefficient estimate but also in the correction factor itself in finite samples. 

Finally, the framework here assumes classical (mean-zero, independent) measurement error. When non-classical error is a concern, identification can still sometimes be recovered under additional conditions; see, for example, \citet{hu2008instrumental}. In settings where external instruments are unavailable, instruments can sometimes be constructed from heteroskedasticity \citep{lewbel2012using}; these ideas complement our split-indicator strategy.

To see that $\hat{\beta}^{*}_\text{IV}$ is equivalent to $\hat{\beta}_\text{OLS}^{*}$ note that $\hat{\beta}_{\text{OLS}, \hat X^{(1)} \sim \hat X^{(2)}}$, the linear regression
slope coefficient estimate when predicting $\hat X^{(1)}$ with $\hat X^{(2)}$, is
equal to
\[
\widehat{\Cor}(\hat X^{(1)},\hat X^{(2)})
\sqrt{\frac{\Var(\hat X^{(1)})}{\Var(\hat X^{(2)})}}
=
\widehat{\Cor}(\hat X^{(1)},\hat X^{(2)}),
\]
since the variances of both $\hat X^{(1)}$ and $\hat X^{(2)}$ are equal to 1 given that each is standardized. This also implies that $\hat{\beta}_{\text{OLS}, \hat X^{(2)} \sim \hat X^{(1)}}=\widehat{\Cor}(\hat X^{(1)},\hat X^{(2)})$ since $\widehat{\Cor}(\hat X^{(1)},\hat X^{(2)})=\widehat{\Cor}(\hat X^{(2)},\hat X^{(1)})$. Therefore, we can write the uncorrected IV slope coefficient using $\hat X^{(1)}$ as our predictor and $\hat X^{(2)}$ as our instrument as
\begin{align*}
\hat{\beta}_{\text{IV}, \hat X^{(1)} \mid \hat X^{(2)}}
&=
\frac{\hat{\beta}_{\text{OLS}, Y \sim \hat X^{(2)}}}
     {\underbrace{\hat{\beta}_{\text{OLS}, \hat X^{(1)} \sim \hat X^{(2)}}}
     _{\substack{=\widehat{\Cor}(\hat X^{(1)},\hat X^{(2)})\\
     \text{because both split scores have variance }1}}} 
\\ & 
\\ &=
\frac{\hat{\beta}_{\text{OLS}, Y \sim \hat X^{(2)}}}
     {\widehat{\Cor}(\hat X^{(1)},\hat X^{(2)})}.
\end{align*}
and hence we have 
\begin{equation}
\begin{aligned}
    \hat{\beta}^{*}_{\text{IV}, \hat X^{(1)} \mid \hat X^{(2)}} &= \hat{\beta}_{\text{IV}, \hat X^{(1)} \mid \hat X^{(2)}} \; \sqrt{\widehat{\Cor}(\hat X^{(1)}, \hat X^{(2)})} \\
    &= \frac{\hat{\beta}_{\text{OLS}, Y \sim \hat X^{(2)}}}{\widehat{\Cor}(\hat X^{(1)},\hat X^{(2)})} \; \sqrt{\widehat{\Cor}(\hat X^{(1)}, \hat X^{(2)})} \\
     &= \frac{\hat{\beta}_{\text{OLS}, Y \sim \hat X^{(2)}}}{\sqrt{\widehat{\Cor}(\hat X^{(1)}, \hat X^{(2)})}} \\
     &= \hat{\beta}^{*}_{\text{OLS},\hat X^{(2)}}
\end{aligned}
\end{equation}
For a single IV direction, the corrected IV estimator with $\hat X^{(1)}$ as the
regressor and $\hat X^{(2)}$ as the instrument equals the corrected OLS estimator
using $\hat X^{(2)}$ as the predictor. Therefore, the symmetrized within-partition
IV average equals the symmetrized within-partition OLS average.

\noindent \textbf{Asymptotic Normality and Bootstrap Validity.}
Although we have discussed consistency, we can also now consider asymptotic Normality. Fix a partition \(p\). More generally, if the estimator is reported for a finite collection of partitions \(\mathcal P=\{p_1,\ldots,p_P\}\), treat \(\mathcal P\) as fixed and not growing with \(n\); if partitions are randomly generated in implementation, the formal argument is conditional on the realized set of partitions, which is then held fixed. Let \(O_i\) denote the full row-level data for unit \(i\): the outcome, any observed covariates, and the complete indicator vector from which the split-specific latent scores are constructed. Let $\hat\zeta_p$ denote the finite-dimensional vector of sample summaries for partition $p$ (e.g., the split-half correlation and uncorrected moments) with population limit $\zeta_{0p}$, and let $\hat\beta_p^*$ denote the resulting correlation-corrected estimator targeting the true slope $\beta_X$.

Now, assume: (i) \(O_1,\ldots,O_n\) are i.i.d.; (ii) the low-dimensional first- and second-stage summaries entering the corrected estimator have finite \(2+\delta\) moments for some \(\delta>0\) (e.g., finite fourth moments suffice); (iii) for the implemented latent-score procedure on split \(p\), after imposing the normalization and sign conventions, the finite-dimensional vector \(\hat\zeta_p\) of partition-specific quantities entering the correction and second-stage estimator admits a joint asymptotically linear expansion:
specifically: 
\[
\sqrt{n}(\hat\zeta_p-\zeta_{0p})
=
\frac{1}{\sqrt{n}}\sum_{i=1}^n \psi_{\zeta,p}(O_i)+o_p(1),
\qquad
\E[\psi_{\zeta,p}(O_i)]=0,
\qquad
\E[\|\psi_{\zeta,p}(O_i)\|^2]<\infty 
\]
for influence function $\psi_{\zeta,p}(\cdot)$;
(iv) every population split-half correlation entering the correction is positive and bounded away from zero; and (v) the relevant population regression/IV moment matrices are full rank, with sample analogues nonsingular with probability approaching one.

Here, $\psi_{\zeta,p}(\cdot)$ represents the influence function, which isolates and quantifies the individual contribution of a single observation $O_i$ to the overall estimation error. We can justify applying this linear representation because the components of $\hat{\zeta}_p$—specifically, the sample variances, covariances, and first-stage regression moments—are standard Z-estimators (or smooth functions of sample moments). Under the regularity conditions specified earlier (such as finite fourth moments), the theory of M-estimation guarantees that such estimators are regular and asymptotically linear.

Consequently, via a stacked estimating-equation expansion, the complex, non-linear error of $\hat{\zeta}_p$ can be mathematically linearized into a simple sum of independent, mean-zero individual components ($\psi$). This linearization is crucial: by expressing the error as a sum of independent variables with finite variance, we satisfy the conditions for the Central Limit Theorem, guaranteeing that the intermediate ingredients $\hat{\zeta}_p$ converge to a joint multivariate Normal distribution. Because the final corrected estimator $\hat{\beta}_p^*$ is a smooth, continuously differentiable function of these Normally distributed ingredients, we can subsequently invoke the Delta Method to establish its asymptotic Normality.

Notably, these high-level conditions does not require the full first-stage latent-score object itself to be finite-dimensional; unit-specific latent scores or other nuisance components may have dimension growing with \(N\), provided the low-dimensional summaries entering the corrected estimator are root-\(N\) regular. In standard finite-dimensional M/Z-estimation settings, this condition follows from a stacked estimating-equation argument, but we use that only as a sufficient special case rather than as a generic characterization of all latent-score procedures.

Under this set of conditions, the generated-regressor first stage affects the corrected estimator only through \(\hat\zeta_p\). The partition-specific corrected estimator can then be written as
\[
\hat\beta_p^*=\phi_p(\hat\zeta_p),
\]
where \(\phi_p\) is continuously differentiable in a neighborhood of \(\zeta_{0p}\), the population limit of the sample summaries for partition $p$, because  \(\phi_p\) is a smooth algebraic function of sample moments and the population split-half correlation is strictly positive (bounded away from zero). In the bivariate OLS case, \(\phi_p\) is a smooth function of sample variances and covariances; in the IV case, it is a smooth function of the corresponding first-stage, reduced-form, and covariance moments; and in either case the square-root split-half correction is smooth because the population split-half correlation is bounded away from zero. Therefore,
\[
\sqrt{n}(\hat\beta_p^*-\beta_X)
=
\nabla\phi_p(\zeta_{0p})^\top
\sqrt{n}(\hat\zeta_p-\zeta_{0p})
+o_p(1)
\overset{d}{\to}
\mathcal{N}(0,\sigma_p^*),
\]
where $(\sigma_p^*)^2$ denotes the asymptotic variance of the corrected estimator for partition $p$, defined as $\nabla\phi_p(\zeta_{0p})^\top \E[\psi_{\zeta,p}(O_i)\psi_{\zeta,p}(O_i)^\top] \nabla\phi_p(\zeta_{0p})$ \citep{stefanski2002calculus}. Hence, the fixed-partition corrected estimator is root-\(N\) consistent and asymptotically Normal. The same argument applies to vector-valued latent-predictor coefficients by replacing derivatives with Jacobians.

Now, let \(\hat\beta_p^{*,b}\) denote the row-bootstrap analog computed from a bootstrap sample \(O_1^{b},\ldots,O_n^{b}\) drawn with replacement from the empirical distribution of the rows. For the bootstrap to target the sampling distribution of the implemented estimator, each bootstrap replicate must rerun the full procedure using the same fixed partition or fixed finite set of partitions: re-estimate the split-specific latent scores from the bootstrap sample, re-center/re-scale and sign-align them using the same convention as in the original sample, recompute the split-half correlation(s), recompute the corrected regression or IV estimate(s), and then apply the same partition aggregation rule. Holding the latent scores, the split-half correlation, or the correction factor fixed across bootstrap samples would condition on part of the estimator and generally understate uncertainty. If, in addition, the ordinary row bootstrap consistently reproduces the joint law of \(\sqrt{n}(\hat\zeta_p-\zeta_{0p})\) for the implemented latent-score procedure under full re-estimation, the same normalization/sign-alignment convention, and the same realized fixed partition set, then the smooth-functional argument yields bootstrap consistency for any fixed partition:
\[
\sup_t
\left|
\Pr^{*}\!\left\{
\sqrt{n}(\hat\beta_p^{*,b}-\hat\beta_p^*)\le t
\right\}
-
\Pr\!\left\{
\sqrt{n}(\hat\beta_p^*-\beta_X)\le t
\right\}
\right|
\overset{p}{\to}0,
\]
where \(\Pr^{*}(\cdot)\) denotes the bootstrap law conditional on the observed sample. We state bootstrap validity for the fixed-partition estimator as an estimator-specific high-level result, rather than as a generic theorem covering all possible latent-score estimators. Under the same positive-correlation assumption, the event that a valid partition is dropped because its estimated split-half correlation is nonpositive has probability \(o(1)\), so such screening does not affect the first-order argument. If \(\mathcal P=\{p_1,\ldots,p_P\}\) is a fixed finite collection of partitions, stacking the corresponding low-dimensional summaries yields joint asymptotic Normality of \((\hat\beta_{p_1}^*,\ldots,\hat\beta_{p_P}^*)\). Hence any aggregation rule \(g(\hat\beta_{p_1}^*,\ldots,\hat\beta_{p_P}^*)\) that is continuously differentiable at \((\beta_X,\ldots,\beta_X)\)---for example, a fixed weighted average over pre-specified partitions---is also asymptotically Normal, and the same ordinary row bootstrap is valid for that smooth aggregation under the same high-level bootstrap condition.

The finite-partition median aggregiation rule requires a caveat. Let
\[
\hat\beta_{\mathrm{med}}
=
\operatorname{med}\{\hat\beta_{p_1}^*,\ldots,\hat\beta_{p_P}^*\},
\]
with \(P\) fixed. When all valid partitions share the same population limit \(\beta_X\), the median map is evaluated at the non-smooth point \((\beta_X,\ldots,\beta_X)\). Accordingly, the preceding smooth delta-method/bootstrap argument does not by itself justify the ordinary \(n\)-out-of-\(n\) row bootstrap for \(\hat\beta_{\mathrm{med}}\). If
\[
\sqrt{n}\bigl(\hat\beta_{p_1}^*-\beta_X,\ldots,\hat\beta_{p_P}^*-\beta_X\bigr)
\overset{d}{\to}
(Z_1,\ldots,Z_P),
\]
with joint Gaussian limit, then
\[
\sqrt{n}(\hat\beta_{\mathrm{med}}-\beta_X)
\overset{d}{\to}
\operatorname{med}(Z_1,\ldots,Z_P),
\]
up to the convention used when \(P\) is even. This limit is generally non-Gaussian, although the estimator remains root-\(N\) consistent. We therefore use the ordinary row bootstrap for the median-aggregated estimator as a practical resampling approximation to the full implemented algorithm, not as a generic theorem implied by the smooth-bootstrap result above. A fully formal generic route is either to replace the median by a smooth aggregation over a fixed finite set of partitions, for which the ordinary row bootstrap applies directly, or to use subsampling or an \(m\)-out-of-\(n\) bootstrap for the finite-partition median under the additional regularity conditions required for nonsmooth functionals \citep{politis1994large}. For further discussion, see \ref{s:Coverage}.

\subsection{Multivariate Corrected IV}
\label{s:multivariate-corrected-iv}

We now give the multivariate version of the split-indicator IV correction. All variables are taken to be centered and an intercept assumed to be present. Let \(\boldsymbol{\beta}_X\in\mathbb R^q\) and consider
\[
Y_i=X_i^\top\boldsymbol{\beta}_X+C_i^\top\gamma+\varepsilon_i,
\]
where \(X_i=(X_{i1},\ldots,X_{iq})^\top\) is a vector of identified latent predictors and \(C_i\) is a vector of $p$ observed background covariates. The components of \(X_i\) need not be mutually independent, and \(X_i\) may be correlated with \(C_i\). The only scale normalization imposed here is componentwise: \(\Var(X_{ik})=1\) for \(k=1,\ldots,q\), with directions aligned to the target latent directions.

For split \(j\in\{1,2\}\), write the unstandardized split measure as
\[
\tilde X_i^{(j)}=X_i+U_i^{(j)},
\]
and the corresponding identified split score (componentwise standardized to unit variance, but not yet corrected) as
\[
\hat X_i^{(j)}=D^{-1}\tilde X_i^{(j)}=D^{-1}(X_i+U_i^{(j)}),
\qquad
D=\operatorname{diag}(d_1,\ldots,d_q),
\qquad
d_k=\sqrt{1+\sigma_{U,k}^2}.
\]
Here \(\sigma_{U,k}^2=\Var(U_{ik}^{(j)})\), assumed equal across \(j=1,2\) for a given component \(k\). Thus, each split score is standardized component by component to have variance of one. 

Then, let
\[
R_i=\begin{pmatrix}\hat X_i^{(1)}\\ C_i\end{pmatrix},
\qquad
Z_i=\begin{pmatrix}\hat X_i^{(2)}\\ C_i\end{pmatrix}.
\]
The split score \(\hat X_i^{(1)}\) is treated as endogenous, while \(C_i\) is included as an exogenous regressor and therefore also appears in the instrument vector. Assume finite second moments; from the population IV conditions:
\begin{align*}
\E[Z_i\varepsilon_i]&=0 \;\;\textrm{\it (exogeneity)}, 
 \qquad
\E[Z_i\{U_i^{(1)}\}^\top]=0 \;\;\textrm{\it (no cross-split contamination)},
\\ & \qquad
Q_{ZR}\equiv \E[Z_iR_i^\top]\ \text{is nonsingular {\it (relevance condition)}}.
\end{align*}
These conditions follow, for example, if the split errors are mean-zero, mutually independent across splits, independent of \(X_i\), \(C_i\), and \(\varepsilon_i\), and the usual first-stage rank condition holds.

Following the standard IV setup, the population IV coefficient vector, \(\kappa^{IV}=(\delta_X^{IV\top},\gamma_C^{IV\top})^\top\), is the unique solution to the moment condition equation: 
\begin{equation}\label{s:IVMultiMoment}
\E\!\left[Z_i\{Y_i-R_i^\top\kappa\}\right]=\mathbf{0}_{(q + p) \times 1}.
\end{equation}
To identify the population IV estimand, evaluate the moment condition at the parameter vector implied by the latent-score rescaling:
\[
\kappa_0=\begin{pmatrix}D\boldsymbol{\beta}_X\\ \gamma\end{pmatrix}.
\]
When we evaluate the IV residual at 
\(\kappa_0=\bigl((D\boldsymbol{\beta}_X)^\top,\gamma^\top\bigr)^\top\), so that 
\(R_i^\top\kappa_0=\{\hat X_i^{(1)}\}^\top D\boldsymbol{\beta}_X+C_i^\top\gamma\),
\[
\begin{aligned}
Y_i-R_i^\top\kappa_0
&=
X_i^\top\boldsymbol{\beta}_X
+
\underbrace{C_i^\top\gamma-C_i^\top\gamma}_{=\,0}
+\varepsilon_i
-
\{\hat X_i^{(1)}\}^\top D\boldsymbol{\beta}_X \\
&=
X_i^\top\boldsymbol{\beta}_X+\varepsilon_i
-
\underbrace{
\{D^{-1}(X_i+U_i^{(1)})\}^\top D\boldsymbol{\beta}_X
}_{\substack{
= (X_i+U_i^{(1)})^\top D^{-1} D\boldsymbol{\beta}_X=(X_i+U_i^{(1)})^\top\boldsymbol{\beta}_X\\
\text{since }D\text{ is diagonal so }(D^{-1})^{\top} = D^{-1}\text{ and }D^{-1}D=I
}} \\
&=
\underbrace{
X_i^\top\boldsymbol{\beta}_X-(X_i+U_i^{(1)})^\top\boldsymbol{\beta}_X
}_{=\;-\{U_i^{(1)}\}^\top\boldsymbol{\beta}_X}
+\varepsilon_i \\
&=
\varepsilon_i-\{U_i^{(1)}\}^\top\boldsymbol{\beta}_X .
\end{aligned}
\]
Substituting this residual into the IV moment function and using the
exogeneity and no-cross-split-contamination assumptions,
\begin{align*}
\mathbb{E}!\left[Z_i\{Y_i-R_i^\top\kappa_0\}\right]
&=
\E\!\left[Z_i\{\varepsilon_i-\{U_i^{(1)}\}^\top\boldsymbol{\beta}_X\}\right] \\
&=
\underbrace{
\E[Z_i\varepsilon_i]
-
\E[Z_i\{U_i^{(1)}\}^\top]\boldsymbol{\beta}_X
}_{\substack{=\,0\text{ by exogeneity}\\
\text{and no cross-split contamination}}} \\
&=
0.
\end{align*}
Thus, \(\kappa_0\) satisfies the population IV moment condition in
Equation \ref{s:IVMultiMoment}. Since \(Q_{ZR}\), the population instrument--regressor cross-moment matrix, is non-singular under the assumptions outlined above, the population IV solution is unique, so
\[
\delta_X^{IV}=D\boldsymbol{\beta}_X
\qquad\text{and}\qquad
\gamma_C^{IV}=\gamma.
\]
Uncorrected IV therefore recovers the correct coefficient on the observed controls here in the multivariate case, but it recovers \(D\boldsymbol{\beta}_X\), not \(\boldsymbol{\beta}_X\), for the standardized latent predictors. The reason is the scale effect: standard IV would recover \(\boldsymbol{\beta}_X\) if the unstandardized latent-scale regressor \(\tilde X_i^{(1)}=X_i+U_i^{(1)}\) were used, but the observed latent score is \(D^{-1}\tilde X_i^{(1)}\), so its IV coefficient is inflated componentwise by \(D\).

It remains to estimate the scale correction \(D^{-1}\). For each component \(k\), define the population split-half correlation
\[
\rho_k\equiv\Cor(\hat X_{ik}^{(1)},\hat X_{ik}^{(2)}),
\]
where the correlation is taken across observations \(i\). Because correlations are unchanged by componentwise affine standardization,
\[
\begin{aligned}
\rho_k
&=
\frac{\Cov(X_{ik}+U_{ik}^{(1)},X_{ik}+U_{ik}^{(2)})}
{\sqrt{\Var(X_{ik}+U_{ik}^{(1)})\Var(X_{ik}+U_{ik}^{(2)})}}\\
&=
\frac{1}{1+\sigma_{U,k}^2}
=
d_k^{-2},
\end{aligned}
\]
where the second equality uses orthogonality of \(X_{ik}\), \(U_{ik}^{(1)}\), and \(U_{ik}^{(2)}\), together with \(\Var(X_{ik})=1\) and \(\Var(U_{ik}^{(1)})=\Var(U_{ik}^{(2)})=\sigma_{U,k}^2\). Thus, the reciprocal scale factor for component \(k\) is
\[
d_k^{-1}=\sqrt{\rho_k}.
\]
Equivalently, the full correction matrix is
\[
D^{-1}
=
\operatorname{diag}(d_1^{-1},\ldots,d_q^{-1})
=
\operatorname{diag}(\sqrt{\rho_1},\ldots,\sqrt{\rho_q}).
\]
These population identities translate directly into the sample estimator. The first step is ordinary 2SLS using the observed, standardized split scores: regressors are \((\hat X^{(1)},C)\), instruments are \((\hat X^{(2)},C)\), and the observed covariates \(C\) are included in both because they are treated as exogenous. The second step is a post-estimation rescaling of only the coefficients on the latent predictors, using the estimated split-half correlations. Formally, let
\[
\hat\kappa^{IV}
=
\begin{pmatrix}
\hat{\boldsymbol{\beta}}^{IV}_{\hat X^{(1)}|\hat X^{(2)}}\\
\hat\gamma_C^{IV}
\end{pmatrix}
\]
denote the sample IV/2SLS estimator using regressors \((\hat X^{(1)},C)\) and instruments \((\hat X^{(2)},C)\). 

In the exactly identified case, there are as many instruments as regressors, so the IV estimator is the solution to the sample analog of the population moment condition,
\[
\mathbb{P}_n\!\left[Z_i\{Y_i-R_i^\top\kappa\}\right]=0, 
\]
where $\mathbb{P}_n$ denotes the empirical averaging operator. Equivalently,
\[
\mathbb{P}_n\!\left[Z_iY_i\right]
-
\mathbb{P}_n\!\left[Z_iR_i^\top\right]\kappa
=
0.
\]
Provided \(\mathbb{P}_n Z_iR_i^\top\) is nonsingular, solving these equations gives
\[
\hat\kappa^{IV}
=
\left(\mathbb{P}_n\!\left[Z_iR_i^\top\right]\right)^{-1}
\mathbb{P}_n\!\left[Z_iY_i\right],
\qquad
\mathbb{P}_n\!\left[f_i\right]\equiv \frac{1}{n}\sum_{i=1}^n f_i.
\]
This is the usual 2SLS estimator written in its identified moment-equation form. Define
\[
\hat D^{-1}
=
\operatorname{diag}
\left(
\sqrt{\widehat{\Cor}(\hat X_{\cdot 1}^{(1)},\hat X_{\cdot 1}^{(2)})},
\ldots,
\sqrt{\widehat{\Cor}(\hat X_{\cdot q}^{(1)},\hat X_{\cdot q}^{(2)})}
\right),
\]
where \(\hat X_{\cdot k}^{(j)}=(\hat X_{1k}^{(j)},\ldots,\hat X_{nk}^{(j)})^\top\) denotes the sample vector of split-\(j\) scores for component \(k\). Set
\[
\hat{\boldsymbol{\beta}}_X^*
=
\hat D^{-1}\hat{\boldsymbol{\beta}}^{IV}_{\hat X^{(1)}|\hat X^{(2)}}.
\]
If the split-half correlations are positive and bounded away from zero, the law of large numbers gives \(\hat D^{-1}\overset{p}{\to}D^{-1}\), while standard IV consistency gives \(\hat{\boldsymbol{\beta}}^{IV}_{\hat X^{(1)}|\hat X^{(2)}}\overset{p}{\to}D\boldsymbol{\beta}_X\). Therefore, by Slutsky's theorem,
\[
\plim \hat{\boldsymbol{\beta}}_X^*
=
D^{-1}D\boldsymbol{\beta}_X
=
\boldsymbol{\beta}_X.
\]
The observed-control coefficient is already correctly scaled:
\[
\plim \hat\gamma_C^{IV}=\gamma.
\]

In practice, the signs of the split-specific latent dimensions should be aligned
to the target latent directions before computing the correlations, and very
small split-half correlations should be treated as a sign of weak-instrument or
weak-measurement dynamics, where caution is warranted.\footnote{Note that the diagonal correction above is appropriate for componentwise unit-variance identification and should not be interpreted as a full whitening or rotation correction. If a measurement procedure applies a general nonsingular non-diagonal transformation \(\hat X=A(X+U)\), the population IV coefficient \(\delta_X^{IV}\) satisfies
\[
A^\top\delta_X^{IV}=\boldsymbol{\beta}_X,\qquad
\delta_X^{IV}=A^{-\top}\boldsymbol{\beta}_X,\qquad
\boldsymbol{\beta}_X=A^\top\delta_X^{IV}.
\]
Componentwise split-half correlations do not identify \(A\); estimating a full rotation or shear correction requires additional information or restrictions.}

\subsection{A Stylized Comparison with the Method of Composition}

In this stylized comparison, suppose the point estimate can be written as
\[
\hat X=\frac{X+U_{\mathrm{MOC}}}{\sqrt{1+\sigma_{U,\mathrm{MOC}}^2}}.
\]
Write the corresponding pre-standardized MOC draw as
\[
\tilde X_{\mathrm{MOC}}^{(t)}=X+U_{\mathrm{MOC}}+\xi^{(t)},
\]
and the standardized MOC regressor as
\[
\hat X_{\mathrm{MOC}}^{(t)}
=
\frac{X+U_{\mathrm{MOC}}+\xi^{(t)}}{\sqrt{1+\sigma_{U,\mathrm{MOC}}^2+\sigma_\xi^2}},
\]
where $\xi^{(t)}$ is mean-zero draw noise independent of $X$, $U_{\mathrm{MOC}}$, and $\varepsilon$. Then
\[
\plim \hat\beta_{\mathrm{MOC}}^{(t)}
=
\frac{\beta_X}{\sqrt{1+\sigma_{U,\mathrm{MOC}}^2+\sigma_\xi^2}},
\]
whereas the regression using the point estimate is centered at
\[
\frac{\beta_X}{\sqrt{1+\sigma_{U,\mathrm{MOC}}^2}}.
\]
Because $\sigma_\xi^2>0$ implies that
\[
\frac{1}{\sqrt{1+\sigma_{U,\mathrm{MOC}}^2+\sigma_\xi^2}}
<
\frac{1}{\sqrt{1+\sigma_{U,\mathrm{MOC}}^2}},
\]
the MOC draw-level regressions are more attenuated in this stylized setting. This result is not a claim about all Bayesian models. It describes the common two-stage MOC implementation in which $X$ is sampled from $p(X\mid W)$ rather than from $p(X\mid W,Y)$. Posterior means or posterior draws from a fully specified Bayesian model need not decompose into a true score plus independent classical noise, so the calculation is a stylized additive-noise comparison (not a general MOC theorem).

\subsection{Alternative Identification Restrictions and Scale Transformations}
\label{s:alternative-identification-restrictions}

Here, we derive the attenuation factor for bivariate linear regression under identifying restrictions other than the mean 0, standard deviation 1 assumption for both the true latent trait $X$ and the latent-trait estimates $\hat{X}$ that is assumed in the paper. 

We do this by noting that if we have (possibly different) sets of identifying restrictions for $X$ and $\hat{X}$ then they can be represented by affine transformations of the paper's standardized measures. Formally, we can write the true latent trait under its new identifying restrictions as $X'$ and the estimated latent trait under its new identifying restrictions as $\hat{X}'$. Then for some constants $a,c$ and positive scale constants $b,d$ after aligning directions, we have:
\begin{equation}
\label{eq:other-identification-x}
X'=\frac{X-a}{b}
\end{equation}
\begin{equation}
\label{eq:other-identification-xhat}
\hat{X}'=\frac{(\hat{X}-c)}{d}
\end{equation}
which allows us to derive the attenuation factor under any identification restrictions. Note that it may seem odd to assume different identifying restrictions for the true latent traits and for the estimates of them, but we derive our results here under the broadest conditions, which  include cases where the same identifying restriction is used on the true and estimated latent traits.

We derive the attenuation factor by noting that the subtraction terms in the numerators of Equations \ref{eq:other-identification-x} and \ref{eq:other-identification-xhat} do not affect the true definition of $\beta_X$ or its estimate $\hat{\beta}_{\hat{X}}$. Dividing $X$ by $b$ would multiply $\beta_X$ by $b$ and dividing $\hat{X}$ by $d$ would multiply $\hat{\beta}_{\hat{X}}$ by $d$. Together, this means that the new attenuation factor is $
\nicefrac{d}{b}$ times the attenuation factor in the paper, so that 
\begin{align}\label{eq:latent-attenuation-factor-otherID}
\plim \hat{\beta}_{\hat X'} = \frac{d}{b\sqrt{1 + \sigma_U^2}} \beta_{X'}.
\end{align}

For example, if one were to use the identifying restriction that two different observations' true latent trait values were -1 and 1, respectively (sometimes referred to as the ``Kennedy-Helms'' identification restriction used to fix the ideology a very liberal and very conservative legislator at these two values), one could simply take the values estimated for the two legislators and find the affine transformation (values of $c$ and $d$ in Equation \ref{eq:other-identification-xhat}) that would place the two legislators at these positions. 

However, if we define the true latent trait $X$ using the same identifying restrictions we apply to its estimate, we recreate the dynamic discussed in the main text. In standard classical measurement error, the observed predictor has a larger variance than the true predictor. In the latent case, applying identical restrictions forces both scales to match. Also, this argument treats $a$, $b$, $c$, and $d$ as fixed constants. If anchors or scale constants are estimated from the data, especially from selected observations, these constants are random and can affect uncertainty calculations and asymptotic approximations.

\renewcommand{\thefigure}{A.II.\arabic{figure}}
\setcounter{figure}{0}  
\renewcommand{\thetable}{A.II.\arabic{table}}
\setcounter{table}{0}  
\renewcommand{\theequation}{A.II.\arabic{equation}}
\setcounter{equation}{0}
\renewcommand{\thesection}{A.II.\arabic{section}}
\setcounter{section}{1}

\renewcommand{\thesubsection}{A.II.\arabic{subsection}}
\setcounter{subsection}{0}
\renewcommand{\thesubsubsection}{A.II.\arabic{subsection}.\arabic{subsubsection}}
\setcounter{subsubsection}{0}

\section*{Appendix II: Evaluating the Sensitivity of the Corrected Estimator to Violations of $\sigma_{U,1}^2 = \sigma_{U,2}^2$}
\addcontentsline{toc}{section}{Appendix II: Evaluating the Sensitivity of the Corrected Estimator to Assumption Violations}

In our main derivations in the paper, particularly in showing how the correlation between the two sets of latent-trait estimates can be used to estimate the measurement error variance, we assumed that the measurement error variance was the same for both of our replicate measures $\hat X^{(1)}$ and $\hat X^{(2)}$ which were obtained by partitioning our indicators into two equally sized sets and estimating the latent trait separately based on each of the two sets. Although this may be roughly true given that each set of estimates is based on the same number of indicators (or nearly the same number in the case of an odd number of indicators), it is unlikely that these variances will be exactly the same in practice. 

Note, for example, that different indicators may contain different amounts of information about the latent trait. To use an example from educational testing, a question that every single student answers incorrectly does not give us \emph{any} information about the relative intelligence of the test takers. For example, even if half of test takers answer a question correctly, their responses may not correlate with overall performance---correct answers might reflect regional familiarity with question content or other non-intelligence factors. And, some highly informative indicators may discriminate very strongly between those with low and high levels of the latent trait.

To the extent that the values of $\sigma_{U,1}^2$ and $\sigma_{U,2}^2$ differ, our estimate of $\rho$ based on $\Cor(\hat X^{(1)}, \hat X^{(2)})$ and hence the implied correction factor applied to the standard OLS or IV estimators will be incorrect which will cause our corrected estimators to be biased. But how large is this bias likely to be in practice if such assumption violations arise? In other words, how different must $\sigma_{U,1}^2$ and $\sigma_{U,2}^2$ be to meaningfully affect the ultimate estimates of $\beta_X$ that our corrected estimators produce? ...and how much are these possible biases mitigated by our strategy of averaging over slope estimates from each of the different partitions of indicators into two sets? In this section, we explore these questions by deriving an expression for the degree to which estimates are affected by given values of the two variances and discussing the implications for our estimators in practice.

Recall that in the main text, we showed that if we estimate a standard linear regression with our observable estimates $\hat{X}$, then the population slope, equivalently the probability limit of the OLS slope estimator, satisfies
\begin{align*}
\plim \hat{\beta}_{\hat{X}}&=\left(\plim \hat{\beta}_{\tildX}\right) \sqrt{1+\sigma^2_{U}} \, = \, \frac{1}{1 + \sigma^2_{U}}\beta_X \sqrt{1+\sigma^2_{U}} =\frac{1}{\sqrt{1+\sigma^2_{U}}}\beta_X.
\end{align*} 
Since as we show, $\Cor(\hat X^{(1)}, \hat X^{(2)})=\frac{1}{1+\sigma^2_U}$ we can obtain a corrected estimator by dividing $\hat{\beta}_{\hat{X}}$ by $\sqrt{\Cor(\hat X^{(1)}, \hat X^{(2)})}$.

But for the correlation between the two latent-trait estimates $\hat X^{(1)}$ and $\hat X^{(2)}$ to appropriately estimate the correction factor in this way, we need to assume that the measurement error variance for both of these estimates of the latent trait are the same, i.e., that $\sigma_{U,1}^2=\sigma_{U,2}^2$, which is why we subscript the variance above with $U$ rather than $U,1$ or $U,2$.

If the variance of the measurement error for $\hat X^{(1)}$ and $\hat X^{(2)}$ is not the same, then the correlation estimates something different. Specifically, we have 
\begin{equation} \label{correlation-correction-different-variances}
\begin{aligned}
    \Cor(\hat X^{(1)}, \hat X^{(2)}) &= \Cor(\tilde X^{(1)}, \tilde X^{(2)}) 
    = \frac{\Cov(\tilde X^{(1)},\tilde X^{(2)})}{\sqrt{\sigma^2_{\tilde X^{(1)}}\sigma^2_{\tilde X^{(2)}}}} \\
    &= \frac{\Cov(X+U^{(1)}, X+U^{(2)})}{\sqrt{(\sigma^2_X+\sigma_{U,1}^2)(\sigma^2_X+\sigma_{U,2}^2)}} \\
    &= \frac{\Cov(X, X)+\Cov(X, U^{(2)})+\Cov(U^{(1)}, X)+\Cov(U^{(1)}, U^{(2)})}{\sqrt{(\sigma^2_X+\sigma_{U,1}^2)(\sigma^2_X+\sigma_{U,2}^2)}} \\
    &= \frac{\sigma^2_X+0+0+0}{\sqrt{(\sigma^2_X+\sigma_{U,1}^2)(\sigma^2_X+\sigma_{U,2}^2)}} \\
    &= \frac{\sigma^2_X}{\sqrt{(\sigma^2_X+\sigma_{U,1}^2)(\sigma^2_X+\sigma_{U,2}^2)}} \\
    &= \frac{1}{\sqrt{(1+\sigma_{U,1}^2)(1+\sigma_{U,2}^2)}},
\end{aligned}
\end{equation}
since as above we assume by identification that $\sigma^2_X=1$.

This implies that when $\sigma_{U,1}^2 \neq \sigma_{U,2}^2$ our proposed correction to the OLS slope estimator will be affected. To see this, note that the naive OLS estimator using $\hat X^{(1)}$ as the predictor has probability limit $\beta_X/\sqrt{1+\sigma_{U,1}^2}$. If we divide by $\sqrt{\Cor(\hat X^{(1)}, \hat X^{(2)})}$ to correct the estimator, then we will not cancel out the attenuation factor and instead the corrected estimator has probability limit
\begin{equation} \label{eq:correlation-correction-different-variances2}
\begin{aligned}
    \plim \hat{\beta}^*_{\hat X^{(1)}}
    &=
    \frac{\beta_X}{\sqrt{1+\sigma_{U,1}^2}}
    \underbrace{
    \left[ (1+\sigma_{U,1}^2)(1+\sigma_{U,2}^2) \right] ^ {\frac{1}{4}}
    }_{\substack{\text{inverse square-root}\\\text{split correlation}}} \\
    &= \beta_X \frac{(1+\sigma_{U,2}^2)^{\frac{1}{4}}}{(1+\sigma_{U,1}^2)^{\frac{1}{4}}}.
\end{aligned}
\end{equation}
The larger the difference between $\sigma_{U,1}^2$ and $\sigma_{U,2}^2$, the farther this fraction will be from 1, and hence the farther the probability limit of the corrected slope will be from $\beta_X$.

\begin{figure}[H]
    \centering
\includegraphics[width=1\linewidth]{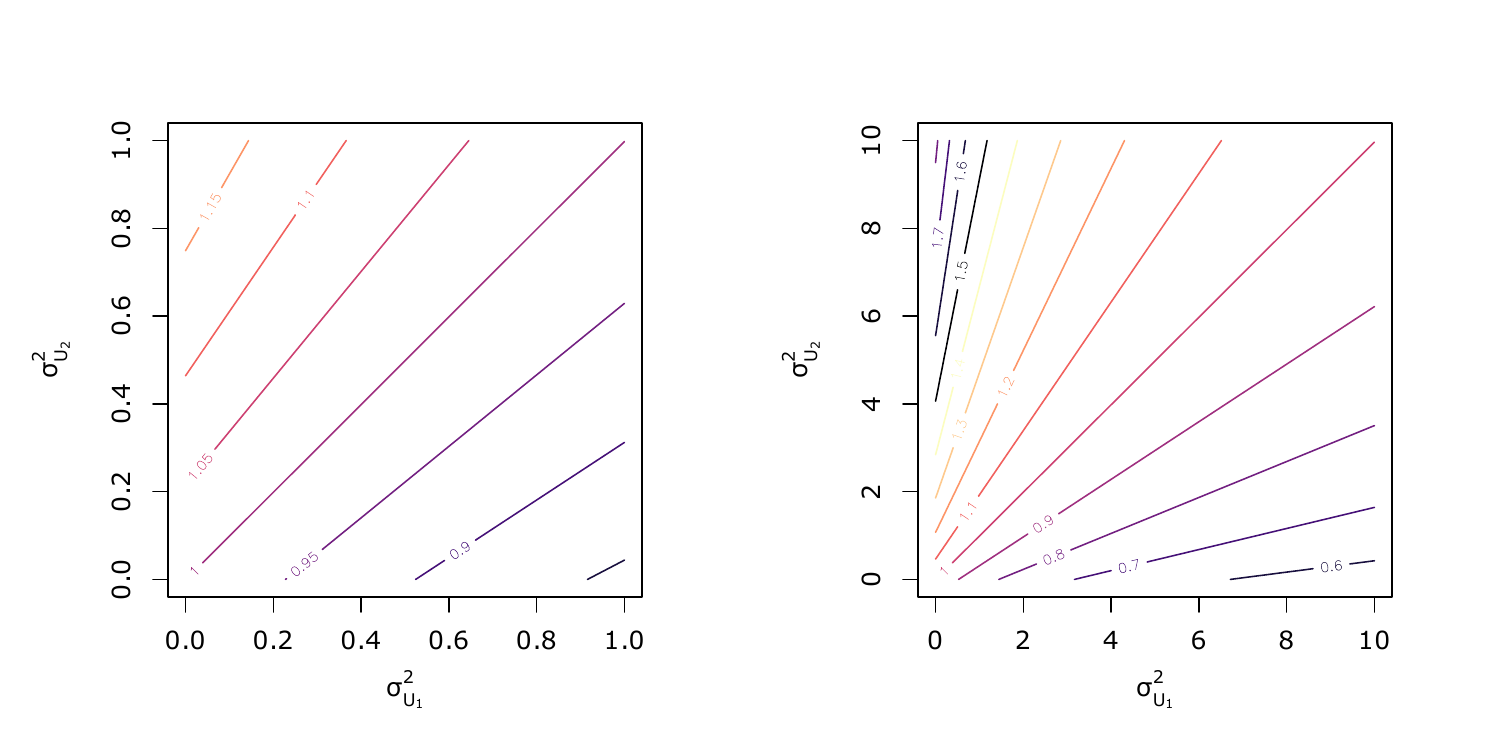}
    \caption{This figure depicts the value of the factor $\frac{(1+\sigma_{U,2}^2)^{\frac{1}{4}}}{(1+\sigma_{U,1}^2)^{\frac{1}{4}}}$ from Equation \ref{eq:correlation-correction-different-variances2} as a function of $\sigma_{U,1}^2$ and $\sigma_{U,2}^2$. The left pane shows the values of this factor when each of these variances ranges between 0 and 1, while the right pane shows values as the variances range between 0 and 10.}
    \label{fig:ME-var-sensitivity-single}
\end{figure}

But how large does the difference between $\sigma_{U,1}^2$ and $\sigma_{U,2}^2$ have to be in order to produce a meaningfully large skew from using this correlation-based correction? In order to assess this Figure \ref{fig:ME-var-sensitivity-single} plots the factor multiplying $\beta_X$ in Equation \ref{eq:correlation-correction-different-variances2} as a function of both measurement error variances. The left pane shows the factor multiplying $\beta_X$ on the right side of Equation \ref{eq:correlation-correction-different-variances2} as the measurement error variances for the two estimates of the latent trait, each of which ranges between 0 and 1. For most applications, we would expect \emph{ex ante} that these parameters would fall into this range, i.e., that the measurement error variance for each of the two sets of estimates would not be larger than the sample variance of the true latent trait itself. Within this range, we observe that the factor is small but not always negligible, reaching values below 0.85 (a 15 percent downward skew in the slope estimate) and above 1.15 (a 15 percent upward skew in the slope estimate) for some instances. But note that these values are ones in which one of the measures has measurement error variance very close to 0, and the other has measurement error variance very close to 1. If we were to partition a set of indicators evenly into two sets, it would seem unlikely that we would be in such a situation. But skewing the slope estimate by smaller amounts, e.g., by 5 or 10 percent in either direction, may still be problematic. 

The right pane shows the impact of this assumption violation for a much wider (perhaps implausibly so) range of values for the two variances, with each going from 0 to 10. We see that this factor reaches values much farther from 1. Note, however, that for many of these values much farther from 1, we again would be dealing with dramatically differing values for $\sigma_{U,1}^2$ and $\sigma_{U,2}^2$, e.g., 1 and 5, which may be unlikely to occur in most settings.

More importantly, recall that our proposed corrected estimator $\hat{\beta}_{\hat X}^{*}$ averages over the two ways to use a given partition, in the case of the IV formulation, with $\hat X^{(1)}$ as the regressor and $\hat X^{(2)}$ as the instrument, and also with $\hat X^{(2)}$ as the regressor and $\hat X^{(1)}$ as the instrument. As discussed in the main text, this would be expected to mitigate the skews that result from differing measurement error variances. This is because when these two measurement errors are not equal, the correlation-based correction factor will be too high when using one of the two sets of estimates as the predictor and too low when using the other one. Specifically, when averaging in this way, we would obtain a factor of 
\begin{equation}
\begin{aligned}
\overbrace{
\frac{1}{2}
\left[
\frac{(1+\sigma_{U,2}^2)^{\frac{1}{4}}}
     {(1+\sigma_{U,1}^2)^{\frac{1}{4}}}
+
\frac{(1+\sigma_{U,1}^2)^{\frac{1}{4}}}
     {(1+\sigma_{U,2}^2)^{\frac{1}{4}}}
\right]
}^{\substack{
\text{average of the two}\\
\text{reciprocal directional factors}
}}
&=
\frac{
\overbrace{r+r^{-1}}^{
{\text{Letting } r
\equiv
\left(\frac{1+\sigma_{U,2}^2}{1+\sigma_{U,1}^2}\right)^{1/4}
>0,}
}
}{2}
\ge
\sqrt{r\cdot r^{-1}}
=
1.
\end{aligned}
\label{eq:factor-averaged}
\end{equation}
The inequality is the arithmetic mean--geometric mean inequality applied to the positive pair \(r\) and \(r^{-1}\). Equality requires \(r=r^{-1}\), or equivalently \(r=1\), which occurs exactly when \(\sigma_{U,1}^2=\sigma_{U,2}^2\). Thus, averaging the two directions removes the first-order imbalance between the reciprocal factors, but it does not eliminate the remaining second-order inflation when the two measurement-error variances differ.

\begin{figure}[H]
    \centering
\includegraphics[width=1\linewidth]{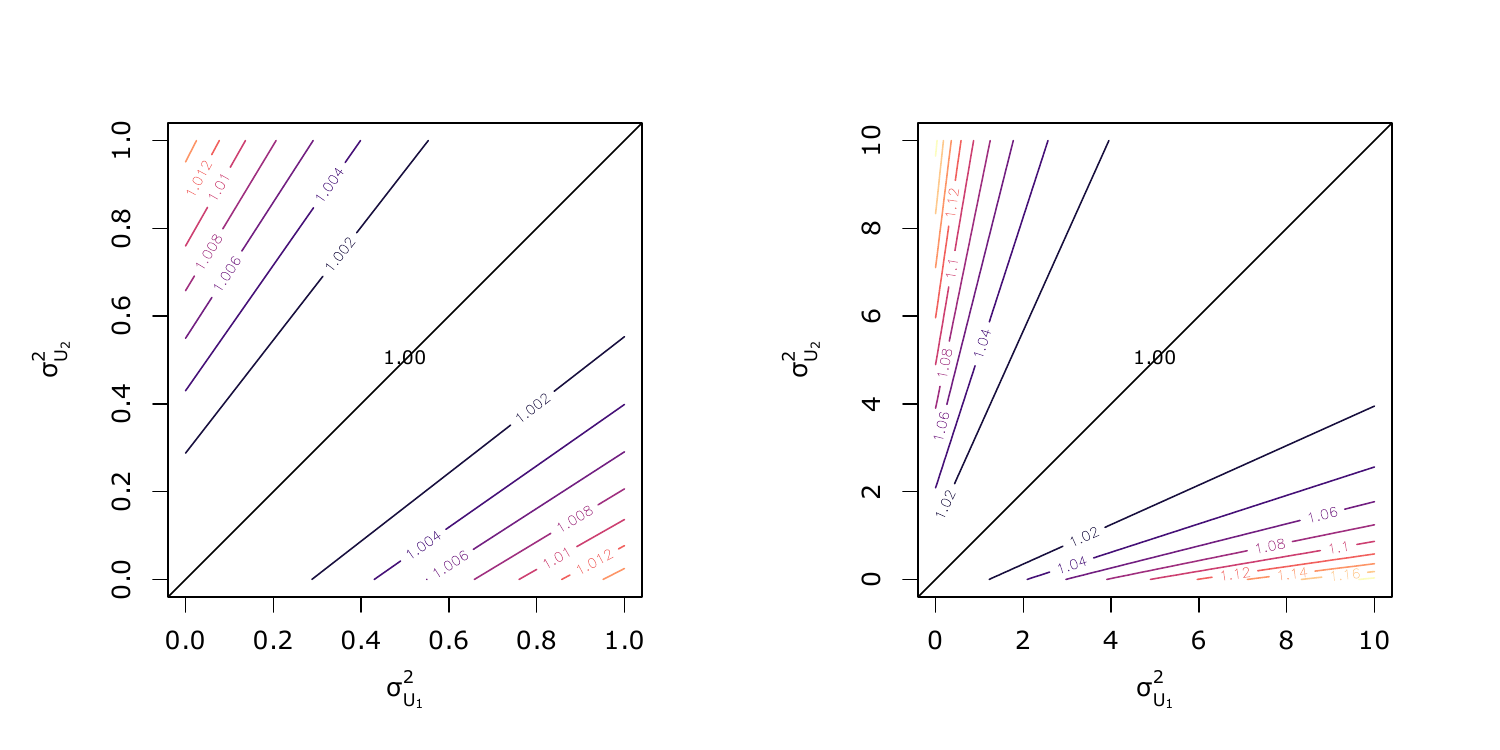}
    \caption{This plot displays the value of the factor  $\nicefrac{ \left[ \frac{(1+\sigma_{U,2}^2)^{\frac{1}{4}}}{(1+\sigma_{U,1}^2)^{\frac{1}{4}}} + \frac{(1+\sigma_{U,1}^2)^{\frac{1}{4}}}{(1+\sigma_{U,2}^2)^{\frac{1}{4}}} \right] }{2}$ from Equation \ref{eq:factor-averaged} as a function of $\sigma_{U,1}^2$ and $\sigma_{U,2}^2$. The left pane shows the values of this factor when each of these variances ranges between 0 and 1, while the right pane shows values as the variances range between 0 and 10. }
    \label{fig:ME-var-sensitivity-averaged}
\end{figure}

Figure \ref{fig:ME-var-sensitivity-averaged} shows how this new averaged factor varies over values of each variance for a correlation-corrected slope estimator that is averaged over using $\hat X^{(1)}$ and using $\hat X^{(2)}$ as the predictor. We see that the factor is much closer to 1 for most plausible parameter values, although it is not exactly 1 when the two variances differ. In the left pane, the factor is between 1 and 1.01 for nearly all possible variances, and is between 1 and 1.002 for a wide range of variances. Even when we let the variances range between 0 and 10, we see that only highly unequal variances produce factors above 1.02. As discussed above, we may be skeptical that measurement error variances would be this unequal in practice when using the same number of indicators to obtain estimates. This suggests that violations of the assumption that $\sigma_{U,1}^2 = \sigma_{U,2}^2$ are unlikely to be severe enough in practice to have a significant impact on our corrected estimates. 

Furthermore, simple and straightforward checks could be used to reassure researchers that a particular partition of indicators into two sets does not result in a large imbalance of information about the latent trait between these two sets of indicators. For example, if using a factor analytic or ideal point model to estimate the latent trait, one could inspect the factor loadings or discrimination parameters from estimating the model using the full set of indicators in order to make sure that one of the two partitions does not contain many that are close to zero and with other containing many that are far from zero. Researchers could also inspect standard errors, posterior standard deviations, or other measures of uncertainty that their chosen measurement model may produce. Even when using simpler strategies, such as additive or averaging, to produce latent-trait estimates, inspecting the correlation matrix of the indicators and the distributions of each individual indicator may be enough to prevent the unexpected use of sets of indicators that each contain dramatically different amounts of information about the latent trait. Many of these things are good practice in any application of measurement models, not just for our specific aims of addressing measurement error in latent predictors.

Also note that the expression in Equation \ref{eq:factor-averaged} holds for a particular partition of the indicators, where the estimates are averaged using one or the other as the predictor. Our proposed corrected estimator $\hat{\beta}^{*}$ takes the median over all such within-partitions averages of the corrected slope estimate, which can further reduce sensitivity to unstable partitions but does not remove the componentwise imbalance factor for any given partition.

Overall, then, we can expect that our correlation-corrected estimators (recall that the corrected OLS and IV estimators are equivalent for the bivariate regression case) should perform well in a wide range of situations and that in practical applications with balanced splits, the assumption that these two measurement error variances are identical is unlikely to meaningfully affect our proposed estimators.

\renewcommand{\thefigure}{A.III.\arabic{figure}}
\setcounter{figure}{0}  
\renewcommand{\thetable}{A.III.\arabic{table}}
\setcounter{table}{0}  
\renewcommand{\theequation}{A.III.\arabic{equation}}
\setcounter{equation}{0}
\renewcommand{\thesection}{A.III.\arabic{section}}
\setcounter{section}{1}

\renewcommand{\thesubsection}{A.III.\arabic{subsection}}
\setcounter{subsection}{0}
\renewcommand{\thesubsubsection}{A.III.\arabic{subsection}.\arabic{subsubsection}}
\setcounter{subsubsection}{0}

\section*{Appendix III: Additional Simulation Details \& Results}\addcontentsline{toc}{section}{Appendix III: Additional Simulation Details \& Results}

\subsection{Simulation Design Details}\label{s:SimDetails}

We assume a two-parameter probit-link IRT model for the indicators, as well as a standard linear regression of the observed dependent variable $Y$ on the latent trait $X$. This setup is common in political science and other literatures, with estimation, often via Gibbs sampling with data augmentation or other methods, being well studied (see e.g. \cite{Clinton2004TheData, Martin2011MCMCpack:R}). 

Formally, we assume that the indicators are generated according to: 
\begin{equation}
 \Pr(W_{ij}=1)=\Phi(X_i \eta_j+\alpha_j).
\label{eq:ideal-point}
\end{equation} 
with independence across individuals $i$ and items $j$. Further, we assume a standard linear regression model for an observed dependent variable $Y$ where the latent trait $X$ is the predictor:
\begin{equation}
 Y_i = \beta_0 + X_i \beta_X + \varepsilon_i.
\label{eq:linreg}  
\end{equation} 
with $\varepsilon_i \sim N(0,\sigma_\varepsilon^2)$. Standard independent conjugate priors are used on all parameters: $X_i \sim N(0,1)$, $\eta_j, \alpha_j, \beta_0, \beta_X \sim N(0, 25)$, $\sigma_\varepsilon^2 \sim \textit{Inverse-Gamma}(.01, .01)$.

We adopt the same data-generating process as above; for each simulation iteration we:
\begin{enumerate}
    \item Generate simulated matrix of indicators $W^{\textrm{Sim}}$ by randomly drawing from the predictive distribution of the ideal point model in Equation \ref{eq:ideal-point}.
    \item Estimate respondent ideal points $\hat X^{\mathrm{Sim}}$ from the simulated indicator matrix $W^{\mathrm{Sim}}$, rescaling these estimated latent scores to have mean 0 and variance 1 with higher values corresponding to more conservative positions.
    \item Estimate the slope coefficient relating $X$ with $Y$ as specified in Equation \ref{eq:linreg} using simple OLS with the true latent predictor, with the estimated predictor, and with the split-indicator IV approach (with uncorrected and corrected estimators). 
\end{enumerate}

\subsection{Ideology Estimation}

Before discussing a larger-scale simulation that varies the number of observations and indicators used in estimation, we first provide a simple illustration of the performance of the estimators discussed above through a set of simulations using a single fixed set of parameter values. To choose these ``true'' data-generating parameter values, we look at data from the 2016 Cooperative Congressional Election Survey (CCES). This survey includes a large number of policy questions on which respondents are asked their positions (e.g., whether respondents support background checks for all gun sales or whether they think women should always be allowed to have an abortion as a matter of choice). We use 30 of these policy questions as our indicators from which we estimate respondents' ideologies (ideal points) and the item parameters based on the Clinton, Jackman, and Rivers (CJR) ideal point model.\footnote{We use the \url{binIRT} function in the \url{emIRT} package. We drop several policy items that contain a large proportion of missing values. We also drop any respondents having missing values on any of these indicators or on the presidential vote question or who stated they did not vote or voted for someone other than Trump or Clinton, and then, we randomly select a subset of respondents to use.} We then use the estimated ideal points to predict 2016 presidential vote (1=Trump, 0=Clinton). 

We then set the true values for 1,000 respondents' ideal points to those point estimates for $X_i$ (after imposing the identifying restriction that these ideal points have mean 0, variance 1, and that higher values correspond to more conservative positions) and set the true values of the discrimination/loading parameters ($\eta_j$) and item intercept/difficulty-style parameters ($\alpha_j$) to those estimated from these indicators in our CCES sample. We also set the true values for the intercept $\beta_0$, the slope $\beta_X$, and the error variance $\Var(\varepsilon_i)$ to roughly correspond to the values estimated in the linear regression predicting presidential vote with the estimated ideal points using the baseline OLS approach.\footnote{The estimated slope using our corrected estimates is slightly under 0.4, but for the purposes of the simulations, we use the value of 0.4 since a round number is easier to use as a baseline for comparison in figures and elsewhere. Note that this value for the slope is likely to be attenuated, i.e., smaller in magnitude than the true slope coefficient, since it was estimated from data that itself includes measurement error in the predictor, but we leave it as is for the true value in our simulations.} 

\begin{figure}[H]
    \centering
\includegraphics[width=0.75\linewidth]{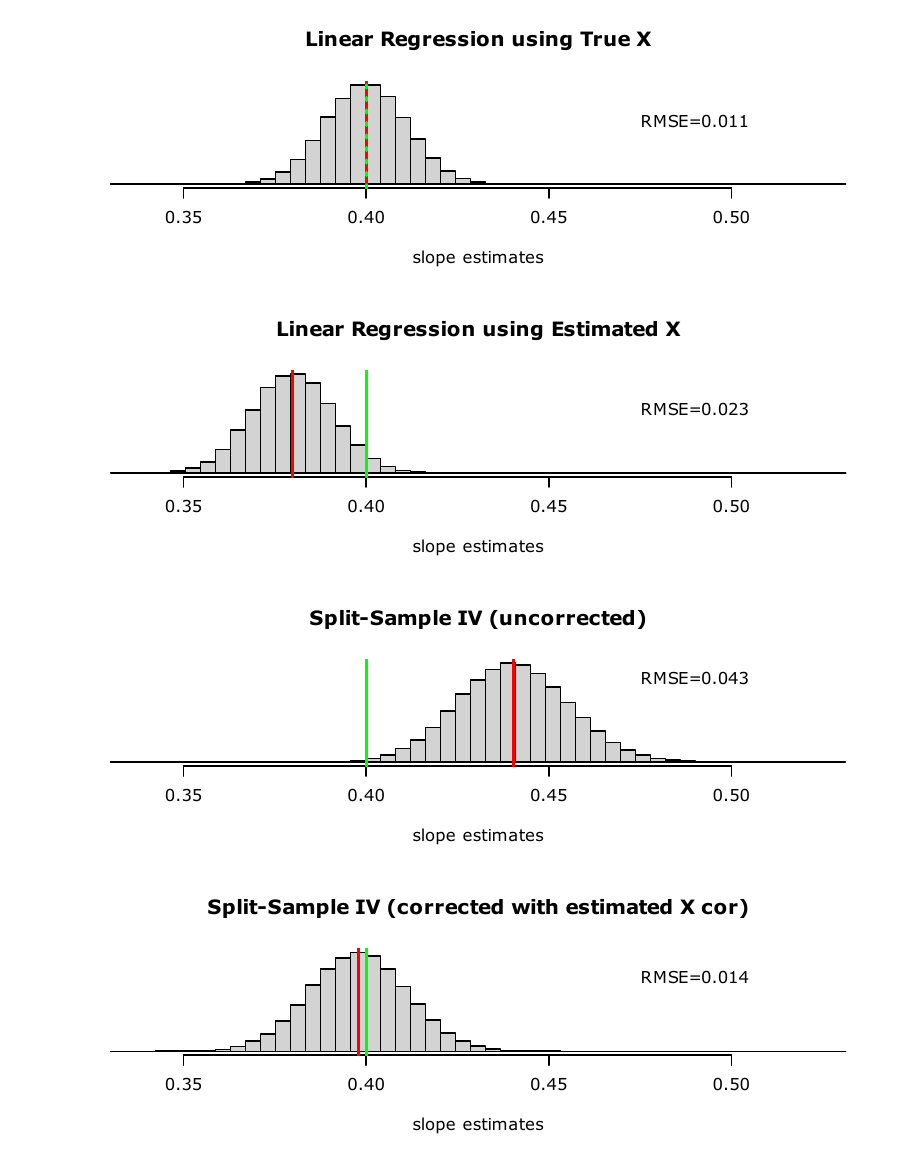}
    \caption{Illustrative simulation results. The true coefficient is highlighted in green (0.40). The mean of the estimates is highlighted in red across the various approaches. }
    \label{fig:IllustrativeSims}
\end{figure}

Although these simulations rely on a single set of true parameters, these values are based on real data of the sort that is often used in latent predictor applications. Furthermore, these simple simulations will illustrate some basic characteristics of the various estimators and their relative performance, at least in this specific example. Figure \ref{fig:IllustrativeSims} shows the distributions of simulated slope estimates ($\beta_X$) based on 100,000 simulation draws for each estimator. 

The top pane of Figure \ref{fig:IllustrativeSims} provides a baseline, showing the simulated sampling distribution for the slope coefficient if we actually knew each respondent's ideal point with certainty (i.e. if there was no measurement error in the predictor) and we estimated a simple bivariate linear regression predicting presidential vote with the ideal points $X$. As expected, the sampling distribution for the slope estimator is centered at the true slope of 0.4. The simulated RMSE is relatively small (0.01). Well over 60 percent of the simulations produced estimates within 0.01 of the true slope, and well over 90 percent produced estimates within 0.02 of the true slope. In a world where the true predictor is measured without error, we can obtain unbiased and relatively precise estimates of the slope coefficient of interest. Of course, with latent predictors, we will never know these true values.

The second pane in Figure \ref{fig:IllustrativeSims} plots the simulated distribution of the uncorrected OLS estimator in the case of measurement error. These simulations estimate the slope coefficient on $X$ predicting the presidential vote, assuming we do not actually observe the true latent trait $X$ but instead observe the set of 30 survey questions (indicators) from which we estimate $X$. In other words, we use \emph{estimated} ideal points $\hat{X}$ as our predictor rather than the true ideal points without any correction or adjustment. As expected, the coefficient estimates are generally smaller in magnitude, with the average estimate falling just below 0.38. The RMSE is more than double that of the known $X$ case.

The third pane of Figure \ref{fig:IllustrativeSims} shows estimates from the standard split-indicator IV approach without any corrections for the latent variable case. In these simulations, after simulating the matrix of indicators, we split the 30 indicators into the first 15 and the second 15, estimating ideal points for all respondents separately using each of these two sets. This produces $\hat X^{(1)}$ and $\hat X^{(2)}$, two separate estimated split scores. We then use 2SLS with $\hat X^{(1)}$ as the regressor and $\hat X^{(2)}$ as the instrument, and symmetrically with $\hat X^{(2)}$ as the regressor and $\hat X^{(1)}$ as the instrument where reported. In the standard case of measurement error in an observable (as opposed to latent) predictor, this IV setup can yield consistent estimates of the slope coefficient. But as shown above, when applied to the latent case, it will \emph{over}-correct for the attenuation bias, producing amplification bias. This is illustrated in the figure, where the average estimate is 0.43; the RMSE is nearly 4 times as large as in the known $X$ case (nearly double the value with uncorrected OLS and estimated $X$).

Finally, the bottom pane of Figure \ref{fig:IllustrativeSims} shows split-indicator IV estimates corrected using the correlation between $\hat X^{(1)}$ and $\hat X^{(2)}$ as described in Equation \ref{IV-cor-corrected}. These estimates are quite similar to those using the true $X$, but with a very slightly higher bias and RMSE. This suggests the feasibility of estimating $\beta_X$ with the corrected IV approach in the more realistic case where we cannot use the true $X$ to estimate the measurement error variance of the split-half estimates. 

In the last two applications (both corrected IV estimates), this simulation suggests favorable bias dynamics. (Appendix I discusses the possibility of a very enduring small amount of downward finite sample bias.)

This example illustrates the attenuation bias induced in estimates of a bivariate linear regression slope by using a noisy measure of a predictor. In our application, the measure is based on a relatively large number (30) of indicators, resulting in a relatively good signal-to-noise ratio. In cases where fewer indicators or less informative indicators are used, the attenuation bias will be larger, as seen below. It is also shown that while the standard IV approach to dealing with measurement error produces \emph{over}-correction, it is possible to correct for this and obtain slope estimates that are nearly unbiased, with variance barely higher than if the true predictor were known without error.

\subsection{Coverage Analysis}\label{s:Coverage}

\begin{figure}[H]
\centering
\includegraphics[width=0.35\linewidth]{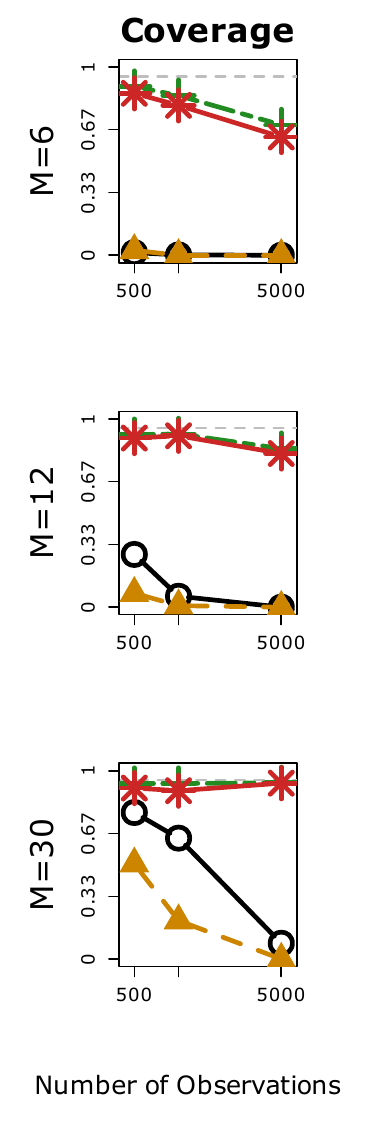}
\par\vspace{0.25em}
\includegraphics[width=0.60\linewidth]{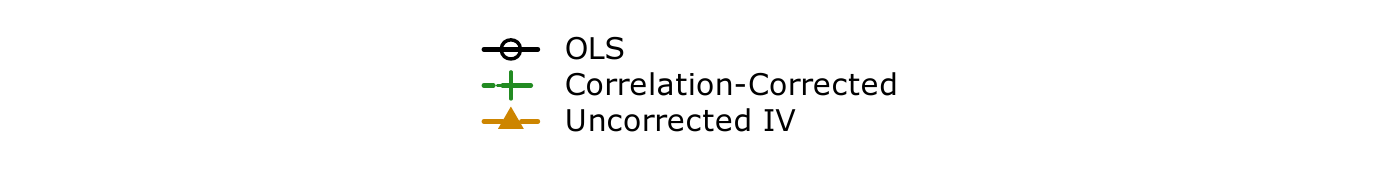}
\caption{Coverage results varying $N$ and $M$. Points show empirical 95\% interval coverage for each estimator across the simulation designs described in Appendix~\ref{s:SimDetails}; the horizontal reference line marks nominal coverage.}
\label{fig:CoverageCombinedVaryND}
\end{figure}

Figure~\ref{fig:CoverageCombinedVaryND} shows that, across most simulated settings, and especially for moderate-to-large $M$, our corrected procedure retains coverage close to nominal while confidence intervals from the uncorrected OLS and IV estimators frequently fail to contain the true parameter value (0.4) even at moderate sample sizes. With the corrected approach, as the number of indicators ($M$) grows, measurement error in the latent predictor diminishes, allowing the corrected estimator's coverage to stay near nominal levels; in the smallest-$M$ settings, coverage can be mildly reduced. In contrast, the uncorrected methods exhibit persistently lower coverage, particularly in smaller $M$ regimes. In this way, statistical inference is improved with the corrections described here. 

A closer look at the small-$M$ regime reveals that even the corrected approach can show mildly reduced coverage there, reflecting finite-sample bias when estimating the latent factor from very little data. One possibility is that, in this low-information setting, random variation in the latent variable estimates via the bootstrap is insufficient to reflect true sampling uncertainty. Another possibility is the presence of small but lingering finite sample bias in this low-information case. Nonetheless, once $N$ and $M$ become sufficiently large, this small-$M$ artifact subsides; the corrected procedure once again not only outperforms the uncorrected OLS or IV, but also yields inferences approaching nominal levels.

In further analyses, we find that subsampling based-boostrap variants \citep{politis1994large} do not markedly improve coverage of the median bootstrap; windorized mean coveage is slightly higher on average. That said, the main contribution to the undercoverage of the corrected estimator seems to be residual finite indicator bias. Pseudo-target coverage, where the target is the cell mean of the estimator (rather than the true parameter value), is much better for the corrected estimator, rising from approximately 0.70 to 0.84 in the $M=6$ case and from 0.936 to 0.944 in the $M=30$ case.

\newcommand{\MultiSimLargestN}{5000}
\newcommand{\MultiSimLargestK}{8}
\newcommand{\MultiSimLatentOLSRMSE}{0.109}
\newcommand{\MultiSimLatentCorrectedIVRMSE}{0.027}
\newcommand{\MultiSimLatentRMSEImprovementRatio}{4.092}
\newcommand{\MultiSimCovariateOLSRMSE}{0.102}
\newcommand{\MultiSimCovariateCorrectedIVRMSE}{0.021}
\newcommand{\MultiSimCovariateRMSEImprovementRatio}{4.767}
\newcommand{\MultiSimMedianSplitCorrelation}{0.465}
\newcommand{\MultiSimMedianFirstStageF}{964.128}

\subsection{Multivariate Simulation Results}\label{s:MultivariateSims}

We also evaluate the multivariate corrected IV extension derived in Appendix~\ref{s:multivariate-corrected-iv}. This simulation focuses on the comparison between uncorrected OLS and the corrected IV estimator because the aim is to assess whether the split-indicator correction carries over when multiple latent predictors and observed covariates enter the same outcome model.

For each Monte Carlo draw, we generate \(K\in\{1,2,4,8\}\) standardized latent predictors with an autoregressive correlation structure and pair them with 10 observed covariates that are correlated with the latent predictors. The outcome is generated as a linear function of the latent predictors, the observed covariates, and independent Gaussian noise. For each latent predictor, we then generate eight indicators and estimate split scores. The primary measurement design uses binary indicators generated from a two-parameter logistic model and estimated by EM. As robustness checks, we also consider continuous equal-loading indicators estimated by averaging and continuous congeneric indicators estimated by PCA. In all designs, we vary \(N\in\{500,1000,5000\}\), use eight split partitions with median aggregation, and apply the correction only to partitions with split-half correlations above 0.01.

\begin{figure}[H]
\centering
\includegraphics[height=0.88\textheight,keepaspectratio]{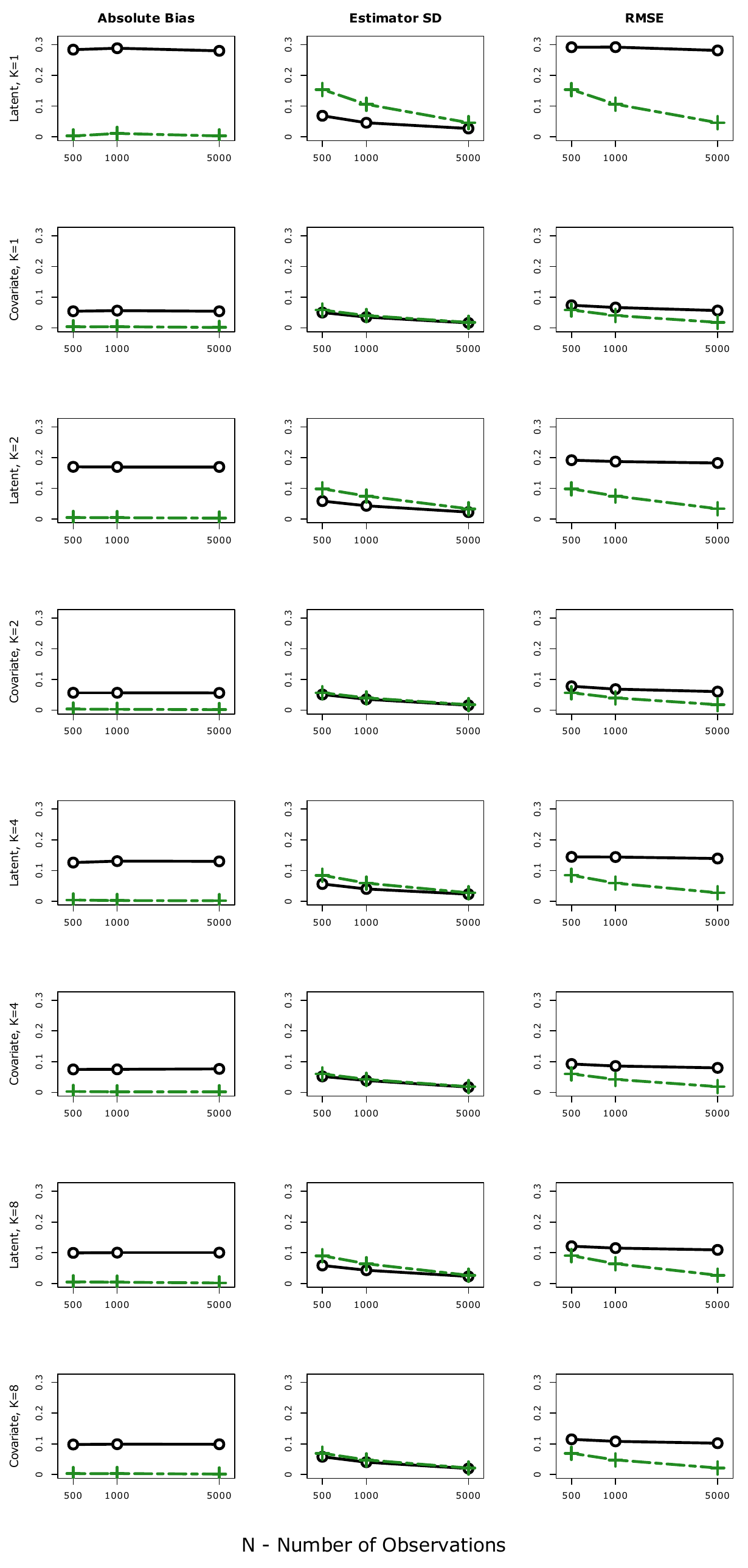}
\par\vspace{0.25em}
\includegraphics[width=0.60\linewidth]{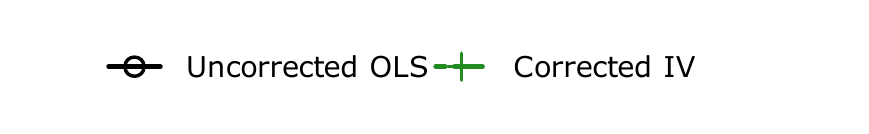}
\caption{Multivariate simulation results for the primary binary 2PL measurement design. Panels report absolute bias, estimator standard deviation, and RMSE for uncorrected OLS and the multivariate corrected IV estimator, separately for latent-predictor and observed-covariate coefficients.}
\label{fig:MultivariateSimsPrimary}
\end{figure}

Figure~\ref{fig:MultivariateSimsPrimary} reports absolute bias, estimator standard deviation, and RMSE for both the latent-predictor coefficients and the observed-covariate coefficients in the primary binary 2PL design. The multivariate results mirror the bivariate simulation patterns. Uncorrected OLS has persistent bias because measurement error in the latent predictors affects not only their own coefficients but also coefficients on observed covariates correlated with them. The corrected IV estimator substantially reduces this bias. In the largest primary design, with \(N=\MultiSimLargestN\) and \(K=\MultiSimLargestK\), latent-coefficient RMSE falls from \MultiSimLatentOLSRMSE\ under uncorrected OLS to \MultiSimLatentCorrectedIVRMSE\ under corrected IV, a roughly \MultiSimLatentRMSEImprovementRatio-fold improvement. For the observed-covariate coefficients, RMSE falls from \MultiSimCovariateOLSRMSE\ to \MultiSimCovariateCorrectedIVRMSE, a roughly \MultiSimCovariateRMSEImprovementRatio-fold improvement.

\begin{figure}[H]
\centering
\includegraphics[width=0.70\linewidth]{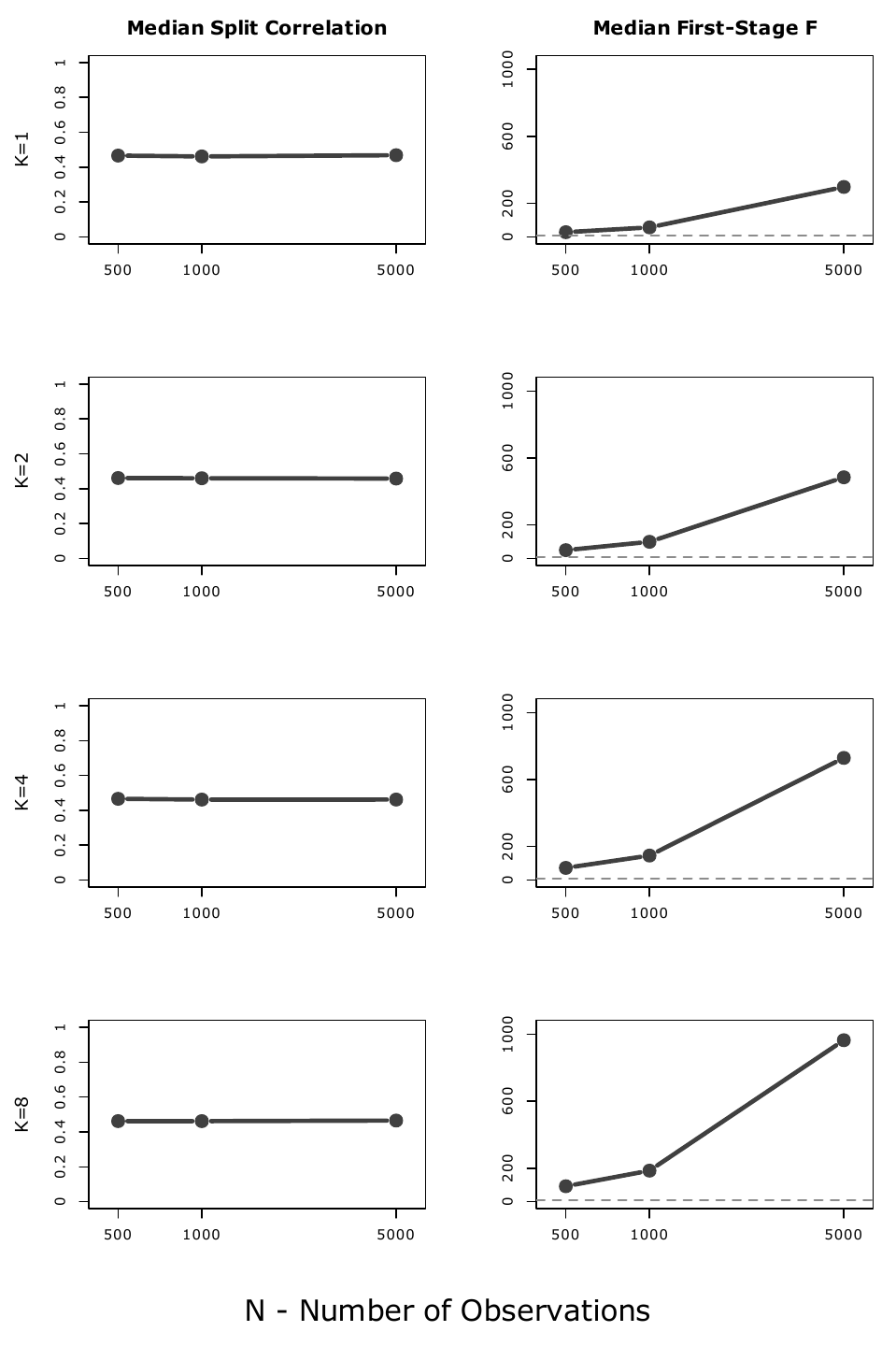}
\caption{Diagnostics for the primary multivariate binary 2PL simulation. Panels report median split-half correlations and median first-stage \(F\)-statistics across values of \(N\) and \(K\); the horizontal reference line marks \(F=10\).}
\label{fig:MultivariateSimsDiagnostics}
\end{figure}

Figure~\ref{fig:MultivariateSimsDiagnostics} reports the corresponding split-half and first-stage diagnostics. The median split-half correlation in the largest primary design is \MultiSimMedianSplitCorrelation, and the median first-stage \(F\)-statistic is \MultiSimMedianFirstStageF. Across the simulated cells, first-stage diagnostics are well above the conventional weak-instrument threshold of 10, consistent with the stable behavior of the corrected estimator in Figure~\ref{fig:MultivariateSimsPrimary}.

\begin{figure}[H]
\centering
\includegraphics[width=0.40\linewidth]{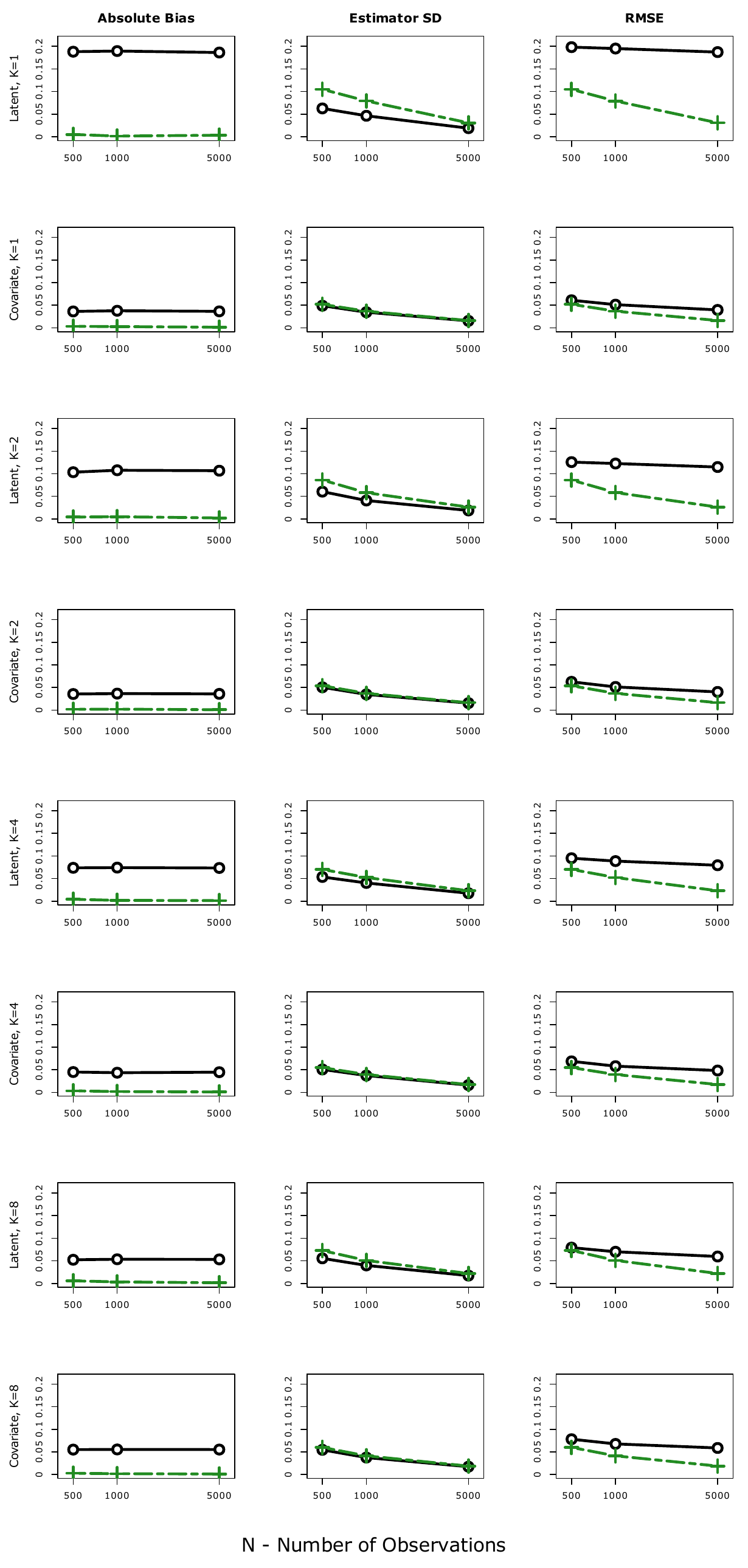}
\hspace{0.8cm}
\includegraphics[width=0.40\linewidth]{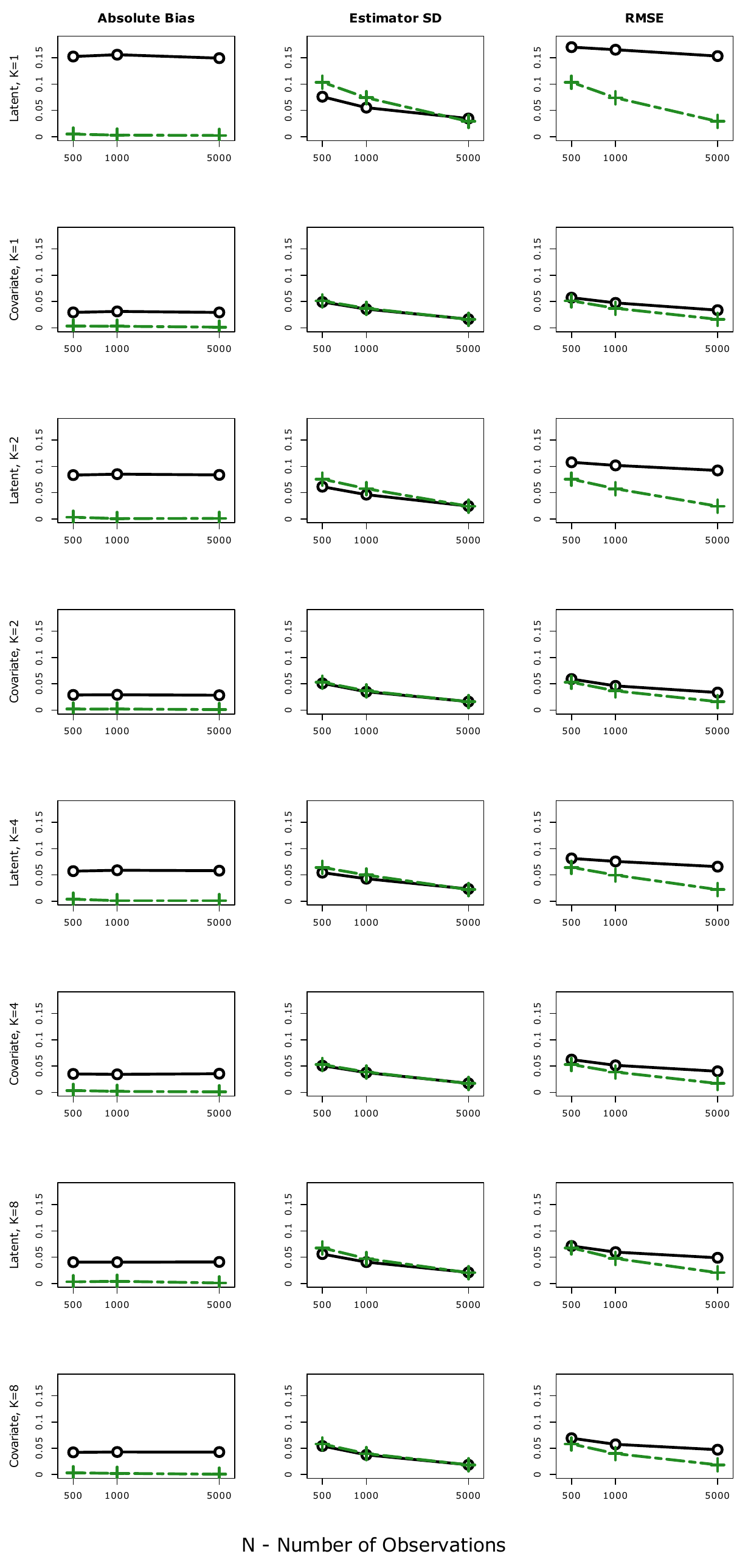}
\caption{Multivariate simulation robustness checks using continuous measurement designs. The left panel uses equal-loading indicators with averaging; the right panel uses congeneric indicators with PCA. Each point represents an values averaged across latent dimensions. The estimator legend is the same as in Figure~\ref{fig:MultivariateSimsPrimary}.}
\label{fig:MultivariateSimsRobustness}
\end{figure}

The same qualitative pattern appears in the continuous-measurement robustness checks. Figure~\ref{fig:MultivariateSimsRobustness} shows that, under both the equal-loading averaging design and the congeneric PCA design, the corrected IV estimator again reduces bias and RMSE relative to uncorrected OLS for the latent-predictor coefficients and for the observed-covariate coefficients affected by latent-predictor measurement error.

\subsection{Comparison with Additional Estimation Strategies}\label{s:Bayesian}

Researchers have increasingly turned to Bayesian strategies for inference regarding latent traits. We examine three such approaches: (1) method of composition, (2) multiple over-imputation, and (3) joint factor and outcome models. We compare these strategies using the same data-generating process outlined in the simulation section (see Section \ref{sec:Simulation}). The first two approaches share an initial measurement-model fit but differ in how the outcome enters the subsequent analysis.

\paragraph{Method of composition.} The method of composition approach draws from the posterior distribution of the latent trait estimates \citep{treier2008democracy, tanner2012tools}. Let $\tilde X_{\mathrm{MOC}}^{(1)},\ldots,\tilde X_{\mathrm{MOC}}^{(T)}$ denote pre-standardized posterior draws from $p(X\mid W)$, where $W$ is the indicator matrix. For each draw, we first place the latent scores on a comparable identified scale, denoting the standardized draw by $\hat X_{\mathrm{MOC}}^{(t)}$. Each draw is standardized separately. We then fit the outcome regression
\[
Y_i=\beta_0^{(t)}+\beta_{\mathrm{MOC}}^{(t)}\hat X_{\mathrm{MOC},i}^{(t)}+\varepsilon_i
\]
for each $t$. In the implementation used for the simulations, the regression is fit by OLS, the covariance matrix is estimated using HC3, and the coefficient vector is optionally drawn from a Normal approximation centered at the OLS coefficient vector with that covariance matrix. The reported MOC slope and uncertainty summarize these coefficient draws across posterior latent-trait draws.

\paragraph{Multiple over-imputation.} Multiple over-imputation, described in \citet{rubin1988overview} and refined in \citet{blackwell2017unified} for outcome and measurement-error settings, treats an error-free variable or latent score as unobserved and imputes it using both measurement-error information and other analysis variables. In our implementation, it starts by fitting the measurement model to the observed indicators. From this first stage, we form a unit-level approximation to each marginal measurement-stage distribution, summarized by the posterior mean and standard deviation,
\[
\mu_i=E(X_i\mid W), \qquad \sigma_i=\operatorname{SD}(X_i\mid W).
\]
We then construct an imputation data set containing the observed outcome $Y$ and a latent-score column initialized at $\mu=(\mu_1,\ldots,\mu_n)$. The latent-score column is over-imputed for all units, rather than treated as fixed at $\mu$, using an Amelia-style routine; $Y$ remains in the imputation model as an observed variable, while $(\mu_i,\sigma_i)$ are supplied as unit-specific prior/uncertainty information for the latent-score column. This produces completed latent-score vectors $X_{\mathrm{imp}}^{(1)},\ldots,X_{\mathrm{imp}}^{(M_{\mathrm{imp}})}$; in our implementation, $M_{\mathrm{imp}}=5$. Each completed vector is then standardized and used in the completed-data regression
\[
Y_i=\beta_0^{(r)}+\beta_X^{(r)}X_{\mathrm{imp},i}^{(r)}+\varepsilon_i.
\]
The over-imputation summary pools the resulting completed-data slope estimates across $r=1,\ldots,M_{\mathrm{imp}}$. Thus, unlike MOC, the outcome enters the imputation step. Unlike full joint estimation, however, the original measurement model is not re-estimated jointly with the outcome equation; instead, the first-stage measurement distribution is approximated through unit-specific summaries supplied to the imputation model. Consequently, these completed-data regressions should generally be interpreted as estimates from a staged approximation to $p(X\mid W,Y)$, not as draws from the full joint posterior $p(\beta_0,\beta_X,\vartheta,X\mid W,Y)$, except under strong specification and compatibility conditions.

In the joint modeling strategy, we fit a Bayesian model that simultaneously uses both outcome and observed factor information to generate estimates of the latent quantities of interest. This approach generates a Bayesian posterior over the coefficient estimates. Again, draws of the latent trait, $X$, are rescaled so that the posterior mean takes on a standard deviation of one.

\begin{figure}[H]
\centering
\includegraphics[width=0.88\linewidth]{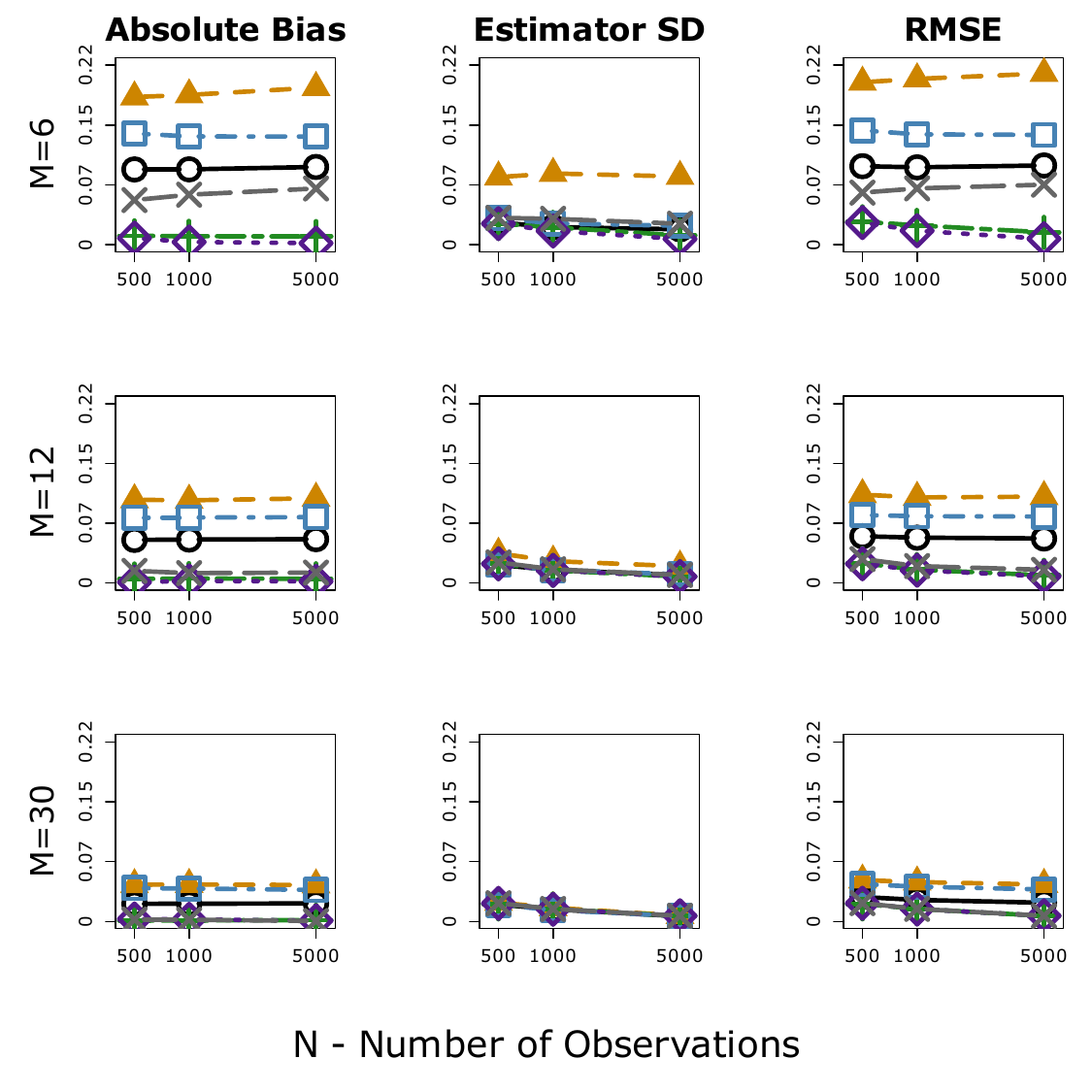}
\par\vspace{0.25em}
\includegraphics[width=0.95\linewidth]{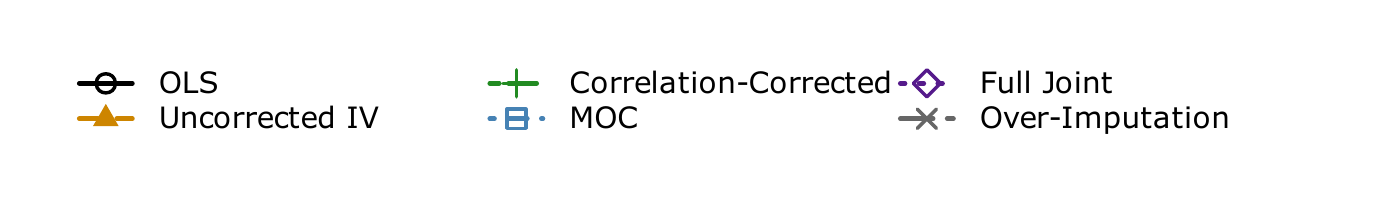}
\caption{Additional estimation strategies compared with the main simulation estimators. Results add uncorrected IV to the comparison of OLS, MOC, over-imputation, full joint estimation, and the split-indicator correlation-corrected estimator.}
\label{fig:BayesianComparison}
\end{figure}

Figure \ref{fig:BayesianComparison} presents the results for these approaches. We observe that in this simulation design, our method performs nearly identically to the most principled method of full joint estimation. In the two-stage MOC implementation, posterior latent-trait draws are noisier than the posterior mean used as a point estimate, which explains the additional attenuation in the stylized calculation in Appendix I. This point does not apply to a correctly specified full joint model in the same way, because the outcome enters the joint posterior for the latent traits and the regression coefficient. Both the joint model and multiple imputations face questions around how to perform standardization, whether that be within or across posterior draws. 

In contrast, simple OLS estimates based on sample splitting, adjusted for measurement error as proposed in this paper, provide a compelling alternative. Beyond improvements in RMSE, this approach does not require fitting a larger-scale Bayesian model, thereby avoiding the increased computational costs associated with generating uncorrelated draws from the posterior distribution.

\renewcommand{\thefigure}{A.IV.\arabic{figure}}
\setcounter{figure}{0}  
\renewcommand{\thetable}{A.IV.\arabic{table}}
\setcounter{table}{0}  
\renewcommand{\theequation}{A.IV.\arabic{equation}}
\setcounter{equation}{0}
\renewcommand{\thesection}{A.IV.\arabic{section}}
\setcounter{section}{1}
\renewcommand{\thesubsection}{A.IV.\arabic{subsection}}
\setcounter{subsection}{0}
\renewcommand{\thesubsubsection}{A.IV.\arabic{subsection}.\arabic{subsubsection}}
\setcounter{subsubsection}{0}
\section*{Appendix IV: Additional Empirical Results}
\addcontentsline{toc}{section}{Appendix IV: Additional Empirical Results}

\begin{figure}[t]
    \centering
\includegraphics[width=0.65\linewidth]{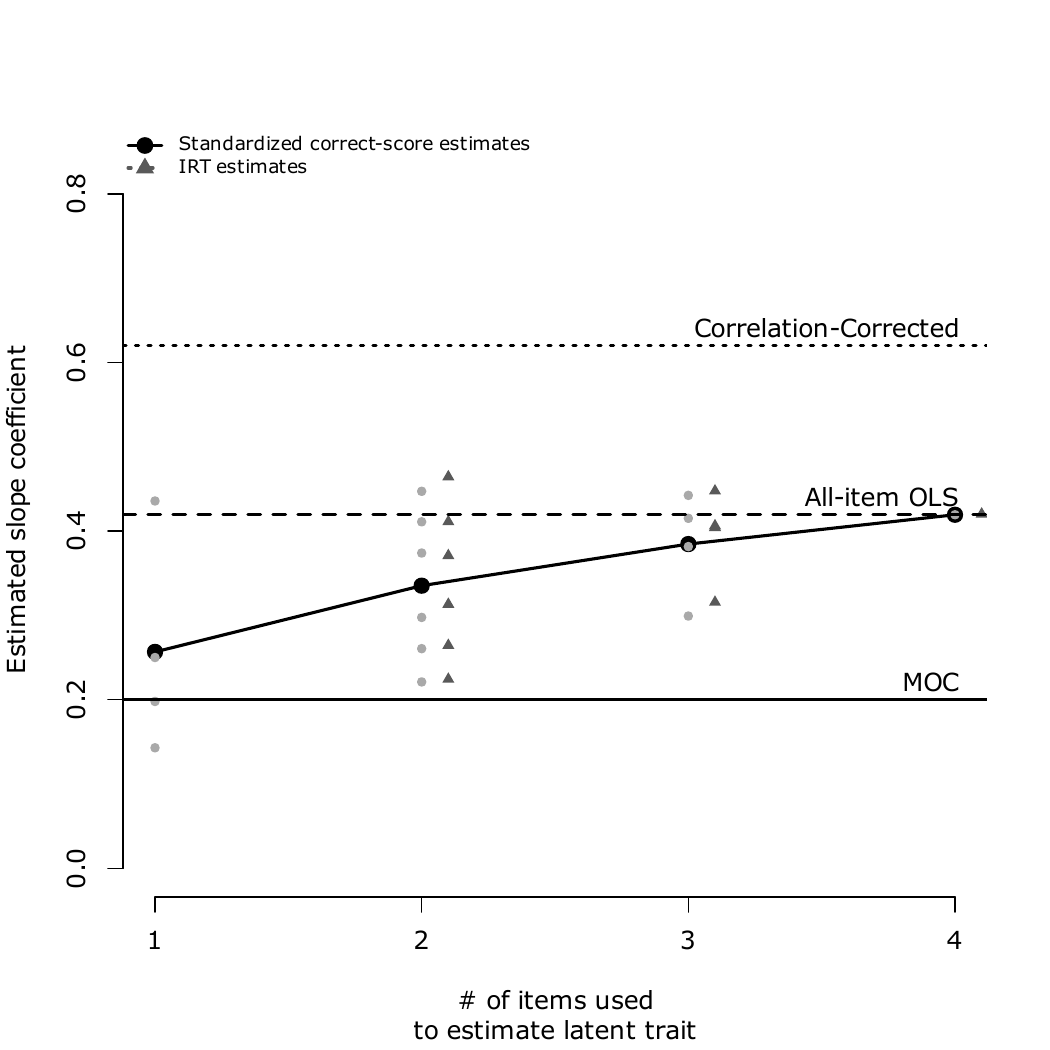}
    \caption{Grey points are uncorrected OLS slope coefficients from regressing duty-to-vote on estimated knowledge, where knowledge is constructed from each nonempty subset of the four indicators. Black dots are the mean slope within each indicator-count. Horizontal lines show the unadjusted OLS estimate using all four items, the split-indicator correlation-corrected estimate, and the MOC estimate. Additional details appear in Appendix~\ref{s:ExPolKnow}.}
    \label{fig:knowledge-voteduty-simple}
  \end{figure}

\subsection{American Politics: Ideology \& Presidential Vote}\label{s:PartisanshipExample}

We here consider an application from American politics, in particular, from the literature on the relationship between latent ideology and partisan vote choice---a relationship that has been extensively studied and shown to be quite strong \citep{jessee2009spatial,tausanovitch2013measuring}. To what extent might measurement error in the estimation of latent ideology lead to over- or under-estimation in estimated relationships with the various strategies previously described?

Using the same design as outlined in Section \ref{s:SimDetails}, we explore the relationship between latent ideology, as measured by 30 questions on the 2016 CCES regarding respondents' policy positions, and vote choice for president. We again vary the number of observations used and policy indicators considered in estimating latent traits, using random sub-samples and computing average estimates. Latent traits are estimated as described above using the expectation-maximization algorithm. 

\begin{figure}[H]
  \centering
\includegraphics[width=0.49\linewidth]{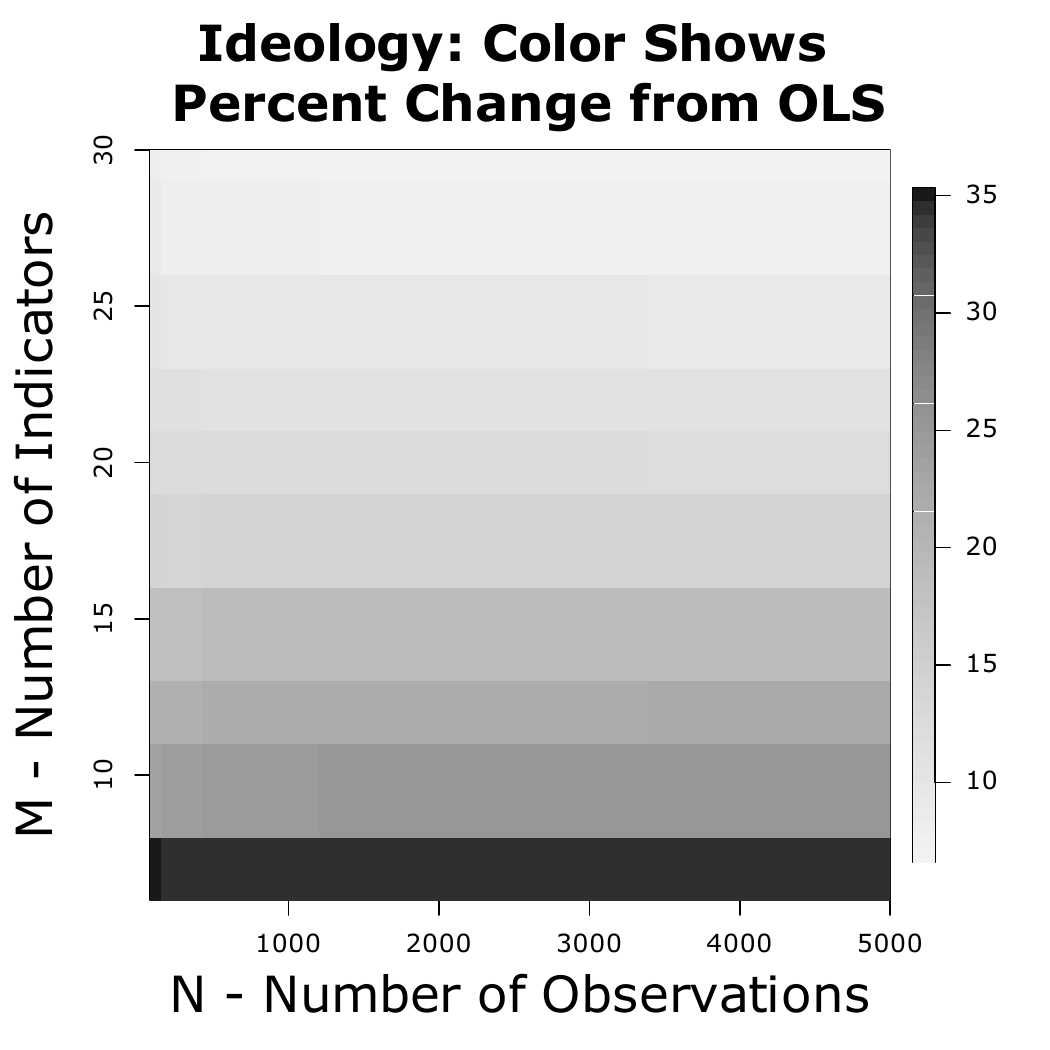}
\includegraphics[width=0.49\linewidth]{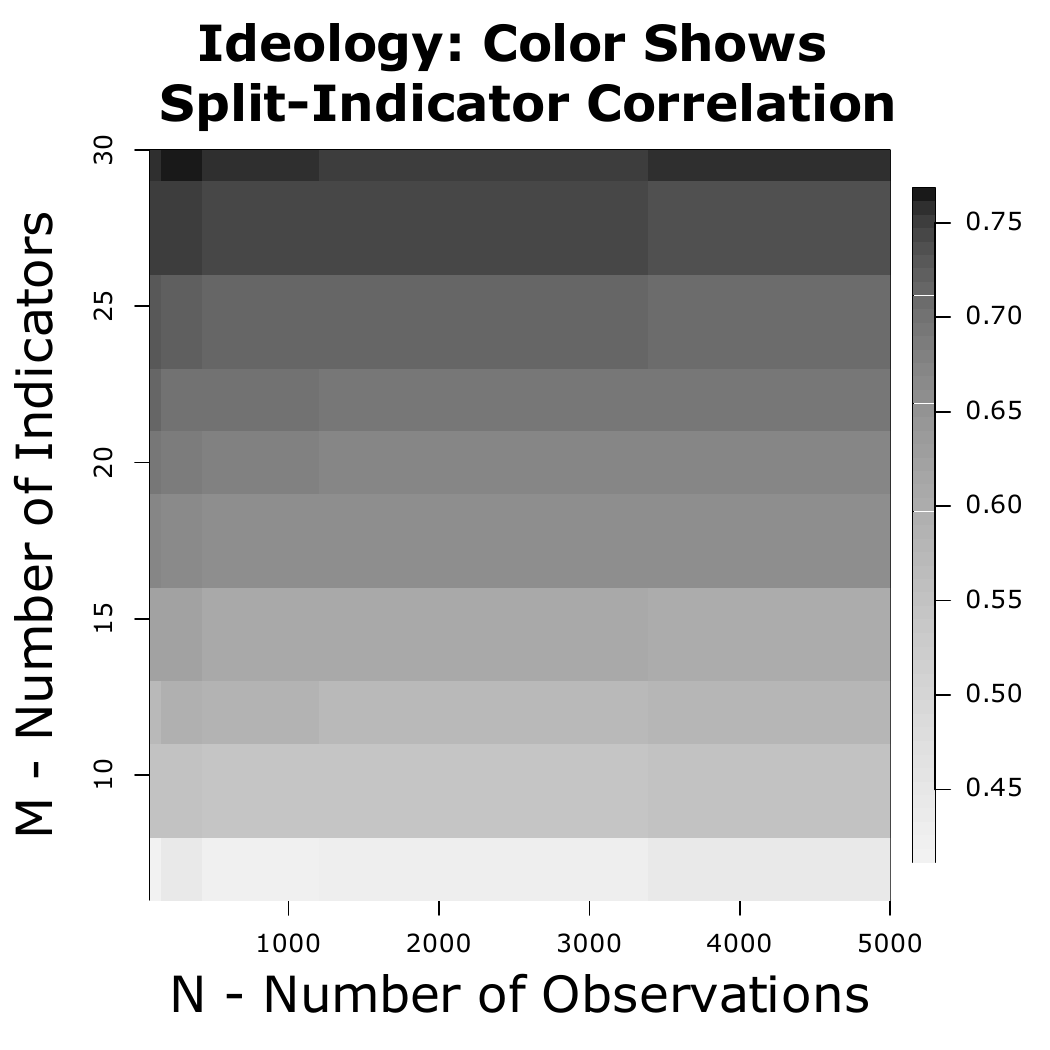}
\caption{
\textbf{Measurement–error amplification for partisan‐ideal‐point predictors.}
Heat maps vary the number of observations ($N$, $x$‑axis) and the number of policy indicators used to estimate the latent trait ($M$, $y$‑axis).
\emph{Left panel}: percentage increase (relative to naïve OLS) in the absolute magnitude of the vote‑share coefficient after applying the split-indicator correction; lighter shades indicate larger upward revisions.
\emph{Right panel}: split-indicator correlations between the two halves of the indicator set, with higher values signaling greater measurement stability.
}
\label{fig:cOLSvOLS_Partisanship_Part1}
\end{figure}

\begin{table}[!htbp] \centering 
  \caption{Percentage change in effect estimate of ideology on vote choice,
                                   corrected OLS vs. OLS. Rows correspond to observation number; columns to indicators used in estimating latent ideology.} 
  \label{tab:cOLSvOLS_ANES-Partisanship_Part1} 
\begin{tabular}{@{\extracolsep{5pt}} ccccccccccc} 
\\[-1.8ex]\hline 
\hline \\[-1.8ex] 
 & $M =$ 6 & 10 & 12 & 14 & 18 & 20 & 22 & 24 & 28 & 30 \\ 
\hline \\[-1.8ex] 
$n =$ 80 & 35.2 & 23.7 & 21.3 & 18.6 & 14.6 & 12.7 & 11.7 & 10.5 & 9.0 & 8.3 \\ 
225 & 34.4 & 24.2 & 21.2 & 18.2 & 14.0 & 12.6 & 11.3 & 9.9 & 8.2 & 7.4 \\ 
632 & 34.5 & 25.0 & 22.0 & 19.0 & 14.2 & 12.5 & 10.9 & 9.7 & 7.8 & 7.0 \\ 
1778 & 34.6 & 25.4 & 22.2 & 19.1 & 14.1 & 12.4 & 10.9 & 9.5 & 7.7 & 6.9 \\ 
5000 & 34.4 & 25.4 & 22.4 & 19.2 & 14.2 & 12.3 & 10.8 & 9.5 & 7.7 & 6.9 \\ 
\hline \\[-1.8ex] 
\end{tabular} 
\end{table}

\subsection{Comparison: Corrected Estimator and Uncorrected IV in Democracy and Growth Application}\label{s:DemGrowth}

We see in Table \ref{tab:cIVvuIV_Freedomhouse_Part1} that, if using the uncorrected IV strategy, we overestimate the effect of democracy on growth relative to the corrected IV strategy using a split-indicator approach. This overestimation is largest when the number of observations and the number of Freedom House sub-indicators used in estimation are small. The overestimation decreases as the number of observations and the number of sub-indicators increase, although again at a faster rate for the number of sub-indicators than for the number of observations. This again highlights the importance of gathering extensive information around the latent trait of interest.

\begin{figure}[ht]
  \centering
\includegraphics[width=0.49\linewidth]{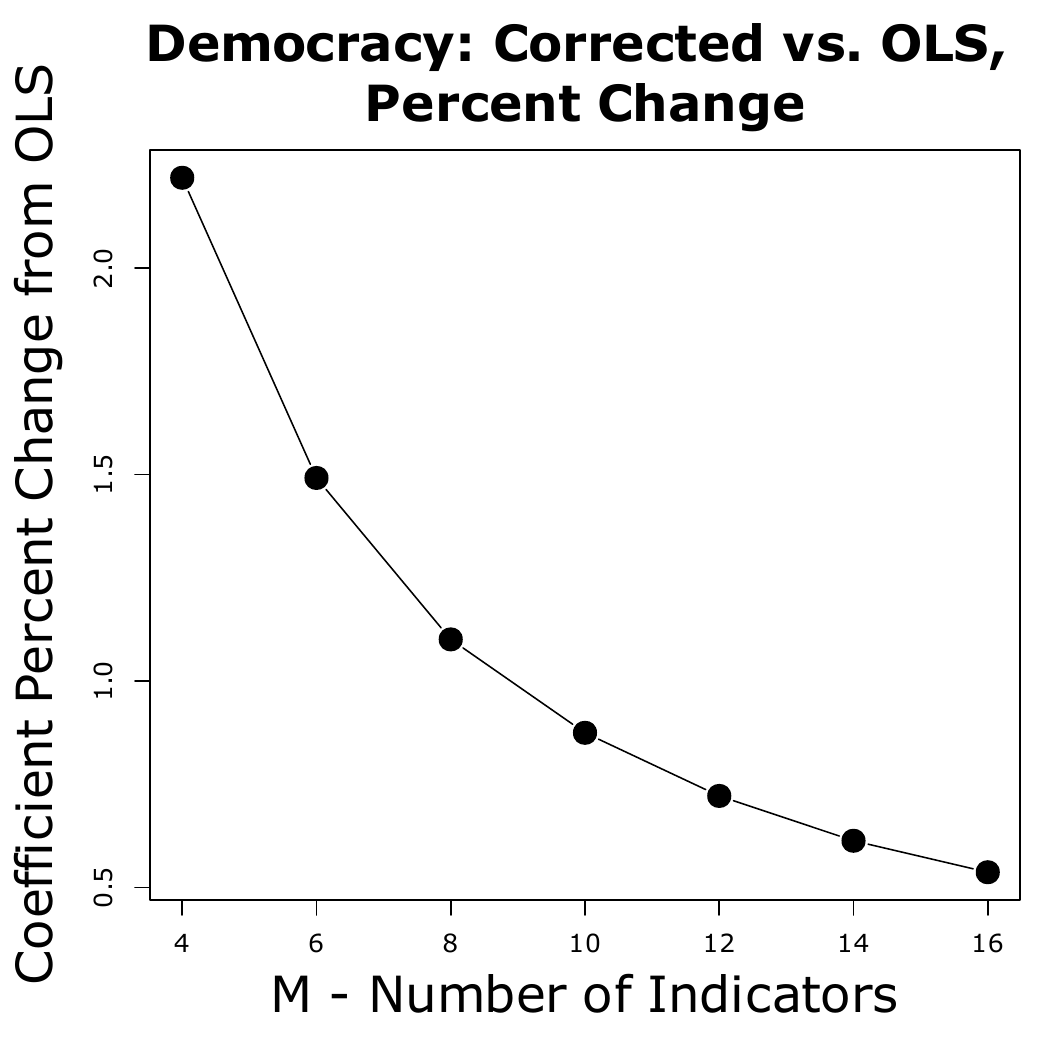}
\includegraphics[width=0.49\linewidth]{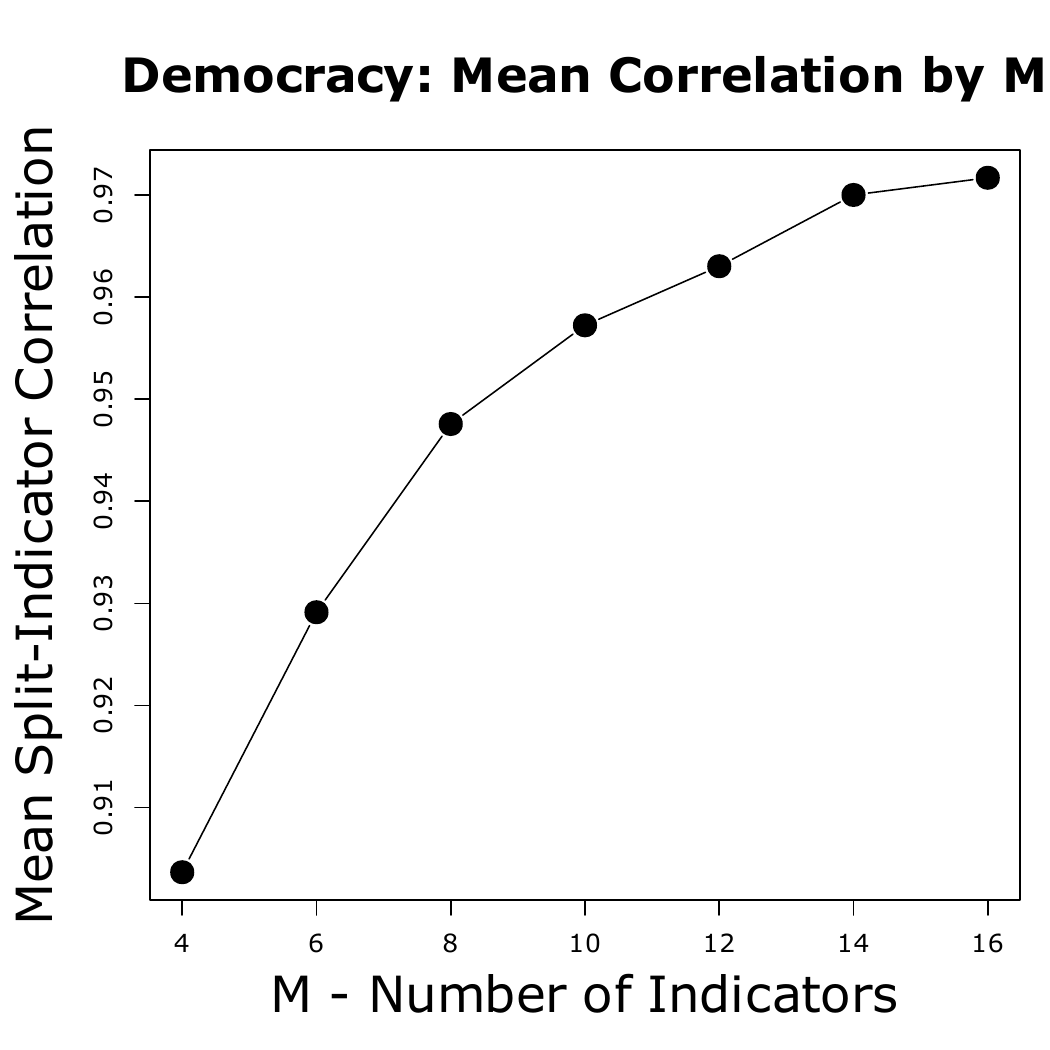}
\caption{
\textbf{Democracy–growth example: Sensitivity to the size of the indicator set.}
\emph{Left panel}: percentage change in the estimated effect of democracy on long‑difference GDP growth after the split-indicator correction, plotted against the number of Freedom‑House sub‑indicators used ($M$).
\emph{Right panel}: mean split-indicator correlations for each value of $M$, showing how measurement error shrinks as more indicators are included.
}
\label{fig:cOLSvOLS_CorSplit_Part1}
\end{figure}

\begin{table}[!htbp] \centering 
  \caption{Percentage change in effect estimate of democracy on growth (long difference), corrected IV vs. uncorrected IV. 
                                      Rows correspond to observation number; columns to indicators used in estimating latent democratic status.} 
  \label{tab:cIVvuIV_Freedomhouse_Part1} 
\begin{tabular}{@{\extracolsep{5pt}} cccccccc} 
\\[-1.8ex]\hline 
\hline \\[-1.8ex] 
 & $M =$ 4 & 6 & 8 & 10 & 12 & 14 & 16 \\ 
\hline \\[-1.8ex] 
$n =$ 184 & -4.8 & -3.5 & -2.6 & -2.1 & -1.7 & -1.5 & -1.3 \\ 
\hline \\[-1.8ex] 
\end{tabular} 
\end{table}

\subsection{Effects of Political Knowledge}\label{s:ExPolKnow}

We here present additional empirical results on the application involving political knowledge.

\begin{table}[!htbp] \centering 
  \caption{Percentage change in effect estimate of knowledge on vote duty, corrected IV vs. OLS.
                                      Rows correspond to observation number; columns to indicators used in estimating latent knowledge.} 
  \label{tab:cIVvOLS_ANES-Knowledge_Part1} 
\begin{tabular}{@{\extracolsep{5pt}} cc} 
\\[-1.8ex]\hline 
\hline \\[-1.8ex] 
 & $M =$ 4 \\ 
\hline \\[-1.8ex] 
$n =$ 80 & 49.6 \\ 
225 & 50.7 \\ 
632 & 48.7 \\ 
1778 & 49.9 \\ 
5000 & 49.8 \\ 
\hline \\[-1.8ex] 
\end{tabular} 
\end{table}

\begin{table}[!htbp] \centering 
  \caption{Percentage change in effect estimate of knowledge on vote duty, corrected IV vs. uncorrected IV.
                                      Rows correspond to observation number; columns to indicators used in estimating latent knowledge.} 
  \label{tab:cIVvuIV_ANES-Knowledge_Part1} 
\begin{tabular}{@{\extracolsep{5pt}} cc} 
\\[-1.8ex]\hline 
\hline \\[-1.8ex] 
 & $M =$ 4 \\ 
\hline \\[-1.8ex] 
$n =$ 80 & -45.0 \\ 
225 & -41.3 \\ 
632 & -41.1 \\ 
1778 & -41.0 \\ 
5000 & -40.8 \\ 
\hline \\[-1.8ex] 
\end{tabular} 
\end{table}

\subsection{Additional Information, Knowledge Example}

We here provide the full text of questions from the American National Election Study (ANES) data used for the political knowledge and duty to vote application. These data were from the pre-release of the 2024 Time Series Study (downloaded 3/11/2025). 

\begin{itemize}
\item[V241612] For how many years is a United States Senator elected - that is, how many years are there in one full term of office for a U.S. Senator? [Respondents specify number of years, 6 is correct answer]

\item[V241613] On which of the following does the U.S. federal government currently spend the least? [Foreign aid (correct), Medicare, National defense, Social Security]

\item[V241614] Do you happen to know which party currently has the most members in the U.S. House of Representatives in Washington? [Democrats, Republicans (correct)]

\item[V241615] Do you happen to know which party currently has the most members in the U.S. Senate? [Democrats (correct), Republicans]

\item[V241218x] V241218x is a summary variable based on V241215, V241216 and V241217.

\item[V241215] Different people feel differently about voting. For some, voting is a duty - they feel they should vote in every election, regardless of their feelings about the candidates and parties. For others, voting is a choice - they feel free to vote or not to vote, depending on their feelings about the candidates and parties. For you personally, is voting mainly a duty, mainly a choice, or neither a duty nor a choice?[Mainly a duty, Mainly a choice, neither a duty nor a choice]

\item[V241216] How strongly do you feel that voting is a duty? [Very strongly, somewhat strongly, or not strongly / Not strongly, somewhat strongly, or very strongly]?

\item[V241217] How strongly do you feel that voting is a choice? [Very strongly, somewhat strongly, or not strongly / Not strongly, somewhat strongly, or very strongly]?

\end{itemize}

\renewcommand{\thefigure}{A.V.\arabic{figure}}
\setcounter{figure}{0}  
\renewcommand{\thetable}{A.V.\arabic{table}}
\setcounter{table}{0}  
\renewcommand{\theequation}{A.V.\arabic{equation}}
\setcounter{equation}{0}
\renewcommand{\thesection}{A.V.\arabic{section}}
\setcounter{section}{1}

\renewcommand{\thesubsection}{A.V.\arabic{subsection}}
\setcounter{subsection}{0}
\renewcommand{\thesubsubsection}{A.V.\arabic{subsection}.\arabic{subsubsection}}
\setcounter{subsubsection}{0}

\section*{Appendix V: Measurement Error and Latent Predictors in the Political Science Literature}
\addcontentsline{toc}{section}{Appendix V: Measurement Error and Latent Predictors in the Political Science Literature}

\begin{figure}[H]
    \centering
\includegraphics[width=0.75\linewidth]{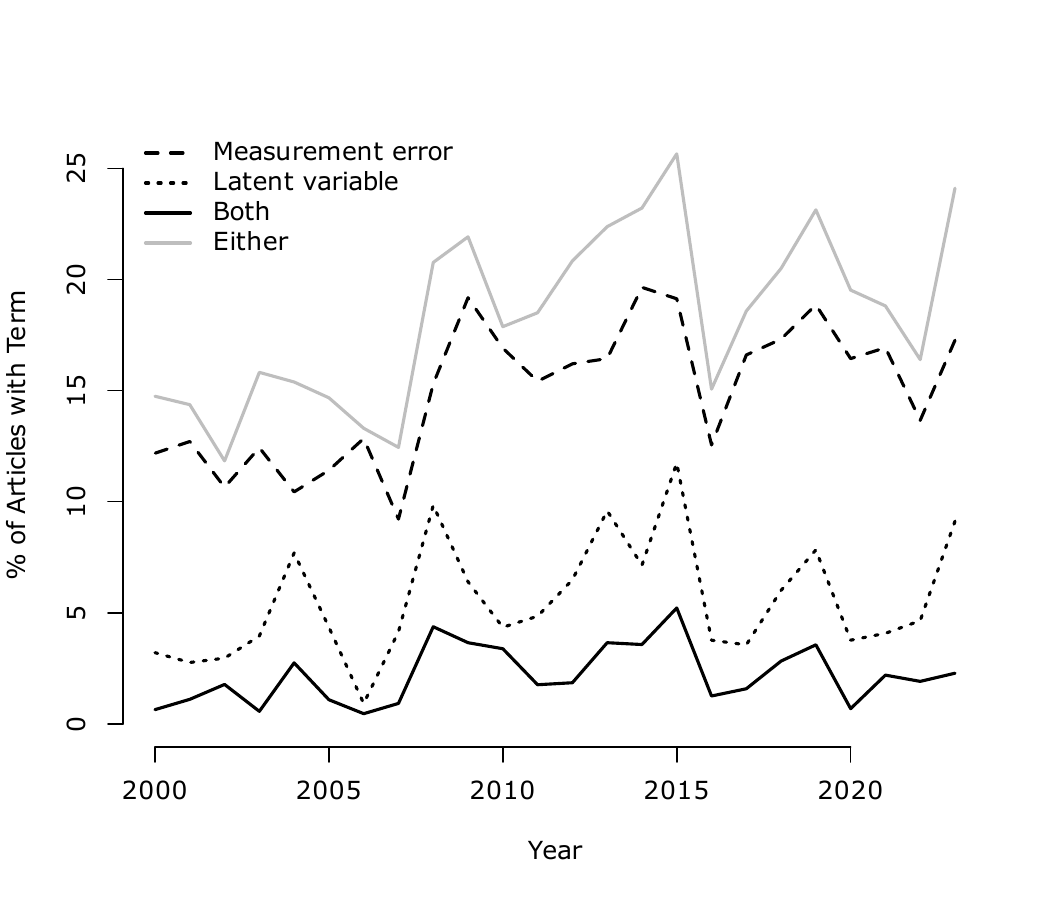}
    \caption{While the overall number of mentions in top political science journals of measurement error and latent variables has been increasing, mentions of the intersection of the two have remained relatively constant in the survey period. The graph shows the number of articles containing each term by year for \it{APSR, AJPS, JOP} and \it{PA} from 2000 through 2023.} 
    \label{fig:Mentions}
\end{figure}

To assess the state of the political science literature regarding measurement error in latent predictors, we conducted an original survey of all articles published in the three top substantive political science journals ({\it APSR, AJPS, JOP}) as well as \emph{Political Analysis (PA)} between 2000 and 2023. We observe in Figure \ref{fig:Mentions} that the overall percentage of articles mentioning latent variables and, especially, measurement error is increasing, while the number of published works mentioning both measurement error and latent variables has remained relatively constant throughout the study period. This survey of recent political science articles demonstrates that while many researchers are aware of and concerned about measurement error, the full extent of its impact on analyses involving latent variables is, perhaps, not yet widely recognized or addressed in applied work.

\renewcommand{\thefigure}{A.VI.\arabic{figure}}
\setcounter{figure}{0}  

\renewcommand{\thetable}{A.VI.\arabic{table}}
\setcounter{table}{0}  

\renewcommand{\theequation}{A.VI.\arabic{equation}}
\setcounter{equation}{0}

\renewcommand{\thesection}{A.VI.\arabic{section}}
\setcounter{section}{1}

\renewcommand{\thesubsection}{A.VI.\arabic{subsection}}
\setcounter{subsection}{0}
\renewcommand{\thesubsubsection}{A.VI.\arabic{subsection}.\arabic{subsubsection}}
\setcounter{subsubsection}{0}

\section*{Appendix VI: A \texttt{lpmec} Package Tutorial}
\addcontentsline{toc}{section}{Appendix VI: A \texttt{lpmec} Package Tutorial}

\noindent \texttt{lpmec} is an R package for latent-predictor measurement error correction under identification restrictions (i.e., when a key regressor is an estimated latent trait from observed indicators, leading to attenuation bias in outcome regressions). Its core workflow repeatedly splits the indicator matrix into independent halves (and can average over many such partitions), estimates the latent score separately in each half (via EM IRT for binary/ordinal items using \texttt{emIRT}, PCA, simple averaging, or a user-supplied estimator), and then applies split-indicator IV and OLS-style corrections that use the cross-half relationship to de-attenuate regression coefficients. The main entry points are \texttt{lpmec\_onerun()} (single partition) and \texttt{lpmec()} (bootstrap + partition aggregation with percentile confidence intervals), with S3 \texttt{print}/\texttt{summary}/\texttt{plot} methods and built-in data (e.g., \texttt{KnowledgeVoteDuty}); the codebase also includes sensitivity reporting via \texttt{sensemakr}'s extreme robustness value.

For Bayesian estimation, \texttt{lpmec} supports MCMC either through an R backend (\texttt{pscl::ideal}) or an optional Python backend (\texttt{numpyro} via \texttt{reticulate}), plus an over-imputation variant using \texttt{Amelia}. The \texttt{mcmc\_joint} option implements a joint Bayesian model that couples an IRT measurement model for the indicators with an outcome model (e.g., $Y_i \sim \mathcal{N}(\beta_0+\beta_X X_i, \sigma)$), so posterior uncertainty in the latent trait propagates directly into inference on the regression slope. When using the NumPyro/JAX backend, this can take advantage of JAX's JIT/XLA compilation (and, optionally, GPU/TPU acceleration) and supports subsampling-based kernels for larger problems.

We average over \texttt{n\_partition} split-indicator partitions:
\begin{verbatim}
# install.packages("devtools")
devtools::install_github(
  "cjerzak/lpmec-software",
  subdir = "lpmec"
)

library(lpmec)
data(KnowledgeVoteDuty)

# Latent error correction method, with partitioning and bootstrap
results <- lpmec::lpmec(
  Y = KnowledgeVoteDuty$voteduty,
  observables = as.matrix(KnowledgeVoteDuty[, 2:5]),
  n_boot = 32L,
  n_partition = 10L,
  estimation_method = "em"
)

print(results)
summary(results)
\end{verbatim}


\renewcommand{\thefigure}{A.VII.\arabic{figure}}
\setcounter{figure}{0}  

\renewcommand{\thetable}{A.VII.\arabic{table}}
\setcounter{table}{0}  

\renewcommand{\theequation}{A.VII.\arabic{equation}}
\setcounter{equation}{0}

\renewcommand{\thesection}{A.VII.\arabic{section}}
\setcounter{section}{1}
\renewcommand{\thesubsection}{A.VII.\arabic{subsection}}
\setcounter{subsection}{0}
\renewcommand{\thesubsubsection}{A.VII.\arabic{subsection}.\arabic{subsubsection}}
\setcounter{subsubsection}{0}

\section*{Appendix VII: Code and Data Availability Statement}
\addcontentsline{toc}{section}{Appendix VII: Code and Data Availability Statement}

Additional data and materials will be made available online. For a code implementation, see 
\begin{itemize}
\item[] \SoftwareURL{}
\end{itemize}
 for open-source software implementing the methods described in the paper. Replication data and scripts will be made available at a Harvard Dataverse Repository. 

\renewcommand{\thefigure}{A.VIII.\arabic{figure}}
\setcounter{figure}{0}
\renewcommand{\thetable}{A.VIII.\arabic{table}}
\setcounter{table}{0}
\renewcommand{\theequation}{A.VIII.\arabic{equation}}
\setcounter{equation}{0}
\renewcommand{\thesection}{A.VIII.\arabic{section}}
\setcounter{section}{1}
\renewcommand{\thesubsection}{A.VIII.\arabic{subsection}}
\setcounter{subsection}{0}
\renewcommand{\thesubsubsection}{A.VIII.\arabic{subsection}.\arabic{subsubsection}}
\setcounter{subsubsection}{0}

\section*{Appendix VIII: Winsorized-Mean Simulation Results}
\addcontentsline{toc}{section}{Appendix VIII: Winsorized-Mean Simulation Results}

\begin{figure}[H]
\centering
\includegraphics[width=0.88\linewidth]{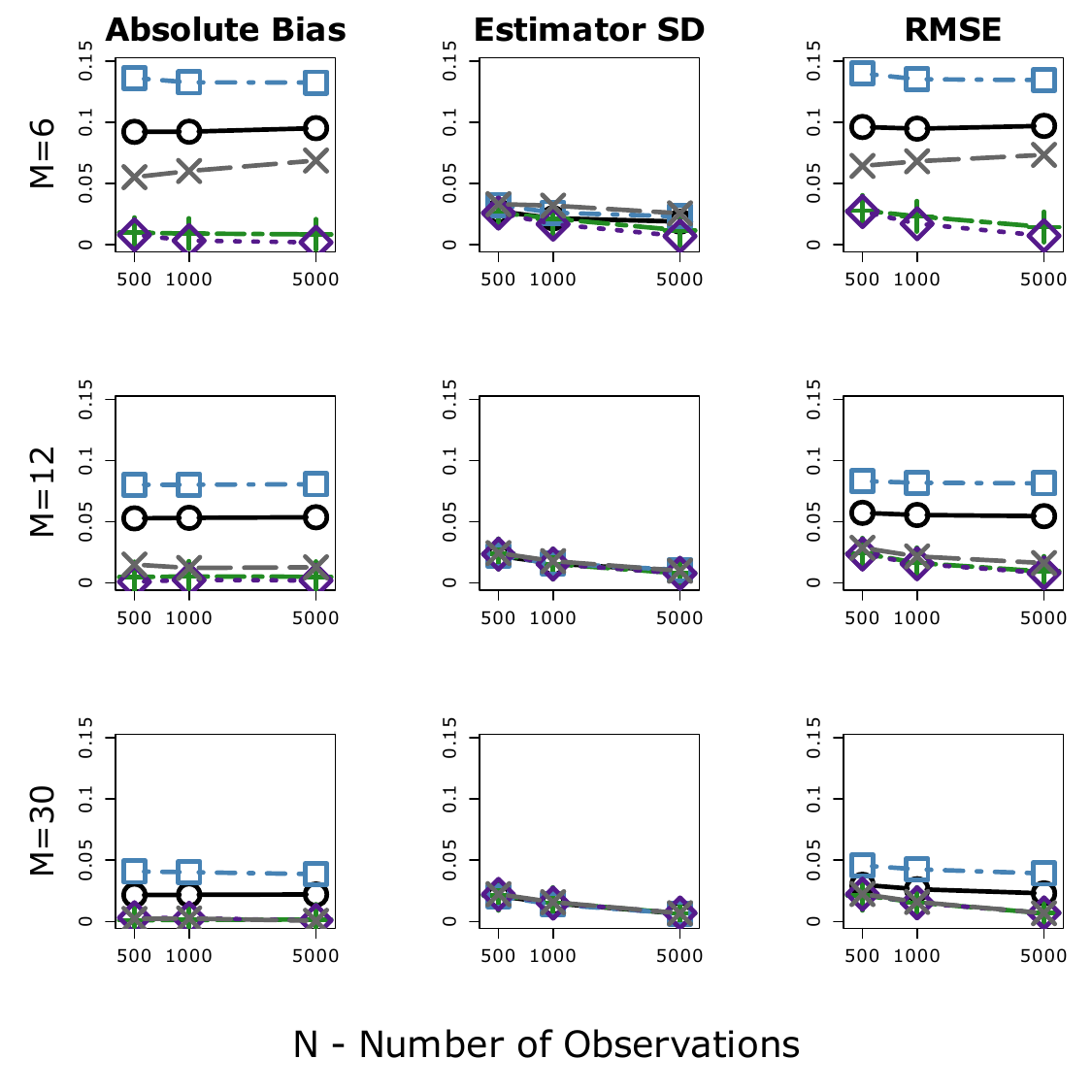}
\par\vspace{0.25em}
\includegraphics[width=0.95\linewidth]{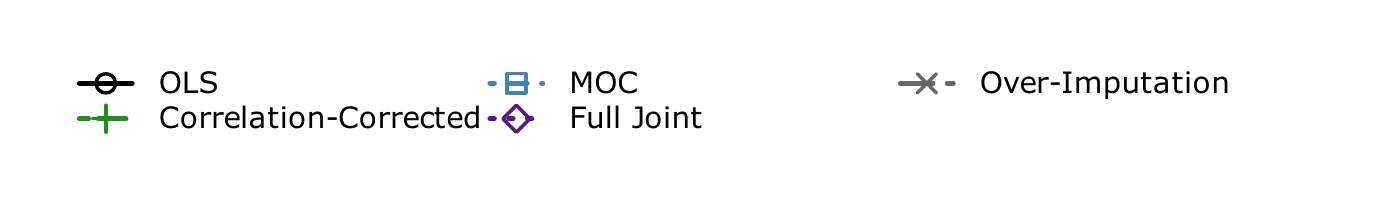}
\caption{Simulation results varying $N$ and $M$ using winsorized-mean aggregation over partitions. Results compare OLS, MOC, over-imputation, full joint estimation, and the split-indicator correlation-corrected estimator; median aggregation remains the baseline specification in the main text.}
\label{fig:CombinedVaryNDWinsorized}
\end{figure}

\begin{figure}[H]
\centering
\includegraphics[width=0.35\linewidth]{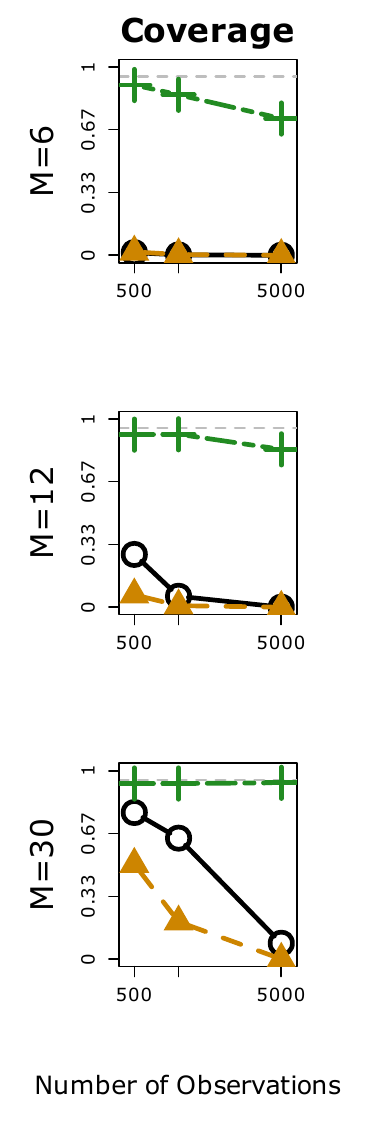}
\par\vspace{0.25em}
\includegraphics[width=0.60\linewidth]{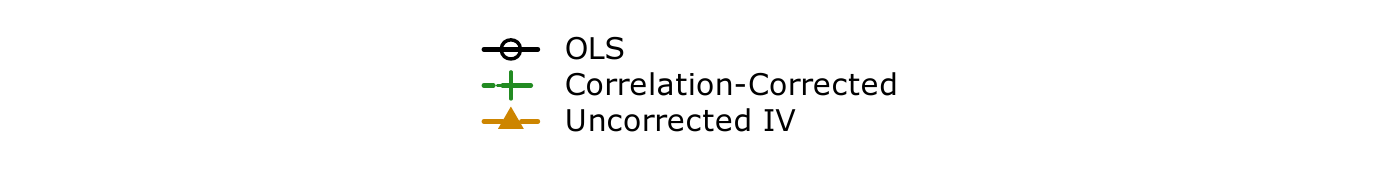}
\caption{Coverage results varying $N$ and $M$ using winsorized-mean aggregation over partitions. Points show empirical 95\% interval coverage; the horizontal reference line marks nominal coverage.}
\label{fig:CoverageCombinedVaryNDWinsorized}
\end{figure}

\begin{figure}[H]
\centering
\includegraphics[width=0.88\linewidth]{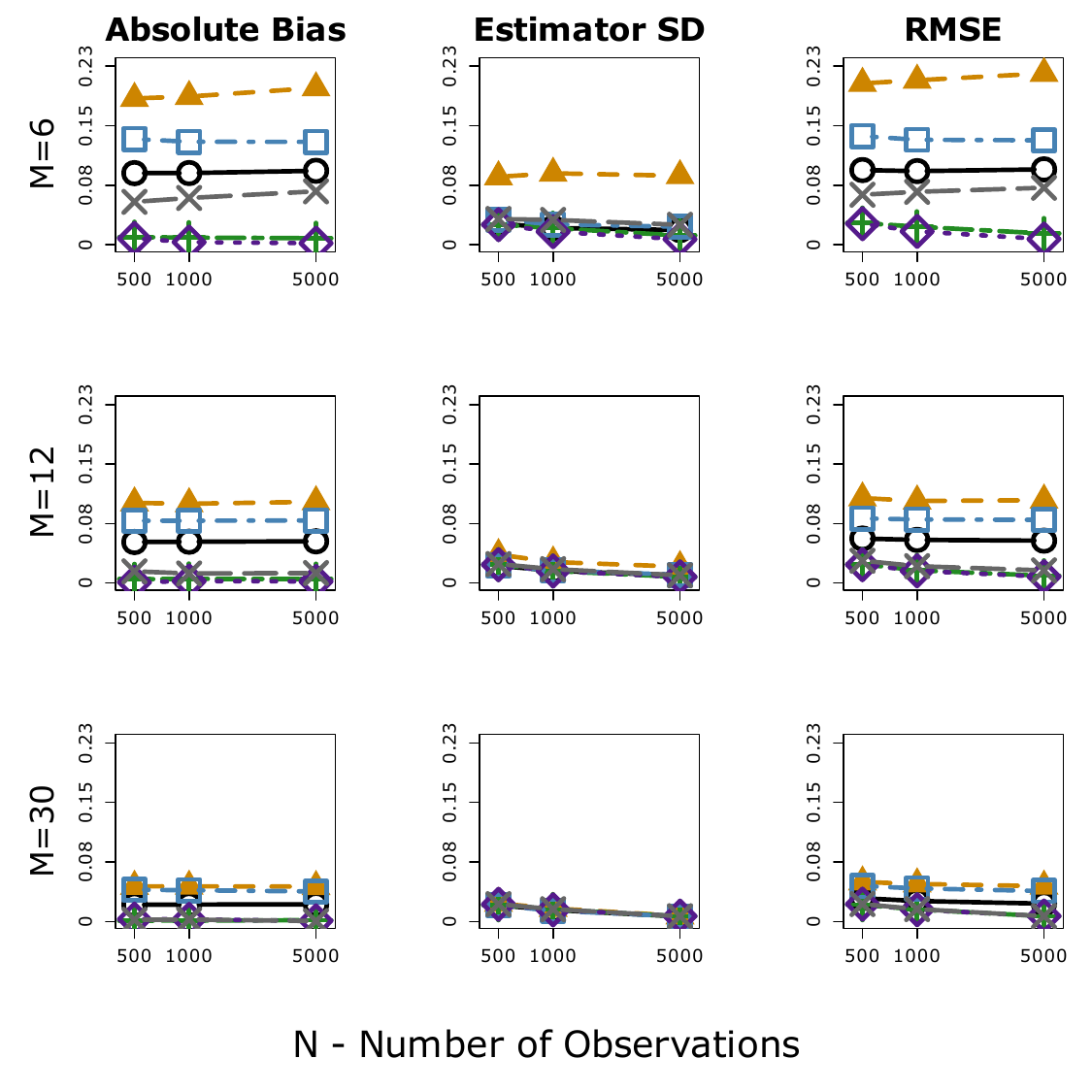}
\par\vspace{0.25em}
\includegraphics[width=0.95\linewidth]{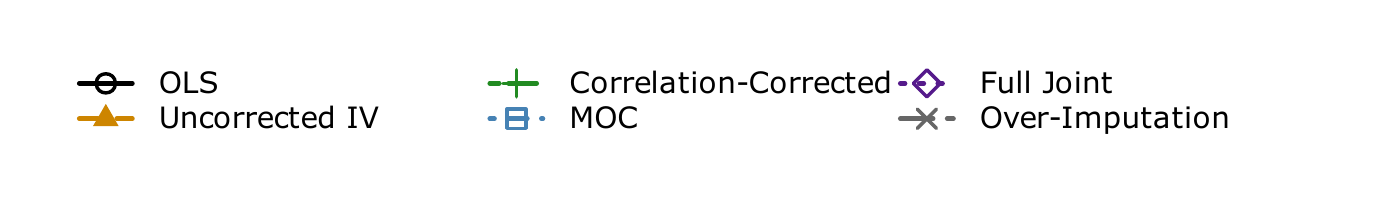}
\caption{Additional estimation strategies using winsorized-mean aggregation over partitions. Results add uncorrected IV to the comparison of OLS, MOC, over-imputation, full joint estimation, and the split-indicator correlation-corrected estimator; median aggregation remains the baseline specification in the main text.}
\label{fig:BayesianComparisonWinsorized}
\end{figure}

Figures~\ref{fig:CombinedVaryNDWinsorized}--\ref{fig:BayesianComparisonWinsorized} repeat the simulation analysis from Appendix~\ref{s:SimDetails} using winsorized-mean aggregation over partitions rather than median aggregation. The median-aggregation results remain the baseline results in the main text and earlier appendix figures.

\end{document}